\documentclass[%
 reprint,
superscriptaddress,
nofootinbib,
 amsmath,amssymb,prd,showpacs
 aps
]{revtex4-2}

\usepackage{graphicx}
\usepackage{dcolumn}
\usepackage{bm}
\usepackage{hyperref}
\usepackage{xcolor} 
\usepackage{braket}
\usepackage[capitalize]{cleveref}
\usepackage[caption=false]{subfig}
\usepackage{enumerate}
\usepackage[normalem]{ulem}
\begin{document}


\title{First T2K measurement of transverse kinematic imbalance in the muon-neutrino charged-current single-$\pi^+$ production channel containing at least one proton}


\newcommand{\INSTHD}{\affiliation{University Autonoma Madrid, Department of Theoretical Physics, 28049 Madrid, Spain}}
\newcommand{\INSTEE}{\affiliation{University of Bern, Albert Einstein Center for Fundamental Physics, Laboratory for High Energy Physics (LHEP), Bern, Switzerland}}
\newcommand{\INSTFE}{\affiliation{Boston University, Department of Physics, Boston, Massachusetts, U.S.A.}}
\newcommand{\INSTGA}{\affiliation{University of California, Irvine, Department of Physics and Astronomy, Irvine, California, U.S.A.}}
\newcommand{\INSTI}{\affiliation{IRFU, CEA, Universit\'e Paris-Saclay, F-91191 Gif-sur-Yvette, France}}
\newcommand{\INSTGB}{\affiliation{University of Colorado at Boulder, Department of Physics, Boulder, Colorado, U.S.A.}}
\newcommand{\INSTFG}{\affiliation{Colorado State University, Department of Physics, Fort Collins, Colorado, U.S.A.}}
\newcommand{\INSTFH}{\affiliation{Duke University, Department of Physics, Durham, North Carolina, U.S.A.}}
\newcommand{\INSTEF}{\affiliation{ETH Zurich, Institute for Particle Physics and Astrophysics, Zurich, Switzerland}}
\newcommand{\INSTIE}{\affiliation{CERN European Organization for Nuclear Research, CH-1211 Genève 23, Switzerland}}
\newcommand{\INSTEG}{\affiliation{University of Geneva, Section de Physique, DPNC, Geneva, Switzerland}}
\newcommand{\INSTHJ}{\affiliation{University of Glasgow, School of Physics and Astronomy, Glasgow, United Kingdom}}
\newcommand{\INSTDG}{\affiliation{H. Niewodniczanski Institute of Nuclear Physics PAN, Cracow, Poland}}
\newcommand{\INSTCB}{\affiliation{High Energy Accelerator Research Organization (KEK), Tsukuba, Ibaraki, Japan}}
\newcommand{\INSTIB}{\affiliation{University of Houston, Department of Physics, Houston, Texas, U.S.A.}}
\newcommand{\INSTED}{\affiliation{Institut de Fisica d'Altes Energies (IFAE) - The Barcelona Institute of Science and Technology, Campus UAB, Bellaterra (Barcelona) Spain}}
\newcommand{\INSTEC}{\affiliation{IFIC (CSIC \& University of Valencia), Valencia, Spain}}
\newcommand{\INSTHH}{\affiliation{Institute For Interdisciplinary Research in Science and Education (IFIRSE), ICISE, Quy Nhon, Vietnam}}
\newcommand{\INSTEI}{\affiliation{Imperial College London, Department of Physics, London, United Kingdom}}
\newcommand{\INSTGF}{\affiliation{INFN Sezione di Bari and Universit\`a e Politecnico di Bari, Dipartimento Interuniversitario di Fisica, Bari, Italy}}
\newcommand{\INSTBE}{\affiliation{INFN Sezione di Napoli and Universit\`a di Napoli, Dipartimento di Fisica, Napoli, Italy}}
\newcommand{\INSTBF}{\affiliation{INFN Sezione di Padova and Universit\`a di Padova, Dipartimento di Fisica, Padova, Italy}}
\newcommand{\INSTBD}{\affiliation{INFN Sezione di Roma and Universit\`a di Roma ``La Sapienza'', Roma, Italy}}
\newcommand{\INSTEB}{\affiliation{Institute for Nuclear Research of the Russian Academy of Sciences, Moscow, Russia}}
\newcommand{\INSTHI}{\affiliation{International Centre of Physics, Institute of Physics (IOP), Vietnam Academy of Science and Technology (VAST), 10 Dao Tan, Ba Dinh, Hanoi, Vietnam}}
\newcommand{\INSTHA}{\affiliation{Kavli Institute for the Physics and Mathematics of the Universe (WPI), The University of Tokyo Institutes for Advanced Study, University of Tokyo, Kashiwa, Chiba, Japan}}
\newcommand{\INSTID}{\affiliation{Keio University, Department of Physics, Kanagawa, Japan}}
\newcommand{\INSTIF}{\affiliation{King's College London, Department of Physics, Strand, London WC2R 2LS, United Kingdom}}
\newcommand{\INSTCC}{\affiliation{Kobe University, Kobe, Japan}}
\newcommand{\INSTCD}{\affiliation{Kyoto University, Department of Physics, Kyoto, Japan}}
\newcommand{\INSTEJ}{\affiliation{Lancaster University, Physics Department, Lancaster, United Kingdom}}
\newcommand{\INSTII}{\affiliation{Lawrence Berkeley National Laboratory, Berkeley, CA 94720, USA}}
\newcommand{\INSTBA}{\affiliation{Ecole Polytechnique, IN2P3-CNRS, Laboratoire Leprince-Ringuet, Palaiseau, France }}
\newcommand{\INSTFC}{\affiliation{University of Liverpool, Department of Physics, Liverpool, United Kingdom}}
\newcommand{\INSTFI}{\affiliation{Louisiana State University, Department of Physics and Astronomy, Baton Rouge, Louisiana, U.S.A.}}
\newcommand{\INSTHB}{\affiliation{Michigan State University, Department of Physics and Astronomy,  East Lansing, Michigan, U.S.A.}}
\newcommand{\INSTCE}{\affiliation{Miyagi University of Education, Department of Physics, Sendai, Japan}}
\newcommand{\INSTDF}{\affiliation{National Centre for Nuclear Research, Warsaw, Poland}}
\newcommand{\INSTFJ}{\affiliation{State University of New York at Stony Brook, Department of Physics and Astronomy, Stony Brook, New York, U.S.A.}}
\newcommand{\INSTGJ}{\affiliation{Okayama University, Department of Physics, Okayama, Japan}}
\newcommand{\INSTCF}{\affiliation{Osaka City University, Department of Physics, Osaka, Japan}}
\newcommand{\INSTGG}{\affiliation{Oxford University, Department of Physics, Oxford, United Kingdom}}
\newcommand{\INSTIC}{\affiliation{University of Pennsylvania, Department of Physics and Astronomy,  Philadelphia, PA, 19104, USA.}}
\newcommand{\INSTGC}{\affiliation{University of Pittsburgh, Department of Physics and Astronomy, Pittsburgh, Pennsylvania, U.S.A.}}
\newcommand{\INSTFA}{\affiliation{Queen Mary University of London, School of Physics and Astronomy, London, United Kingdom}}
\newcommand{\INSTGD}{\affiliation{University of Rochester, Department of Physics and Astronomy, Rochester, New York, U.S.A.}}
\newcommand{\INSTHC}{\affiliation{Royal Holloway University of London, Department of Physics, Egham, Surrey, United Kingdom}}
\newcommand{\INSTBC}{\affiliation{RWTH Aachen University, III. Physikalisches Institut, Aachen, Germany}}
\newcommand{\INSTFB}{\affiliation{University of Sheffield, Department of Physics and Astronomy, Sheffield, United Kingdom}}
\newcommand{\INSTDI}{\affiliation{University of Silesia, Institute of Physics, Katowice, Poland}}
\newcommand{\INSTBB}{\affiliation{Sorbonne Universit\'e, Universit\'e Paris Diderot, CNRS/IN2P3, Laboratoire de Physique Nucl\'eaire et de Hautes Energies (LPNHE), Paris, France}}
\newcommand{\INSTEH}{\affiliation{STFC, Rutherford Appleton Laboratory, Harwell Oxford,  and  Daresbury Laboratory, Warrington, United Kingdom}}
\newcommand{\INSTCH}{\affiliation{University of Tokyo, Department of Physics, Tokyo, Japan}}
\newcommand{\INSTBJ}{\affiliation{University of Tokyo, Institute for Cosmic Ray Research, Kamioka Observatory, Kamioka, Japan}}
\newcommand{\INSTCG}{\affiliation{University of Tokyo, Institute for Cosmic Ray Research, Research Center for Cosmic Neutrinos, Kashiwa, Japan}}
\newcommand{\INSTHF}{\affiliation{Tokyo Institute of Technology, Department of Physics, Tokyo, Japan}}
\newcommand{\INSTGI}{\affiliation{Tokyo Metropolitan University, Department of Physics, Tokyo, Japan}}
\newcommand{\INSTHG}{\affiliation{Tokyo University of Science, Faculty of Science and Technology, Department of Physics, Noda, Chiba, Japan}}
\newcommand{\INSTF}{\affiliation{University of Toronto, Department of Physics, Toronto, Ontario, Canada}}
\newcommand{\INSTB}{\affiliation{TRIUMF, Vancouver, British Columbia, Canada}}
\newcommand{\INSTDJ}{\affiliation{University of Warsaw, Faculty of Physics, Warsaw, Poland}}
\newcommand{\INSTDH}{\affiliation{Warsaw University of Technology, Institute of Radioelectronics and Multimedia Technology, Warsaw, Poland}}
\newcommand{\INSTFD}{\affiliation{University of Warwick, Department of Physics, Coventry, United Kingdom}}
\newcommand{\INSTGH}{\affiliation{University of Winnipeg, Department of Physics, Winnipeg, Manitoba, Canada}}
\newcommand{\INSTEA}{\affiliation{Wroclaw University, Faculty of Physics and Astronomy, Wroclaw, Poland}}
\newcommand{\INSTHE}{\affiliation{Yokohama National University, Department of Physics, Yokohama, Japan}}
\newcommand{\INSTH}{\affiliation{York University, Department of Physics and Astronomy, Toronto, Ontario, Canada}}

\INSTHD
\INSTEE
\INSTFE
\INSTGA
\INSTI
\INSTGB
\INSTFG
\INSTFH
\INSTEF
\INSTIE
\INSTEG
\INSTHJ
\INSTDG
\INSTCB
\INSTIB
\INSTED
\INSTEC
\INSTHH
\INSTEI
\INSTGF
\INSTBE
\INSTBF
\INSTBD
\INSTEB
\INSTHI
\INSTHA
\INSTID
\INSTIF
\INSTCC
\INSTCD
\INSTEJ
\INSTII
\INSTBA
\INSTFC
\INSTFI
\INSTHB
\INSTCE
\INSTDF
\INSTFJ
\INSTGJ
\INSTCF
\INSTGG
\INSTIC
\INSTGC
\INSTFA
\INSTGD
\INSTHC
\INSTBC
\INSTFB
\INSTDI
\INSTBB
\INSTEH
\INSTCH
\INSTBJ
\INSTCG
\INSTHF
\INSTGI
\INSTHG
\INSTF
\INSTB
\INSTDJ
\INSTDH
\INSTFD
\INSTGH
\INSTEA
\INSTHE
\INSTH

\author{K.\,Abe}\INSTBJ
\author{N.\,Akhlaq}\INSTFA
\author{R.\,Akutsu}\INSTHA
\author{A.\,Ali}\INSTCD
\author{C.\,Alt}\INSTEF
\author{C.\,Andreopoulos}\INSTEH\INSTFC
\author{M.\,Antonova}\INSTEC
\author{S.\,Aoki}\INSTCC
\author{T.\,Arihara}\INSTGI
\author{Y.\,Asada}\INSTHE
% ...

\collaboration{The T2K Collaboration}\noaffiliation

\date{\today}

\begin{abstract}
This paper reports the first T2K measurement of the transverse kinematic imbalance in the single-$\pi^+$ production channel of neutrino interactions. We measure the differential cross sections in the muon-neutrino charged-current interaction on hydrocarbon with a single $\pi^+$ and at least one proton in the final state, at the ND280 off-axis near detector of the T2K experiment. 
The extracted cross sections are compared to the predictions from different neutrino-nucleus interaction event generators. 
Overall, the results show a preference for models which have a more realistic treatment of nuclear medium effects including the initial nuclear state and final-state interactions.

\end{abstract}

\maketitle


\section{Introduction}
In recent years, neutrino oscillation measurements have reached unprecedented precision~\cite{PhysRevLett.124.161802,Abe:2017aap,PhysRevLett.123.151803,PhysRevLett.121.241805,DoubleChooz:2019qbj,PhysRevD.98.012002,PhysRevLett.120.071801}. The next generation of long-baseline (LBL) neutrino oscillation experiments, such as DUNE~\cite{Abi:2020qib} and Hyper-Kamiokande~\cite{Abe:2015zbg}, aim to measure important neutrino properties such as the CP-violating phase and mass ordering~\cite{10.1143/PTP.28.870,Pontecorvo:1967fh}. This requires unprecedented constraints on the neutrino flux, neutrino cross sections and interaction model, and detector response. 
Amongst all the systematic uncertainties, the limited knowledge of neutrino-nucleus interactions, especially those related to nuclear medium effects, is particularly concerning because it can cause biases in event classification and neutrino energy reconstruction. In the latest T2K oscillation analysis~\cite{10.1038/s41586-020-2177-0}, the uncertainty in nucleon removal energy in charged current quasielastic (CCQE) interactions is the dominant systematic component. In order to reduce its value, a more refined analysis is necessary to avoid potential biases in the next measurement of $\Delta m^2_{32}$.

In the range of energies of current LBL experiments, neutrinos interact predominantly with nucleons. The initial state nucleon can be described by Fermi motion together with nucleon-nucleon correlations in a mean field potential. After a neutrino interacts with a nucleon, the residual nucleus may be left in a simple one-particle-one-hole (1p1h) excited state, or collective 1p1h excitations described by random phase approximations (RPA)~\cite{SINGH1992587,GIL1997543,PhysRevC.70.055503,PhysRevC.73.025504,PhysRevC.80.065501}. It is also possible to have two-particle-two-hole (2p2h) excitations due to meson-exchange currents (MEC) or short-range correlations~\cite{PhysRevC.80.065501,DELORME1985263,MARTEAU200076,PhysRevC.81.045502,PhysRevC.83.045501,NIEVES201272,PhysRevC.84.055502}. However, in most generators, these correlations are only implemented in the  quasielastic (QE) channel,  not for the resonant production (RES) or deep inelastic scattering (DIS) channels.

Moreover, after the primary neutrino-nucleon interaction, the outgoing hadrons have to propagate through the nuclear remnant before they can be detected. Final-state interactions (FSI) may cause energy dissipation and hadron absorption, or conversely induce the emission of additional hadrons. 
As a result, FSI can change the final-state topology of a neutrino-nucleon interaction, making the identification of primary neutrino-nucleon interaction and the measurement of primary hadronic kinematics difficult. Neutrino cross sections are often measured in terms of experimentally accessible final-state topologies, e.g. in charged-current (CC) interactions, the CC0$\pi$ topology has only one charged lepton, any number of nucleons and no other particles; the CC1$\pi^+$ topology has only one charged lepton, one $\pi^+$, any number of nucleons and no other particles.

To achieve the designed sensitivity of future LBL experiments, nuclear effects have to be modelled accurately and consistently amongst all interaction channels. Experimental studies probing nuclear effects in carbon, through the measurement of transverse kinematic imbalance (TKI) in CC interactions~\cite{Lu:2015hea,Lu:2015tcr}, have been performed in T2K~\cite{Abe:2018pwo} and MINER$\nu$A~\cite{Lu:2018stk,Cai:2019hpx,Coplowe:2020yea}. TKI  explores the lepton-hadron correlations on the plane that is transverse to the initial neutrino direction and helps precisely identify intranuclear dynamics~\cite{Lu:2015tcr, Furmanski:2016wqo, Abe:2018pwo, Dolan:2018sbb, Lu:2018stk, Dolan:2018zye, Lu:2019nmf, Harewood:2019rzy, Cai:2019jzk, Cai:2019hpx, Coplowe:2020yea, Bourguille:2020bvw}, or the absence thereof~\cite{Lu:2015hea, Duyang:2018lpe, Duyang:2019prb, Munteanu:2019llq, Hamacher-Baumann:2020ogq}, in neutrino-nucleus interactions. These measurements, in particular, either focused on final-state topologies without any pions, or final-state topologies with at least one neutral pion. These studies suggest that modelling nuclear effects with Fermi gas initial state models is insufficient, but more data is needed to draw solid conclusions.

In this paper, we describe the first measurement of the $\nu_\mu$ cross section on hydrocarbon as a function of TKI variables in CC production of exactly one $\pi^+$ and no other mesons, and at least one proton. We introduce TKI in \cref{sec:tki} and the T2K experiment in \cref{sec:t2k}. The event simulation and event selection of the analysis are described in \cref{sec:simulation} and \cref{sec:selection} respectively. Then, the analysis procedure is explained in \cref{sec:analysis}, followed by the interpretation of results in \cref{sec:results}. We conclude in \cref{sec:conclusion}.

\section{Observables}\label{sec:tki}
In a $\nu_{\mu}$ CC RES $\pi^+$ interaction on a free proton p, 
\begin{equation}\label{eq:res_reaction}
    \nu_\mu+\textrm{p}\rightarrow \mu^-+\pi^++\textrm{p},
\end{equation}
a $\nu_{\mu}$ interacts with an initial-state p to produce a final-state $\mu^-$, $\pi^+$ and p.
This is the most important channel that produces $\pi^+$ with the T2K neutrino beam which is narrowly peaked at 0.6 GeV. 
However, in most neutrino experiments, the target involves some nucleus, A, heavier than hydrogen. In general, a  $\nu_{\mu}$ CC1$\pi^+$ interaction with at least one proton in the final-state can be written as
\begin{equation}\label{eq:cc1pi_reaction}
    \nu_\mu+\textrm{A}\rightarrow \mu^-+\pi^++\textrm{p}+\textrm{A}',
\end{equation}
where $\textrm{A}'$ is the final-state hadronic system consisting of the nuclear remnant and other possible knocked-out nucleons. Apart from the RES interaction in \cref{eq:res_reaction}, this topology also includes DIS interactions where multiple pions are produced and some are subsequently absorbed through FSI, leaving only one $\pi^+$ visible in the detector. Alternatively, CCQE interactions can be included in this topology when an additional $\pi^+$ is produced through FSI. The kinematics of the $\mu^-$, $\pi^+$ and p tracks are used to construct the TKI. If there is more than one proton observed in the final state, only the highest momentum one is considered.

The set of three TKI variables, $\delta p_{TT}$, $p_N$ and $\delta\alpha_T$, were first introduced in Refs. \cite{Lu:2015hea,Lu:2015tcr,Furmanski:2016wqo,Lu:2019nmf}.
These observables are designed to characterize the nuclear effects that are most relevant to oscillation experiments: the initial nuclear state, such as the Fermi motion of initial state nucleon and the nucleon removal energy, and the FSI of outgoing hadrons.
The term ``transverse" refers to the fact that all these observables are closely related to the transverse momentum component $\vec{p}^{\textrm{ }i}_T$ (with respect to the incoming neutrino direction) of the final-state particle $i$. In this analysis,  the relevant transverse momenta are  the transverse momenta of the muon, $\vec{p}^{\textrm{ }\mu}_T$, pion, $\vec{p}^{\textrm{ }\pi}_T$, and proton, $\vec{p}^{\textrm{ p}}_T$. 

\begin{figure}
\centering
    \subfloat[$\delta p_{TT}=p^\pi_{TT}+p^\textrm{p}_{TT}$.] {\includegraphics[width=0.7\linewidth]{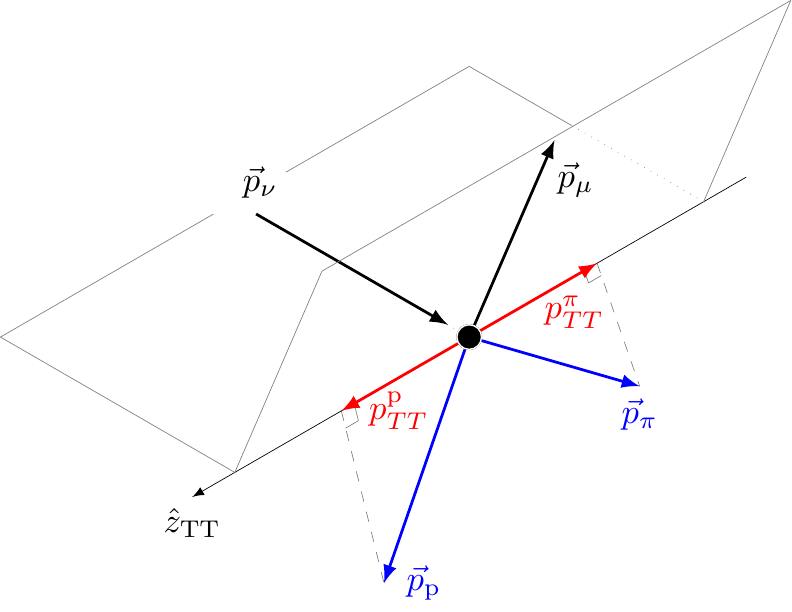}\label{fig:dptt_fig}}
    \hfill
    \subfloat[$\delta\vec{p}_T$ and $\delta\alpha_{T}$.] {\includegraphics[width=0.7\linewidth]{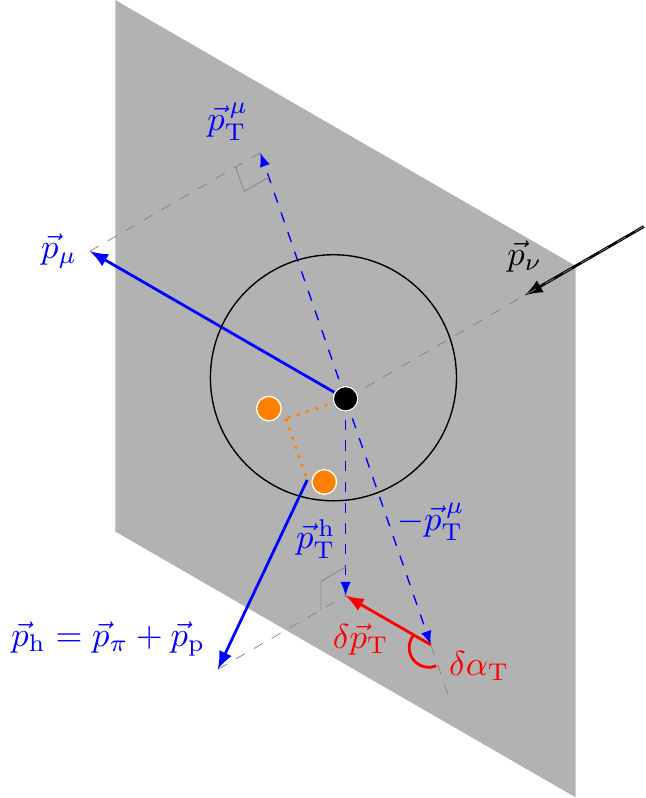}\label{fig:pN_dat_fig}}\hfill
\caption{Schematic illustration of the TKI variables. The total momentum of particle $i$ is given by $\vec{p}_i$, while its transverse component with respect to the neutrino direction is represented by $\vec{p}_T^{\textrm{ }i}$. 
In (b), the black circle represents the initial nucleon; the gray plane shows the transverse plane; the orange circles and dashed lines indicate possible FSI experienced by the outgoing hadrons. Figures adapted from Refs.~\cite{Lu:2015tcr,Coplowe:2020yea}.} \label{fig:tki_fig}
\end{figure}

The first observable $\delta p_{TT}$ is the double-transverse momentum imbalance~\cite{Lu:2015hea}, illustrated in \cref{fig:dptt_fig}. A double-transverse axis is defined as
\begin{equation}
    \hat{z}_{TT}\equiv\frac{\vec{p}_{\nu}\times\vec{p}_{\mu}}{|\vec{p}_{\nu}\times\vec{p}_{\mu}|},
\end{equation}
and the pion and proton momenta are projected onto this axis:
\begin{equation}
    \begin{split}
        p^\pi_{TT}&=\hat{z}_{TT}\cdot\vec{p}_\pi, \\
        p^\textrm{p}_{TT}&=\hat{z}_{TT}\cdot\vec{p}_\textrm{p}.
    \end{split}
\end{equation}
The imbalance is defined as
\begin{equation}
    \delta p_{TT}=p^\pi_{TT}+p^\textrm{p}_{TT}.
\end{equation}
In the absence of nuclear effects, $\delta p_{TT}=0$ is expected due to momentum conservation. Inside a nuclear medium, an imbalance is caused by the initial state of the bound nucleon and the FSI experienced by the outgoing pion and proton. 

The second observable $p_N$ is the initial nucleon momentum. Assuming the target nucleus is at rest and there are no FSI, $p_N$ can be computed following the steps in Ref.~\cite{Lu:2019nmf}. The transverse component of $p_N$ is equal to $\delta\vec{p}_T$ which is the sum of the transverse momenta~\cite{Lu:2015tcr} (\cref{fig:pN_dat_fig}):
\begin{equation}
    \delta\vec{p}_T=\vec{p}^{\textrm{ }\mu}_T+\vec{p}^{\textrm{ }\pi}_T+\vec{p}^\textrm{ p}_T. \label{eq:dpt}
\end{equation}
The longitudinal component of $p_N$ is given by~\cite{Furmanski:2016wqo}
\begin{equation}
\begin{split}
    p_L=&\frac{1}{2}(M_\textrm{A}+p_L^\mu+p_L^\pi+p_L^\textrm{p}-E_\mu-E_\pi-E_\textrm{p})\\&-\frac{1}{2}\frac{\delta p_T^2+M_{\textrm{A}'}^2}{M_\textrm{A}+p_L^\mu+p_L^\pi+p_L^\textrm{p}-E_\mu-E_\pi-E_\textrm{p}}, \label{eq:dpl}
\end{split}
\end{equation}
where $p_L^i$ and $E_i$ are the longitudinal momentum and the energy of the final-state particles. The target nucleus mass $M_\textrm{A}$ and the residual nucleus mass $M_{\textrm{A}'}$ are related by 
\begin{equation}
    M_{\textrm{A}'}=M_\textrm{A}-M_\textrm{p}+\braket{\epsilon}_\textrm{p}, \label{eq:ma}
\end{equation}
where $M_\textrm{p}$ is the proton mass, and $\braket{\epsilon}_\textrm{p}=26.1$ MeV~\cite{Bodek:2018lmc} is the proton mean excitation energy for carbon. The total initial nucleon momentum $p_N$ is given by~\cite{Furmanski:2016wqo}
\begin{equation}
    p_N=\sqrt{\delta p_T^2+p_L^2}, \label{eq:pn}
\end{equation}
which probes the Fermi motion inside the nucleus. Smearing by FSI can shift the peak position of $p_N$, and cause a long tail in the region of large imbalance (similarly for $\delta p_{TT}$).

The third observable $\delta\alpha_T$ is the transverse boosting angle~\cite{Lu:2015tcr}:
\begin{equation}
    \delta\alpha_T=\arccos\left(\frac{-\vec{p}_T^{\textrm{ }\mu}\cdot\delta\vec{p}_T}{p_T^{\mu}\delta p_T}\right),
\end{equation}
as illustrated in \cref{fig:pN_dat_fig}. This observable quantifies whether the hadronic system is accelerated or decelerated by nuclear effects. 
Without FSI, the isotropic Fermi motion of the initial-state nucleon would produce a rather flat $\delta\alpha_T$ distribution. 
However, FSI usually slows down the outgoing hadrons, making $\delta\alpha_T>90^{\text{o}}$. Therefore, the strength of FSI can be inferred from the shape of $\delta\alpha_T$.

In the case where there are multiple nucleons emitted, these nucleons are not included in the above calculation and very likely result in a large imbalance in all the TKI variables.

\section{The T2K experiment}\label{sec:t2k}
The Tokai-to-Kamioka (T2K) experiment~\cite{ABE2011106} is a LBL accelerator-based neutrino experiment measuring oscillations with a $\nu_\mu$ ($\bar{\nu}_\mu$) beam. The neutrino beam is produced at the Japan Proton Accelerator Research Complex (J-PARC) which is located on the East coast of Japan in T$\bar{\textrm{o}}$kai, Ibaraki. The neutrino beam is discussed in more detail in \cref{subsec:beam}. 
J-PARC is also home to a suite of near detectors used to measure the properties of the unoscillated beam. 

The near detector complex is located at 280 m from the neutrino beam production target and consists of several detectors. INGRID~\cite{ABE2012211} is an on-axis detector consisting of an array of 16 iron/scintillator modules, which precisely measures the beam direction and intensity. The detector of primary interest for this analysis is the Near Detector at $280$ m (ND280) which is placed $2.5^{\circ}$ away from the beam axis and measures neutrino interactions for the off-axis flux. It is discussed in more detail in \cref{subsec:ND280}. The WAGASCI~\cite{Kin:2017way} and BabyMIND~\cite{Antonova:2017thk} detectors are located in the same near detector complex but are situated $1.5^{\circ}$ off-axis. 

The far detector Super-Kamiokande~\cite{FUKUDA2003418} is a 50~kt water Cherenkov detector located at a distance of 295 km away from the J-PARC facility on the West coast of Japan in Hida, Gifu. Super-Kamiokande is on the same off-axis angle as ND280. Neutrino CC interaction events can be classified into $\nu_\mu$ and $\nu_e$ like, according to the shape of Cherenkov rings of the outgoing leptons.

\subsection{Neutrino Beam}
\label{subsec:beam}
The J-PARC facility utilizes a $30$ GeV proton beam as the primary beamline. A proton spill consists of eight bunches spaced $580$~ns apart and is produced every $2.48$~s. The beam power has increased over time and reached $520$~kW during the latest data-taking period in 2019. To produce a neutrino beam, the proton beam is collided with a $91.4$ cm graphite target to produce a secondary beam which is primarily composed of pions and kaons. Three magnetic horns are used to focus positively (negatively) charged hadrons which then decay to produce a beam dominated by $\nu_{\mu}$ ($\bar{\nu}_{\mu}$). The magnetic horns are operated with a current of $250$ kA ($-250$ kA) to produce a $\nu_{\mu}$ ($\bar{\nu}_{\mu}$) beam. The data taken while producing a $\nu_\mu$ ($\bar{\nu}_\mu$) beam is qualified as neutrino-mode (antineutrino-mode). 
The focused beam of hadrons then enters a helium-filled, 96~m long decay volume where they decay to produce neutrinos. At the end of the decay volume there is a beam dump and, behind this, a muon monitor~\cite{MATSUOKA2010385,MATSUOKA2010591} which is used to monitor the stability of the secondary beam. 
   
\begin{figure}[h]
\centering
    \includegraphics[width=1.0\linewidth]{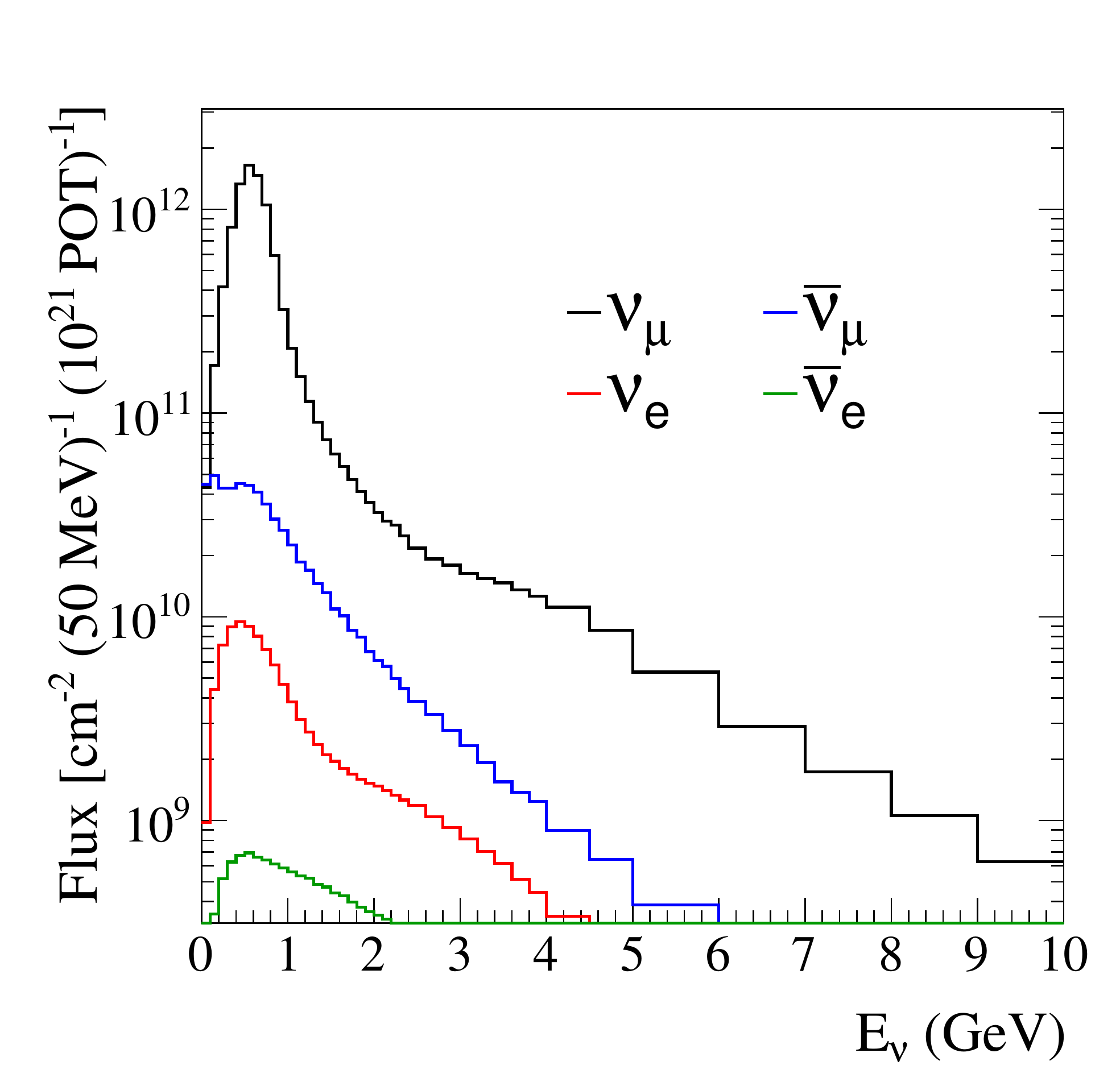}
    \caption{The flux prediction for ND280 in neutrino-mode is shown as well as the contributions from different neutrino flavours. } \label{fig:neutrino_fluxes}
\end{figure}

The neutrino beams are made up of $\nu_{\mu}$, $\bar{\nu}_{\mu}$, $\nu_{e}$ and $\bar{\nu}_{e}$ components. The neutrino flux predictions and the different flavour components at ND280 are shown in \cref{fig:neutrino_fluxes}~\cite{Abe:2012av}. The off-axis configuration allows a narrow energy spectrum with a peak energy of around 0.6 GeV. 

\subsection{The off-axis Near Detector}
\label{subsec:ND280}
\begin{figure}[h]
\includegraphics[width=\linewidth]{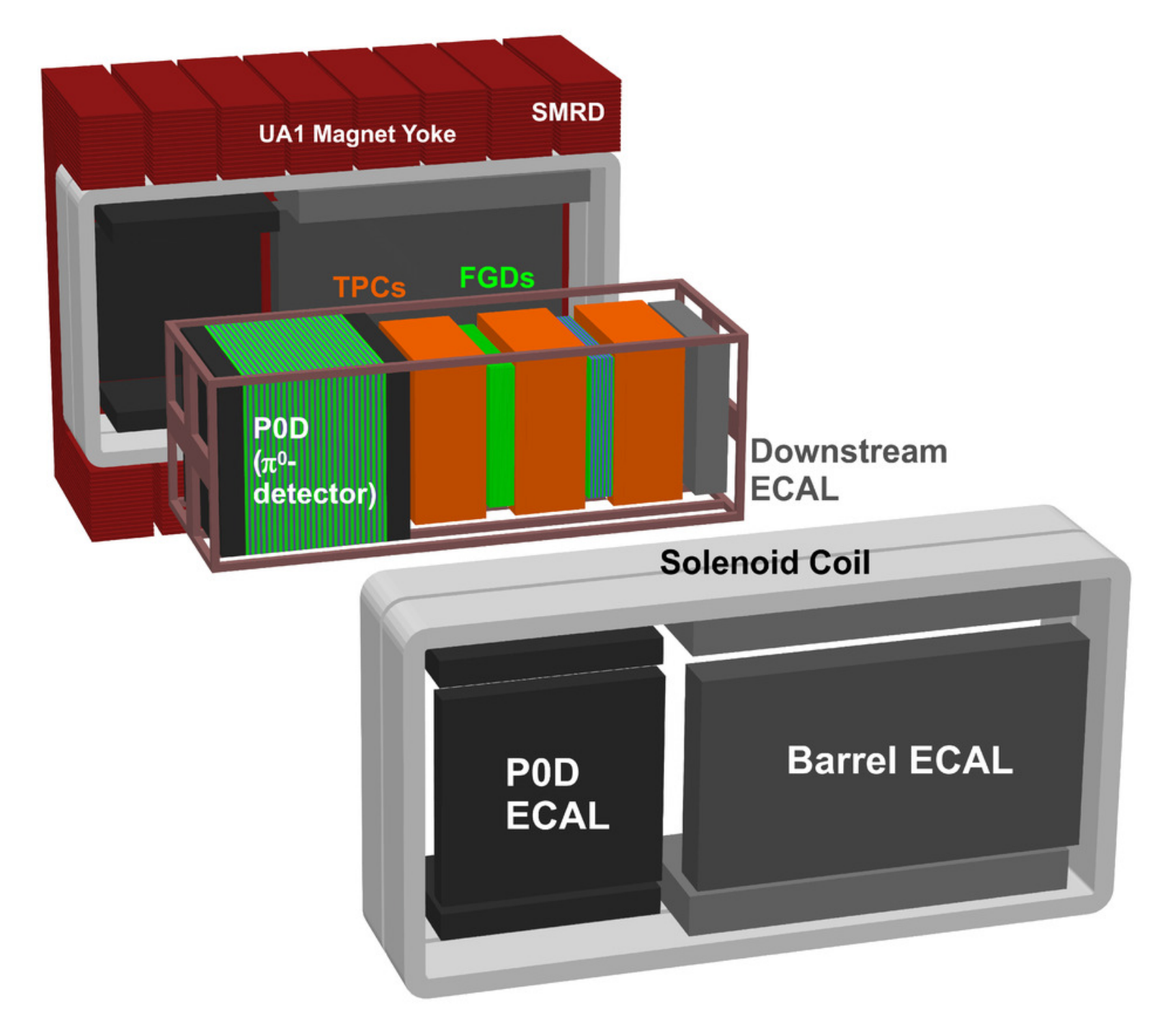}
\caption{Schematic showing an exploded view of the ND280 off-axis detector. Each subdetector is labeled using the acronyms given in the text. FGD1 is placed upstream of FGD2 and is shown in light green. The neutrino beam enters from the left of the figure.}
\label{fig:nd280}
\end{figure}

In this analysis, we measure the $\nu_\mu$ differential cross sections as a function of TKI variables at the off-axis detector ND280. 
As shown in \cref{fig:nd280}, ND280 is composed of an upstream $\pi^0$ detector (P\O{}D)~\cite{ASSYLBEKOV201248}, followed by a central tracker region, all surrounded by an electromagnetic calorimeter (ECal)~\cite{Allan_2013}. The outermost component is the former UA1/NOMAD magnet, which provides a 0.2~T dipole field, and contains scintillator modules in the air gaps acting as the side muon range detector (SMRD)~\cite{AOKI2013135}.

The central tracker region contains two fine grained detectors (FGD1 and FGD2)~\cite{AMAUDRUZ20121} and three time projection chambers (TPCs)~\cite{ABGRALL201125}. The FGDs are instrumented with finely segmented scintillator bars which act as both the target mass and particle tracker. The scintillator bars are made of $86.1\%$ carbon, $7.4\%$ hydrogen and $3.7\%$ oxygen by mass.
The bars are oriented alternately along the two detector coordinate axes (XY axes) transverse to the incoming neutrino beam (Z axis), 
and allow 3D tracking of charged particles. The most upstream FGD (FGD1) is composed of $15$ XY planes of scintillator with each plane having $2 \times 192$ bars. The downstream FGD (FGD2) has seven XY planes of scintillator with six $2.54$ cm thick layers of water in between, which allows cross section measurements to be made on water. This study focuses on carbon interactions and only events occurring in FGD1 are analyzed. For charged particles entering the TPCs, the curvature of the particle's track and thus its momentum can be determined in the presence of the magnetic field with a resolution of 10\% at 1 GeV. In combination with the measurement of energy loss per unit distance, TPCs provide high quality particle identification (PID) for charged particles. 

The ECal is a sampling calorimeter consisting of three key parts: the P\O{}D ECal which surrounds the P\O{}D; the Barrel ECal which surrounds the FGDs and TPCs; and the Downstream ECal which is located downstream of the FGDs and TPCs. The Barrel ECal and Downstream ECal together are referred to as the tracker-ECal. All ECals use layers of plastic scintillator bonded to lead sheets, and each alternating scintillator layer is rotated by $90^{\circ}$ to give 3D reconstruction. The tracker-ECal is designed to complement the FGDs and TPCs by giving detailed reconstruction of electromagnetic showers and a secondary PID, with an energy resolution of 10\% at 1 GeV.

\section{Event simulation}\label{sec:simulation}
For all T2K analyses, we need a reference Monte Carlo (MC) simulation to get a prediction based on the nominal neutrino flux, neutrino interaction model and detector effects. Data are then compared to MC to extract the physics quantities of interest and estimate the systematic uncertainties.

\begin{figure}[h!]
    \centering
    \includegraphics[width=1.0\linewidth]{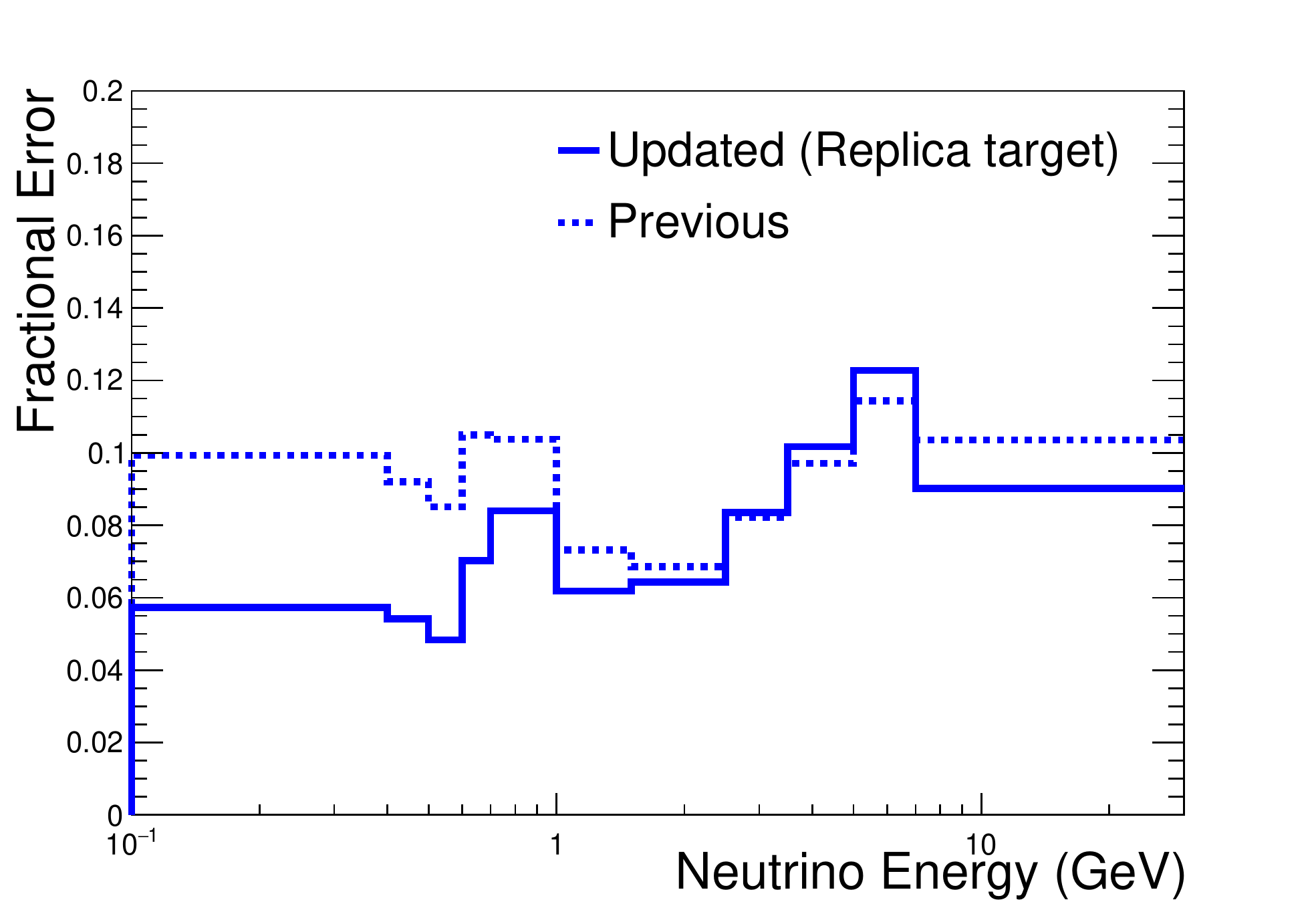}
    \caption{The fractional error on the muon neutrino flux at ND280 as a function of energy used in this analysis (solid) and previous T2K analyses (dashed). 
    }
    \label{fig:fluxerror}
\end{figure}

The modelling of the T2K neutrino flux~\cite{Abe:2012av} starts with the modelling of interactions of protons with the graphite target, which is done using the FLUKA 2011 package~\cite{Ferrari:2005zk,BOHLEN2014211}. 
Outside the target, the simulation of hadronic interactions and decays is done using the GEANT3~\cite{Brun:1994aa} and GCALOR~\cite{Zeitnitz:1992vw} software packages.
Hadronic interactions are further tuned with the recent measurements of $\pi^\pm$ yields performed by NA61/SHINE experiment using a T2K replica target~\cite{Berns:2018tap}. 
The conditions of the proton beam, magnetic horn current and neutrino beam axis direction are continuously monitored and incorporated into the simulation. This data-driven strategy helps to reduce the neutrino flux uncertainty near the flux peak (0.5 - 0.6 GeV) to 5\%. This results in a significant improvement with respect to previous T2K cross-section analyses~\cite{PhysRevD.101.112001,PhysRevD.101.112004} where the uncertainty was around 8.5\%~\cite{Abgrall:2015hmv}. A comparison of the flux uncertainty used in this analysis and the flux uncertainty used in previous T2K analyses is shown in \cref{fig:fluxerror}.

Neutrino-nucleus interactions and FSI of the outgoing particles are simulated using the neutrino event generator NEUT version 5.4.0~\cite{HAYATO2002171,Hayato:2009zz}.  
NEUT uses the spectral function (SF) in Ref.~\cite{BENHAR1994493} to describe the CCQE cross section. 
The modelling of 2p2h interactions is based on the model from Nieves \textit{et al.}~\cite{PhysRevD.85.113008}. The RES pion production process is described by the Rein-Sehgal model~\cite{REIN198179} with updated nucleon form-factors~\cite{PhysRevD.77.053001} and an axial mass ($M_A^{RES}$) of 1.07 GeV/c$^2$. The model contains contributions from non-resonant, $I_{1/2}$ pion-production channels. The nuclear model used for RES is a relativistic global Fermi gas (RFG)~\cite{SMITH1972605}, without a removal energy and with a Fermi momentum of 217 MeV/c. DIS interactions are modelled using the GRV98~\cite{Gluck:1998xa} parton distribution functions with corrections from Bodek and Yang ~\cite{doi:10.1063/1.1594324}. In the low invariant hadronic mass, W, region ($1.3 <$ W ${\leq}$ ${2.0}$ GeV/c$^2$) a custom hadronisation model~\cite{Aliaga:2020rqb} is used with suppressed single pion production to avoid double counting RES interactions. 
For W~\textgreater~2~GeV/c$^2$, PYTHIA/JETSET~\cite{Sjostrand:1993yb} is used as the hadronization model. The FSI, describing the transport of hadrons produced in elementary neutrino interaction through the nucleus, are simulated using a semi-classical intranuclear cascade model. The NEUT cascade model has been tuned to external pion-scattering data, which is described in Ref.~\cite{PhysRevD.99.052007}.

Outside the nucleus, final-state particles are then propagated through the detector material using GEANT4 version 4.9.4~\cite{AGOSTINELLI2003250}. The physics list~\cite{Allison:2016lfl} \texttt{QGSP\_BERT} is used for the hadronic physics, \texttt{emstandard\_opt3} for the electromagnetic physics  and \texttt{G4DecayPhysics} for the particle decays. The pion secondary interactions are handled by the cascade model in NEUT. The detector readout is simulated with a custom electronics simulation~\cite{ABE2011106}.  

\section{Data and event selection}\label{sec:selection}
In this analysis, the neutrino-mode data collected between 2010 and 2017 is used, which corresponds to $11.6\times 10^{20}$ protons on target (POT) and an integrated muon neutrino flux of $2.2\times 10^{13}/\textrm{cm}^2$. 
Events are required to have an interaction vertex in the FGD1 fiducial volume (FV), which includes all the XY planes of scintillator except for the most upstream one, and excludes the outermost five bars on either end of each layer. This leaves the FV with $2\times 182\times 14$ bars, and a total mass of approximately 973 kg. The MC contains simulated data equivalent to $195.1\times 10^{20}$ POT.

\subsection{Signal definition}\label{sec:signal_def}
The goal of this analysis is to characterize nuclear effects in $\nu_\mu$ CC1$\pi^+$ interactions on carbon using neutrino interactions inside FGD1, which is a hydrocarbon (CH) target. 
Since the CC1$\pi^+$ production on carbon and on hydrogen cannot be clearly separated, the combined cross section on CH is measured, with the TKI variables on hydrogen calculated in the same way as on carbon: for hydrogen signal events, in which there are no nuclear effects, it is expected that $\delta p_{TT}=0$ and $p_N\approx 26$~MeV/c. $\delta\alpha_T$ is undefined for interactions on hydrogen because $\delta p_T=0$. A flat distribution across $0$--$180^\text{o}$ is assigned because it resembles the real $\delta\alpha_T$ distribution due to the small but non-vanishing isotropic Fermi motion of a free proton.

To ensure the cross section results are not dependent on the signal model used in the reference T2K simulation, extensive precautions are taken in the analysis. A crucial one is to have the signal definition only be reliant on observables experimentally accessible to ND280. Therefore, the signal is defined as any event with one $\mu^-$, one $\pi^+$ and no other mesons, and at least one proton in the final state, so that there is need to account for the pion and proton FSI. Hereafter, the signal topology is denoted as CC1$\pi^+$Xp, where X$\geq$1. 
In order to mitigate model dependence in the efficiency correction, phase-space restrictions are applied in the signal definition to restrict the measurement to the regions of kinematic phase space ND280 is sensitive to. These restrictions are defined in \cref{tab:signal_ps_cut}. However, the consideration of three-particle final states in this analysis necessitates the inclusion of a high dimensional kinematic phase space over which the efficiency cannot be kept entirely flat with simple phase-space constraints. This leads to a potential source of bias from the input neutrino interaction model predictions. To alleviate this concern, additional model uncertainties are added (discussed in \cref{sec:sys_err}) to allow a variation of the input simulation in regions of the underlying particle kinematics where the efficiency is not flat. The size of this uncertainty roughly double the largest variation in the efficiency seen from a wide variety of different generator predictions (broadly spanning those shown in \cref{sec:nugen}).

\begin{table}[h!]
    \centering
	\caption{\label{tab:signal_ps_cut}%
		CC1$\pi^+$Xp signal phase-space restrictions for the post-FSI final-state particles. The angle $\theta$ is relative to the neutrino direction. For events with multiple protons, only the highest momentum proton is considered, and other protons are ignored.
	}
		\begin{tabular}{l@{\hskip 10pt}c@{\hskip 10pt}c}
		\hline\hline
			Particle & Momentum $p$ & Angle $\theta$ \\
			\hline
			$\mu^-$    & 250-7000 MeV/c & $<70^\text{o}$ \\
			$\pi^+$  & 150-1200 MeV/c & $<70^\text{o}$ \\
			p      & 450-1200 MeV/c & $<70^\text{o}$ \\
			\hline\hline
		\end{tabular}
\end{table}	

\begin{figure*}
\centering
    \includegraphics[width=\linewidth]{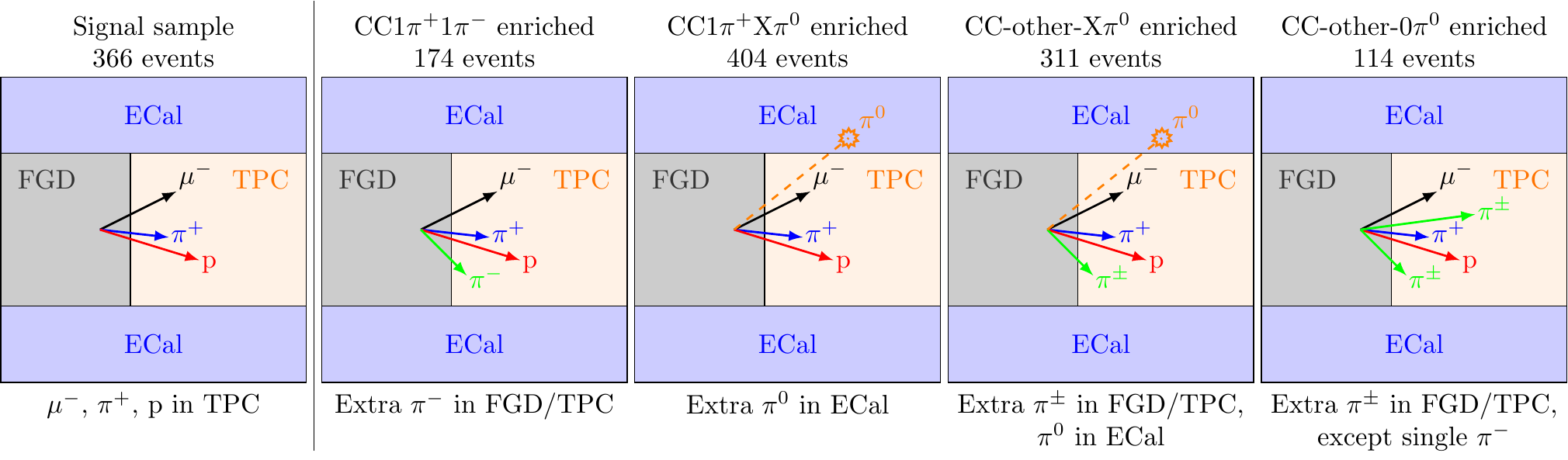}
\caption{Schematic representation of the signal sample (left) and control samples (right) selection, together with the number of events observed in data. Details of the selection criteria are described in \cref{sec:signal_sample,sec:control_sample}.} \label{fig:sample_schematics}
\end{figure*}

We select one signal sample for the events of interest, and four control samples to constrain the number of background events in the signal sample. The five samples are shown schematically in \cref{fig:sample_schematics}.

\subsection{Signal sample selection}\label{sec:signal_sample}
The signal sample contains neutrino events with exactly one $\mu^-$ track, one $\pi^+$ track, and at least one proton track, maximizing the number of signal events selected with minimal background.

The selection starts by searching for a good quality $\mu^-$ track. Events within a 120 ns time window around one of the eight bunch centers per 5~$\mu$s spill structure  of  the beam are considered. The highest momentum, negatively charged track originating from the FGD1 FV and making a long track through the downstream TPC is chosen to be a $\mu^-$ candidate. Other detector activities in or around FGD1 are used as a veto to ensure the $\mu^-$ track is not a broken segment of another track from outside the FV. Then a muon PID cut is applied based on the energy loss and momentum measurement in the TPC as in Ref.~\cite{PhysRevD.98.012004}. After this step a $\nu_\mu$ CC sample of 90.3\% purity is obtained.

Next, all other tracks originating from the FGD1 FV with a long segment in the TPC are classified by the TPC PID. For positively charged tracks, three particle hypotheses are considered: $\pi^+$, $e^+$ and proton; for negatively charged tracks, only two particle hypotheses are considered: $\pi^-$ and $e^-$. Events with exactly one $\pi^+$ track, and at least one proton track are selected. Those with $\pi^-$ or $e^{\pm}$ are rejected because they are likely to be the products of DIS or other background interactions. 

Additional pions are identified in FGD1 and the tracker-ECal. Tracks fully contained inside FGD1 are classified into pions or protons if the energy deposition and range are consistent with the corresponding particle hypotheses. Michel electrons~\cite{Michel_1950} are also identified by looking for a time-delayed FGD1 hit cluster, and are regarded as products of the pion-muon-electron decay chain. The tracker-ECal is employed to identify isolated objects that are consistent with a photon shower, and tags these as products of $\pi^0\rightarrow 2\gamma$ decay. Events with additional charged pions in FGD1 or $\pi^0$ in the ECal are rejected.

In the final step, events with additional tracks in FGD1 (either the fully contained tracks that are not classified, or the non-fully contained tracks without TPC PID) are rejected to reduce the low energy pion backgrounds that are missed by the pion selection processes. Then we require the $\mu^-$, $\pi^+$ and p tracks to have their starting positions to be within a box of 50~mm$\times$50~mm$\times$30~mm in the XY and Z planes.
This ensures the tracks are coming from the same interaction vertex. 
Events that are not reconstructed to have matched the kinematic requirements in \cref{tab:signal_sample_ps_cut} are put into an out-of-phase-space (OOPS) bin. Compared to the signal definition in \cref{tab:signal_ps_cut}, the kinematic cuts have slightly larger ranges in momenta to compensate for the finite momentum resolution. The extremely good angular resolution (about 1$^\circ$) allows us to use the same angular restriction.
\begin{table}[h!]
	\caption{
	    Kinematic cuts for the reconstructed particles in the analysis samples. The particle type and kinematics are the reconstructed quantities. The angle $\theta$ is relative to the neutrino direction. For events with multiple reconstructed protons, only the highest momentum proton is considered, and other protons are ignored.
	}\label{tab:signal_sample_ps_cut}%
		\begin{tabular}{l@{\hskip 8pt}c@{\hskip 8pt}c}
		\hline
		\hline
			Particle & Momentum $p$ & Angle $\theta$ \\
			\hline
			$\mu^-$    & 225-7700 MeV/c & $<70^\text{o}$ \\
			$\pi^+$  & 135-1320 MeV/c & $<70^\text{o}$ \\
			p      & 405-1320 MeV/c & $<70^\text{o}$ \\
			\hline\hline
		\end{tabular}
\end{table}	

Following the signal sample identification, the selected events (except the OOPS bin) are binned in one of the reconstructed TKI variables and the reconstructed highest proton momentum, $p_\textrm{p}$. The binning in TKI variables is the same as that used in the cross section extraction in \cref{sec:analysis}. The binning in $p_\textrm{p}$ helps to correct for the bias in estimating selection efficiencies. The binning in $p_\textrm{p}$ is chosen over other kinematic variables because nucleon emission from neutrino interactions is less understood than pion or muon emission. In addition, the TPC proton detection threshold is around 400 MeV, which might significantly affect the efficiency. \cref{tab:signal_sample_binning} summarizes the signal sample binning. The CC1$\pi^+$Xp cross sections are measured as a function of a single TKI variable only, thus the number of reconstructed bins is much more than the number of cross-section bins. 
For example, in the $\delta p_{TT}$ measurement,
there are six $p_\textrm{p}$ bins for each of the five $\delta p_{TT}$ bins in the signal sample. In total there are $6\times 5=30$ signal sample bins to extract the differential cross sections in five bins of $\delta p_{TT}$.
\begin{table}[h!]
    \centering
	\caption{\label{tab:signal_sample_binning}%
	    Analysis bin edges for the CC1$\pi^+$Xp cross sections as a function of the TKI variables. The signal sample is binned  in one of the reconstructed TKI variables vs. reconstructed $p_\textrm{p}$. The control samples are binned in the reconstructed TKI variable only.}
		\begin{tabular}{l@{\hskip -1pt}c@{\hskip -4.5pt}c}
		\hline\hline
			Variable & Number of bins & Bin edges \\
			\hline
			$\delta p_{TT}$ (MeV/c)    & 5 & -700,-300,-100,100,300,700 \\
			$p_N$ (MeV/c)  & 4 & 0,120,240,600,1500 \\
			$\delta\alpha_T$ (deg)      & 3 & 0,60,120,180 \\
			\hline
			$p_\textrm{p}$ (MeV/c) & 6 & 405,575,700,825,950,1075,1320 \\
			\hline\hline
		\end{tabular}
\end{table}

\begin{figure*}
\centering
    \subfloat{\includegraphics[width=0.49\linewidth]{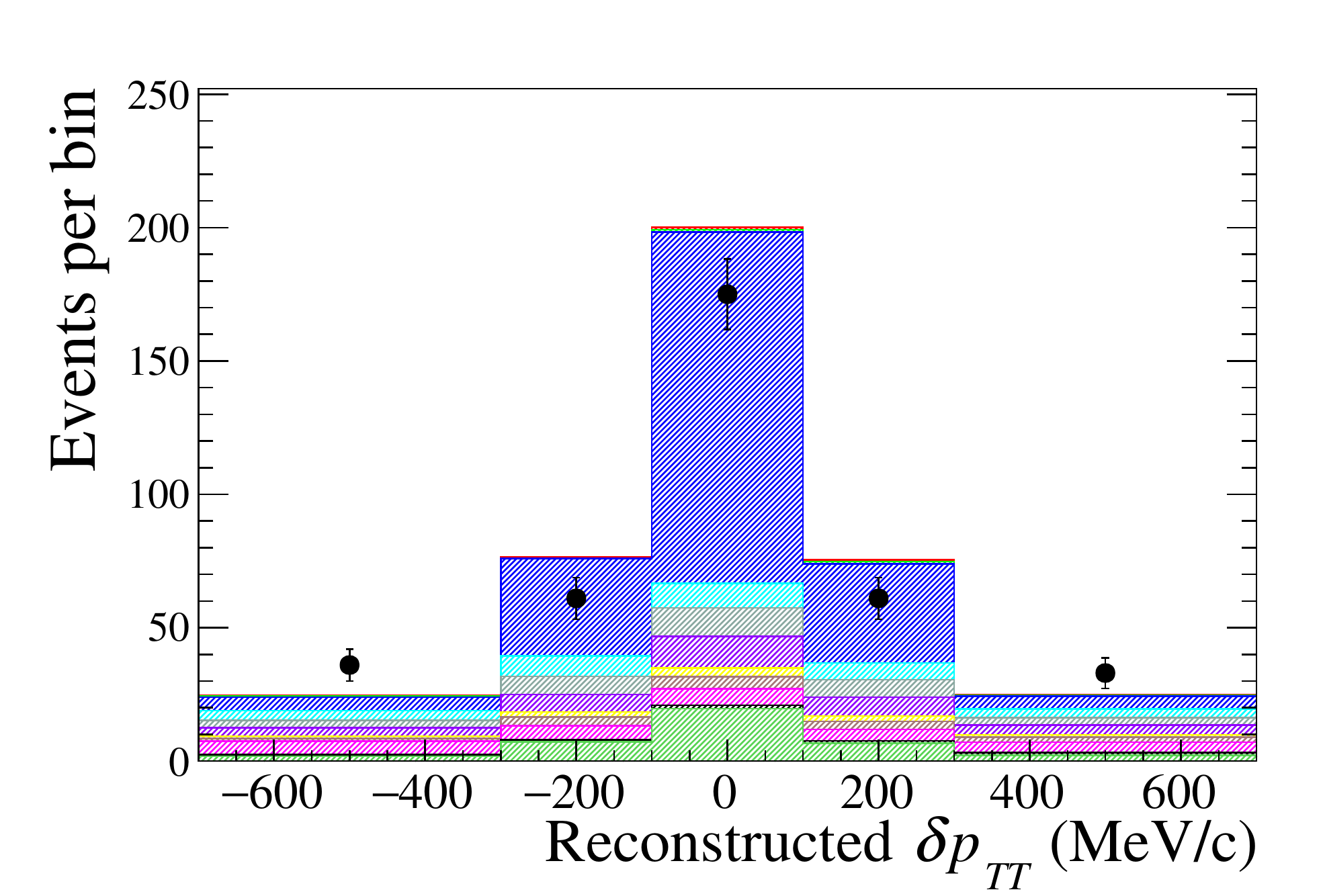}}
    \subfloat{\includegraphics[width=0.49\linewidth]{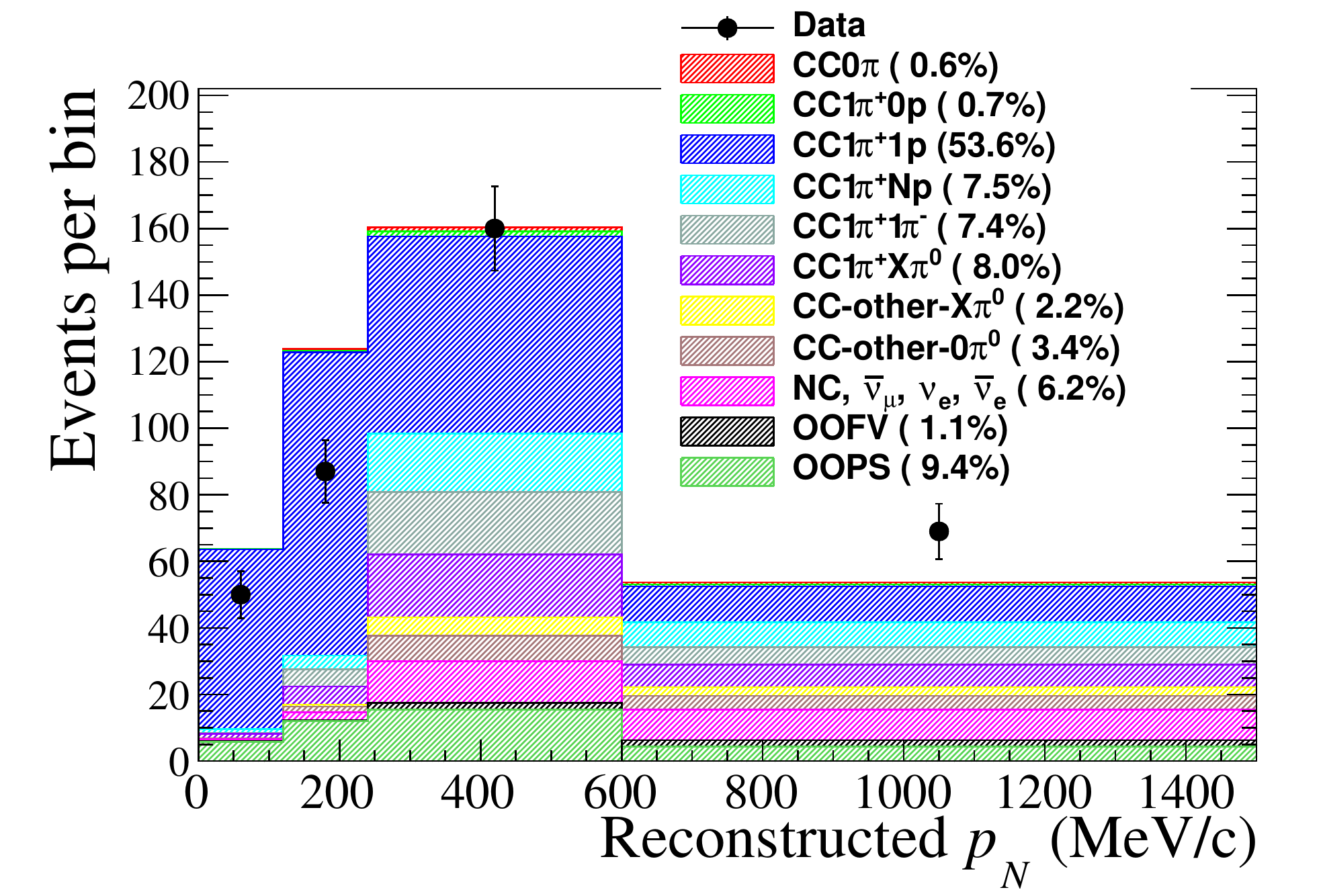}}\hfill
    \subfloat{\includegraphics[width=0.49\linewidth]{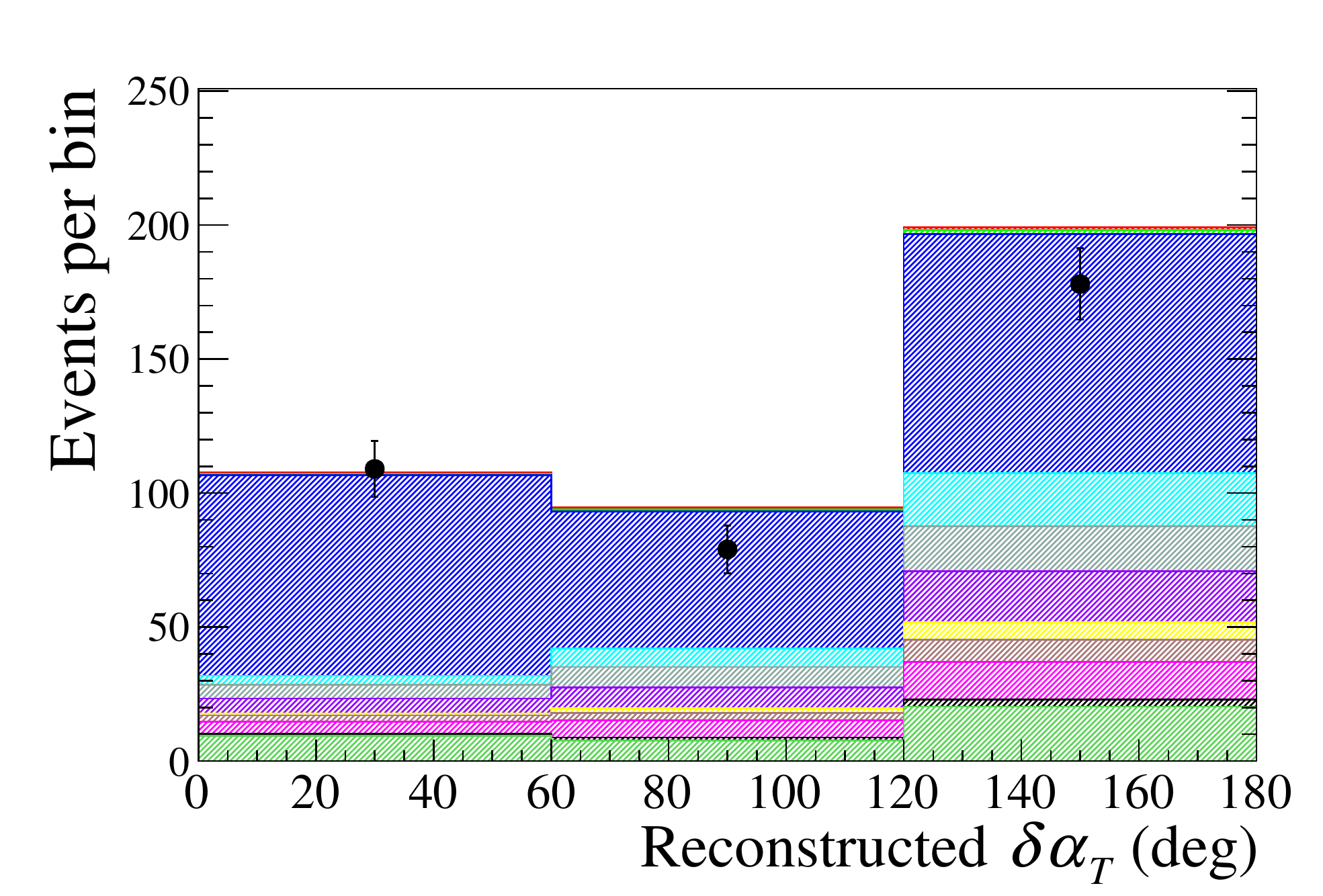}}
    \subfloat{\includegraphics[width=0.49\linewidth]{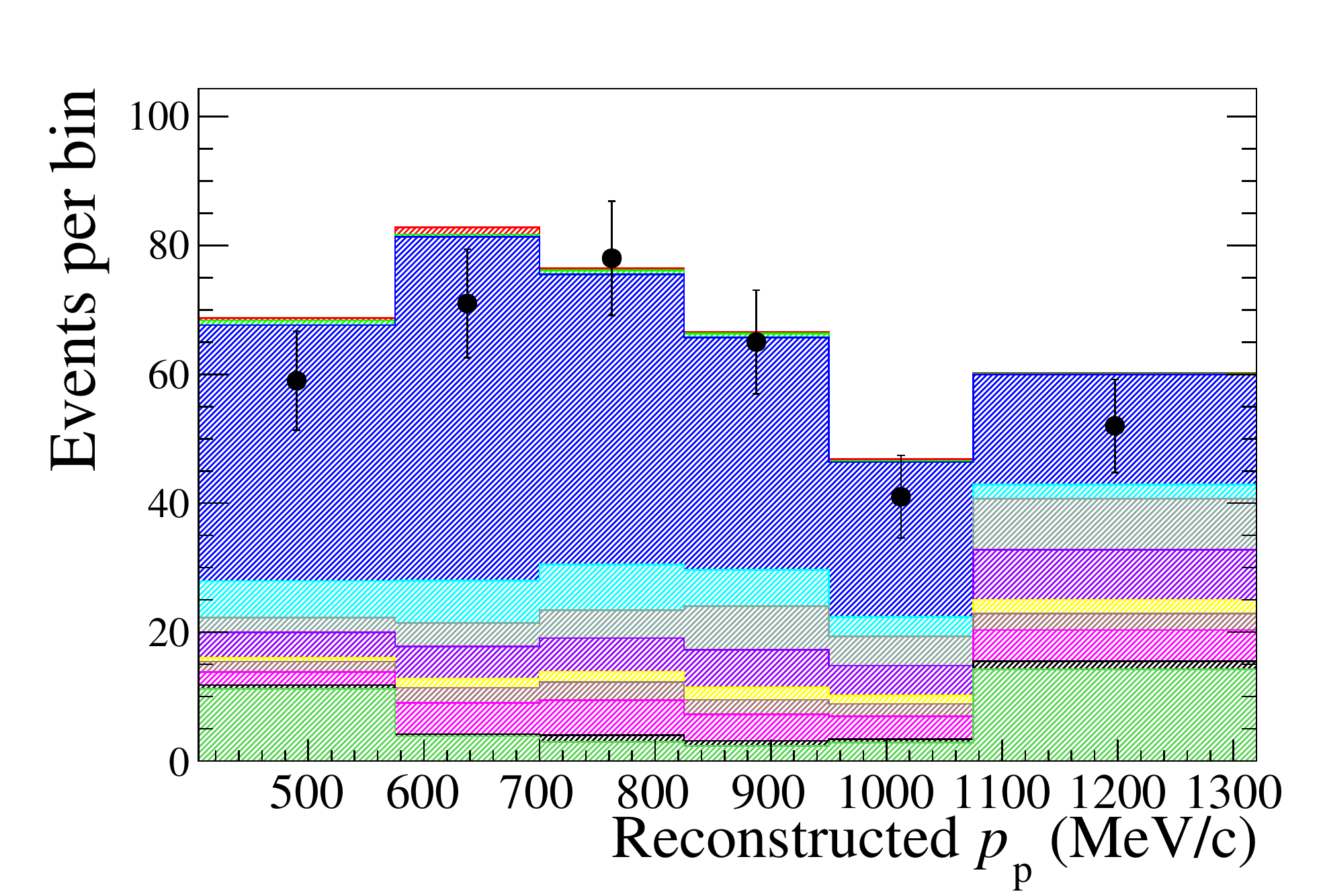}}
\caption{Distribution of events in the signal sample as a function of the reconstructed TKI variables and highest proton momentum, broken down into true final-state topology predicted by the nominal MC. The legend shows the fraction of events in all plots.  Histograms are stacked. The MC has been normalized to 11.6$\times 10^{20}$ POT, the equivalent number of POT collected for the data. The error bars show the statistical uncertainty in data.} \label{fig:signal_sample}
\end{figure*}

\cref{fig:signal_sample} shows the distribution of the reconstructed TKI variables and $p_\textrm{p}$ in the signal sample (without the OOPS bin). A total of 366 events are observed in data. The overall signal selection efficiency is around 14\%. When broken down by final-state topology, the total CC1$\pi^+$1p (one proton) and CC1$\pi^+$Np (multiple protons) signal purity is 61.1\%. The four categories of CC-other events with multiple pions in the final-state, CC$1\pi^+$1$\pi^-$, CC$1\pi^+$X$\pi^0$, CC-other-X$\pi^0$ and CC-other-0$\pi^0$, are mostly produced by DIS interactions and are the dominant backgrounds. Details on how to constrain these backgrounds are described in Sec.\ref{sec:control_sample}. 
There are also small amounts of neutral current (NC) and $\bar{\nu}_\mu/\nu_e/\bar{\nu}_e$ events where a $\pi^-/e^-$ is misidentified as a $\mu^-$. In most cases the misidentification comes from NC interactions. The contribution from out of fiducial volume (OOFV) events is almost negligible. 
The OOPS background in \cref{fig:signal_sample} refers to CC1$\pi^+$Xp events which do not satisfy the phase-space restrictions in \cref{tab:signal_ps_cut}, and the separated OOPS bin is used to constrain this background.

\subsection{Control sample selection}\label{sec:control_sample}
\begin{figure*}
\centering
    \subfloat{\includegraphics[width=0.27\linewidth]{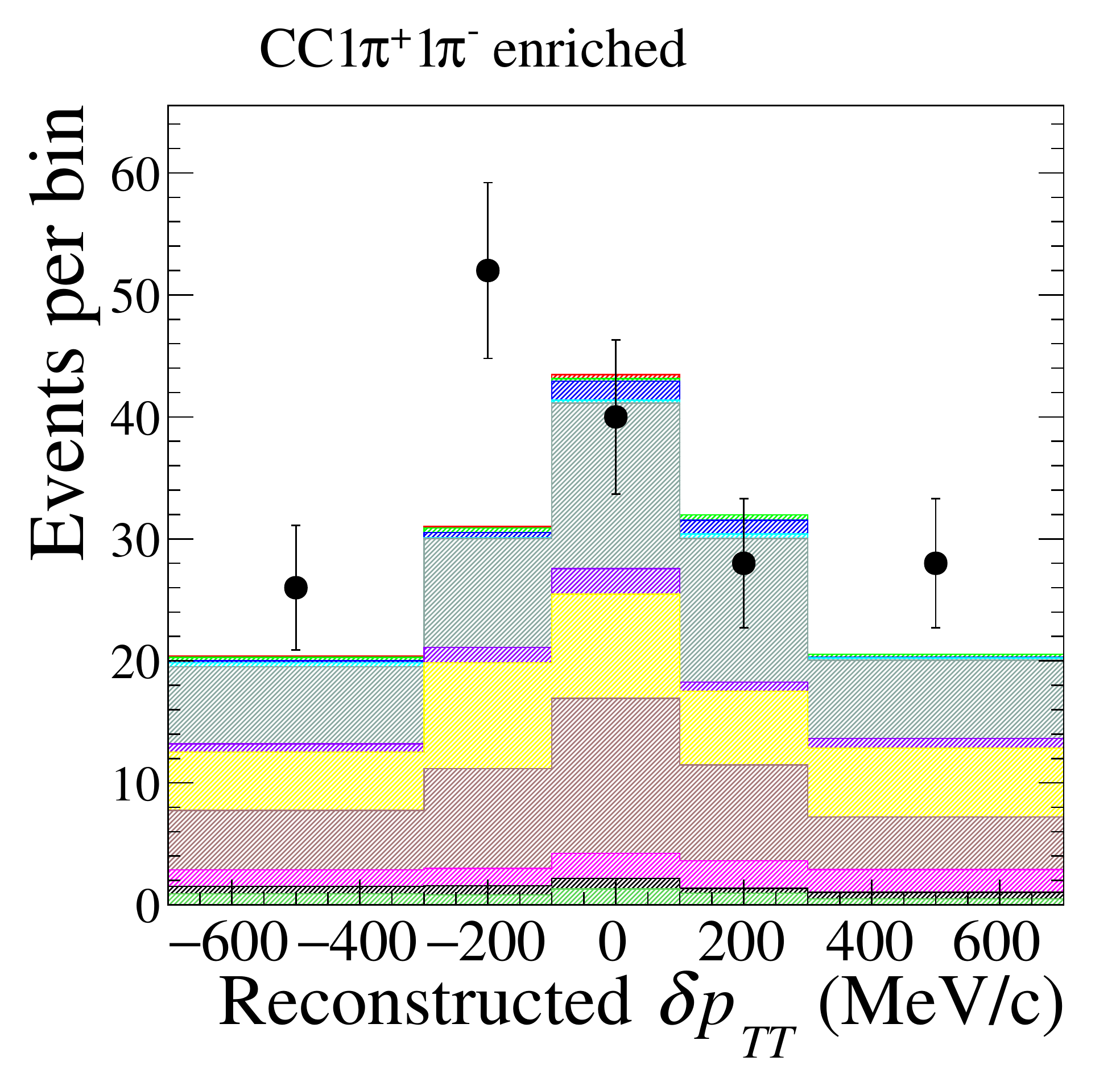}}
    \subfloat{\includegraphics[width=0.27\linewidth]{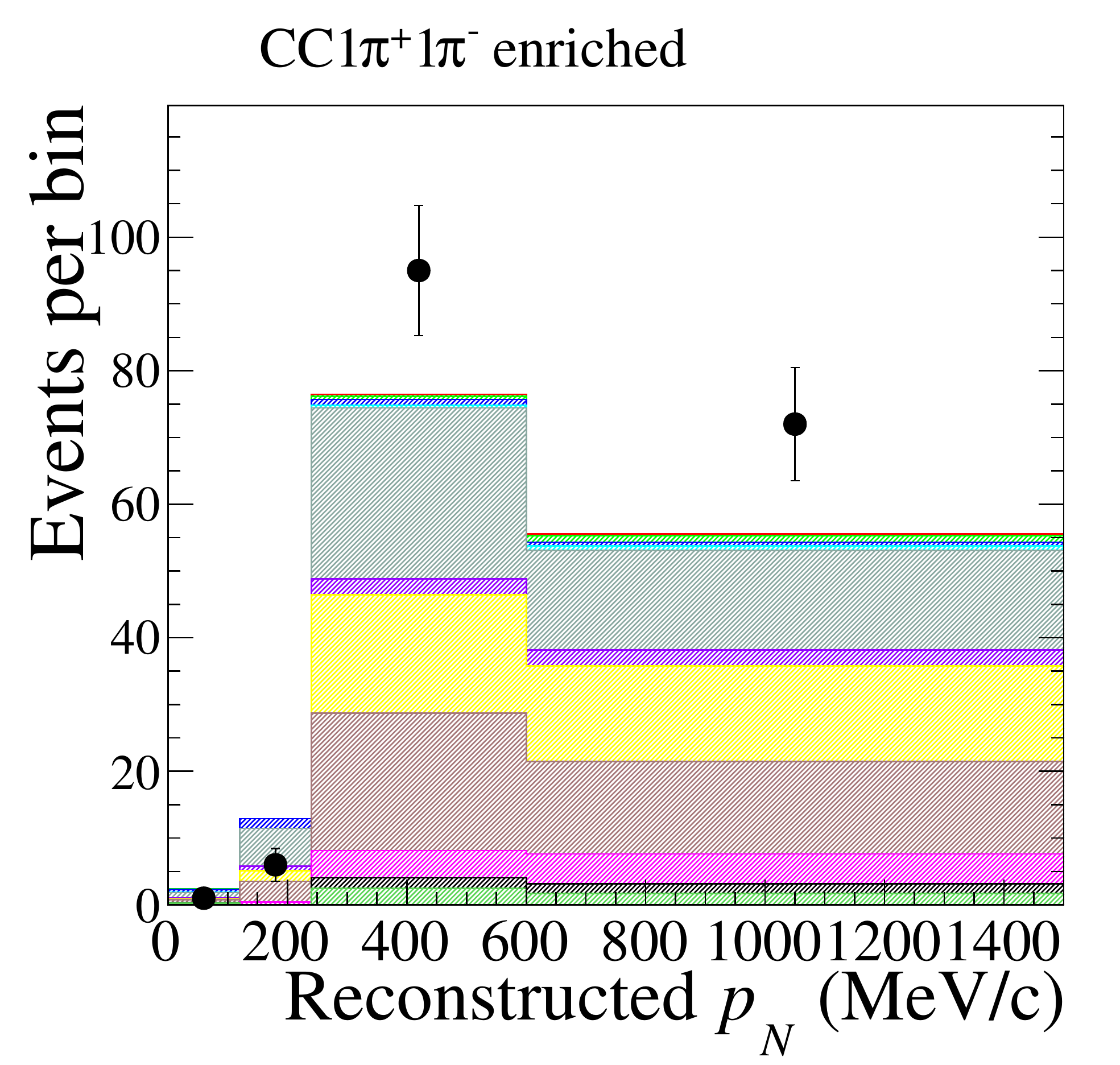}}
    \subfloat{\includegraphics[width=0.27\linewidth]{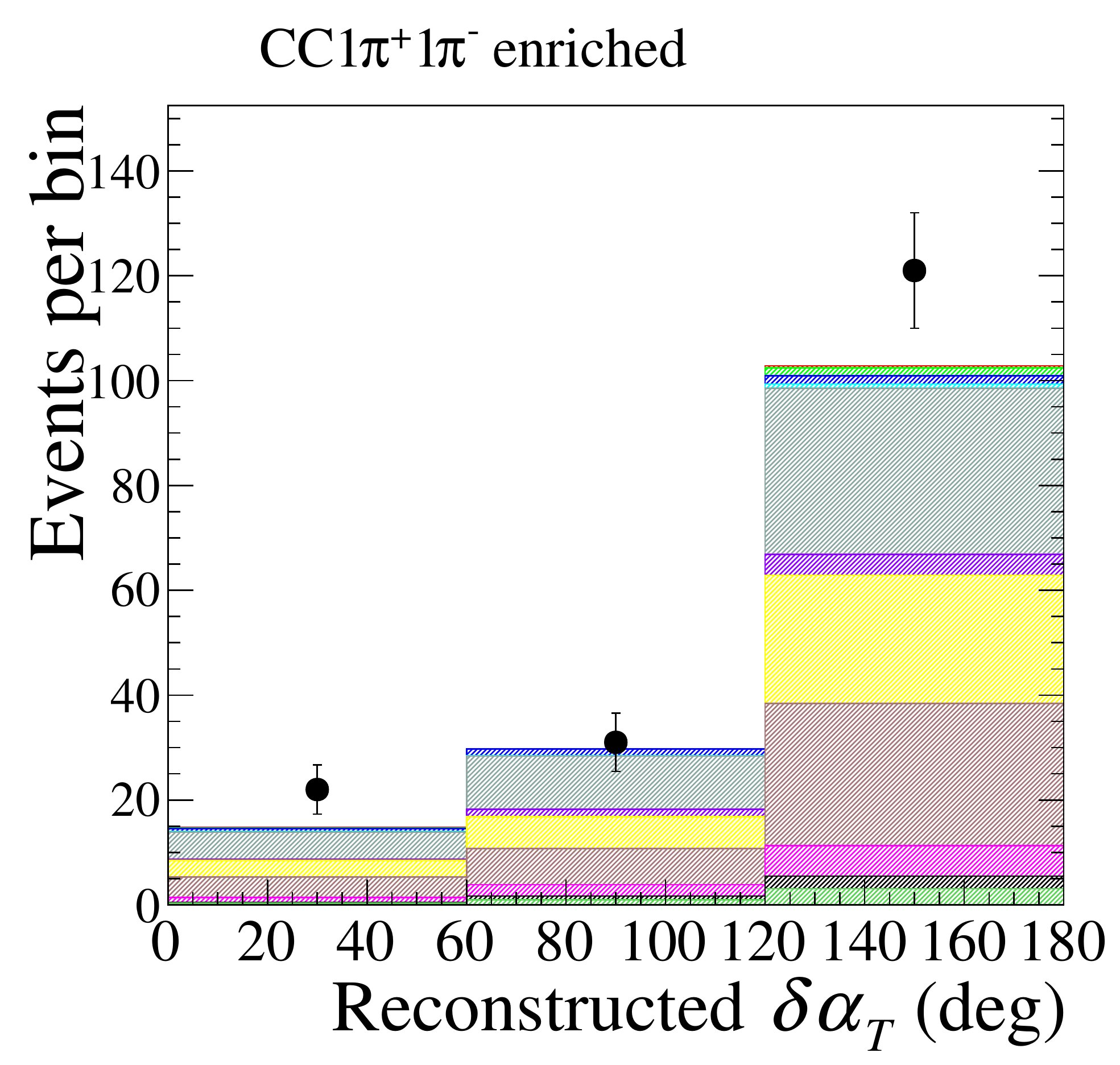}}
    \subfloat{\includegraphics[width=0.19\linewidth]{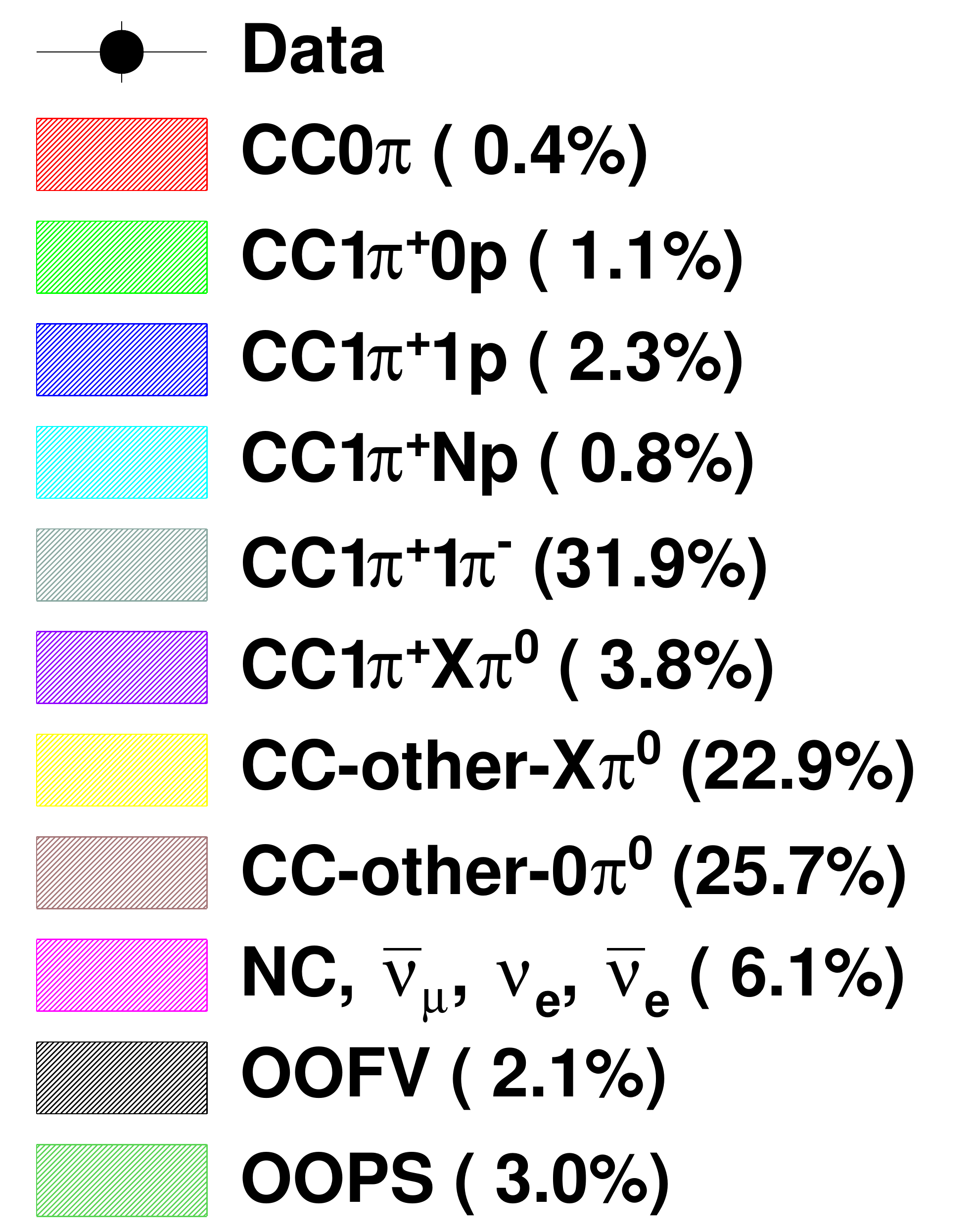}}
    \hfill
    \subfloat{\includegraphics[width=0.27\linewidth]{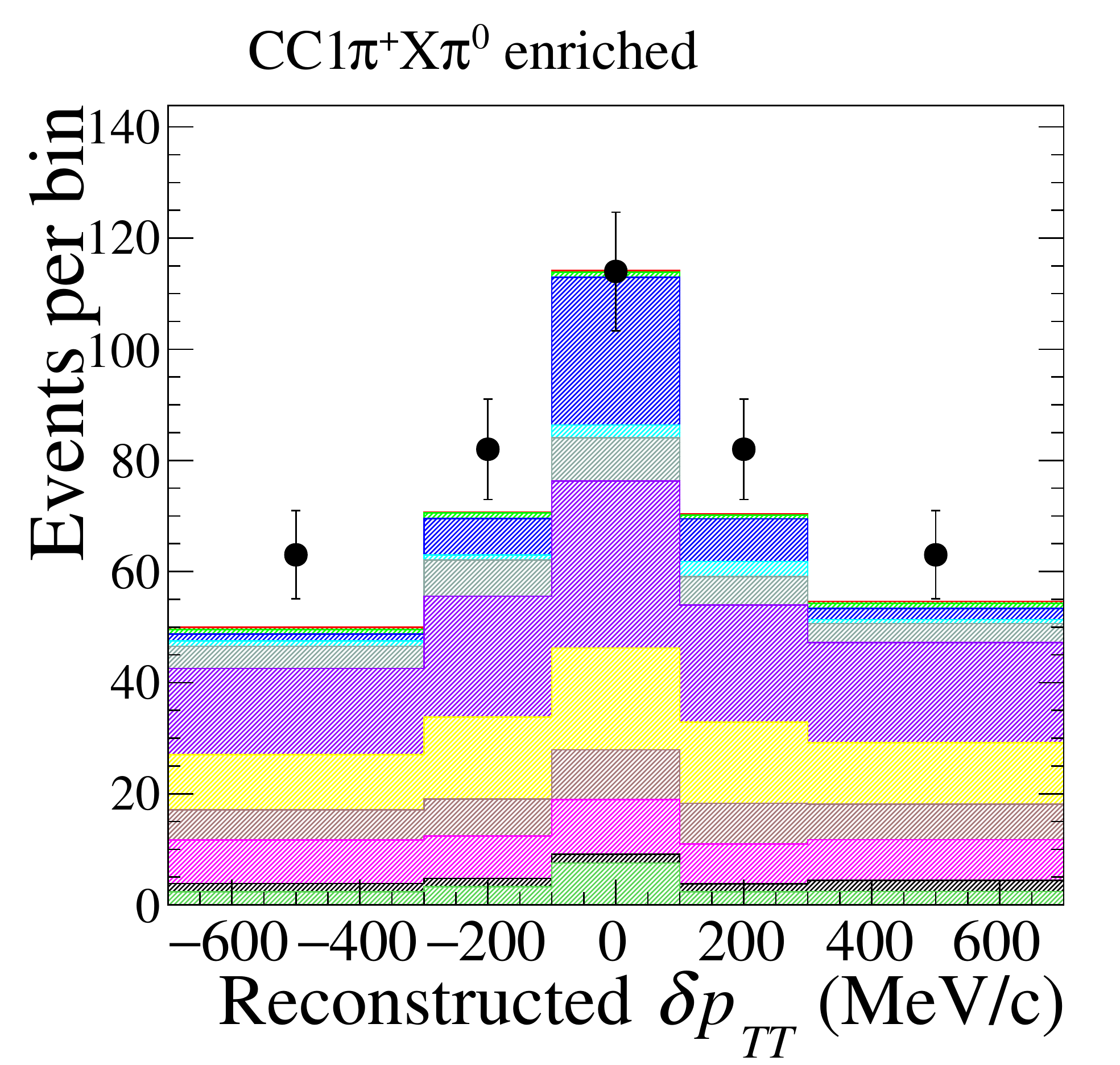}}
    \subfloat{\includegraphics[width=0.27\linewidth]{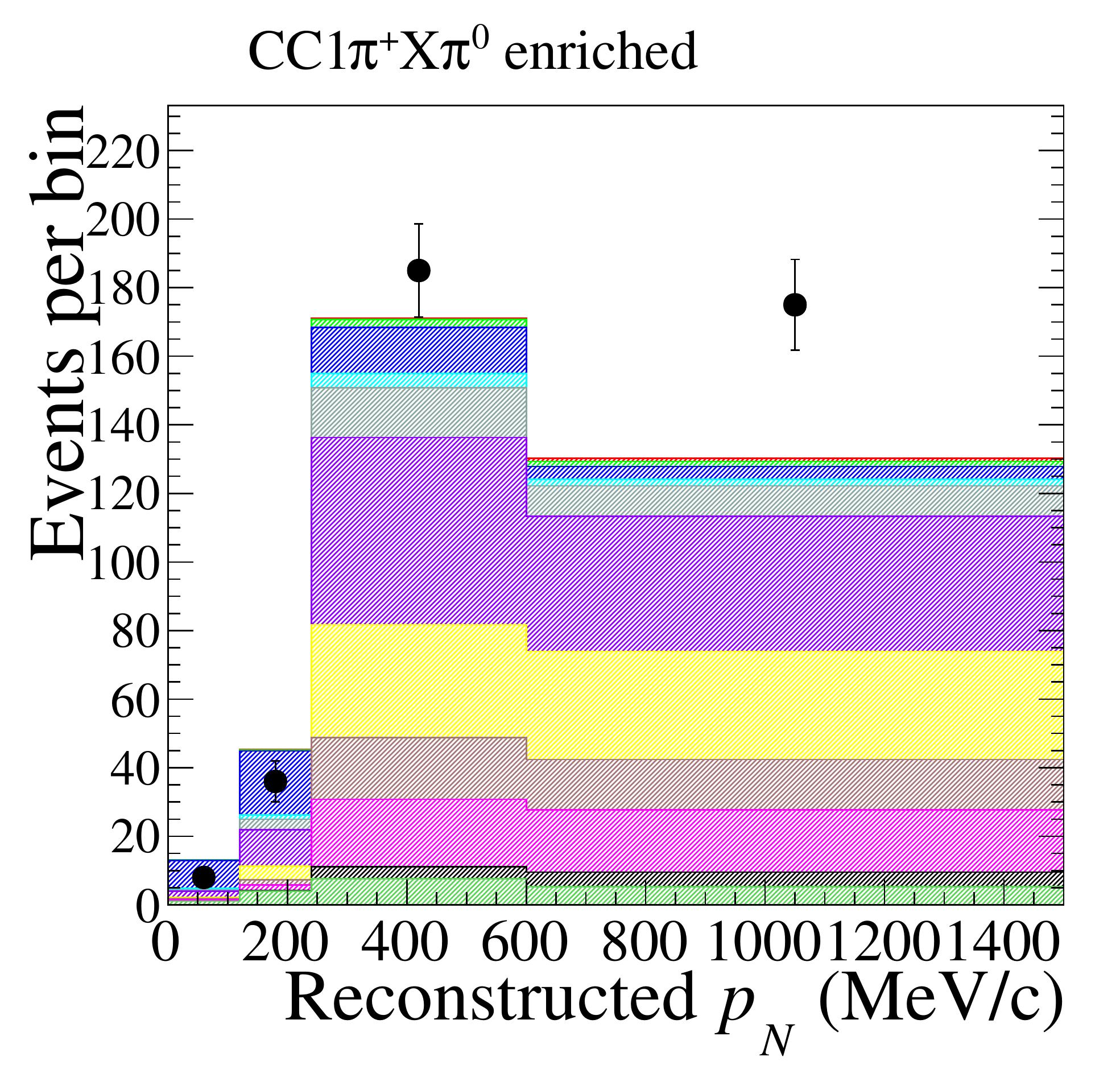}}
    \subfloat{\includegraphics[width=0.27\linewidth]{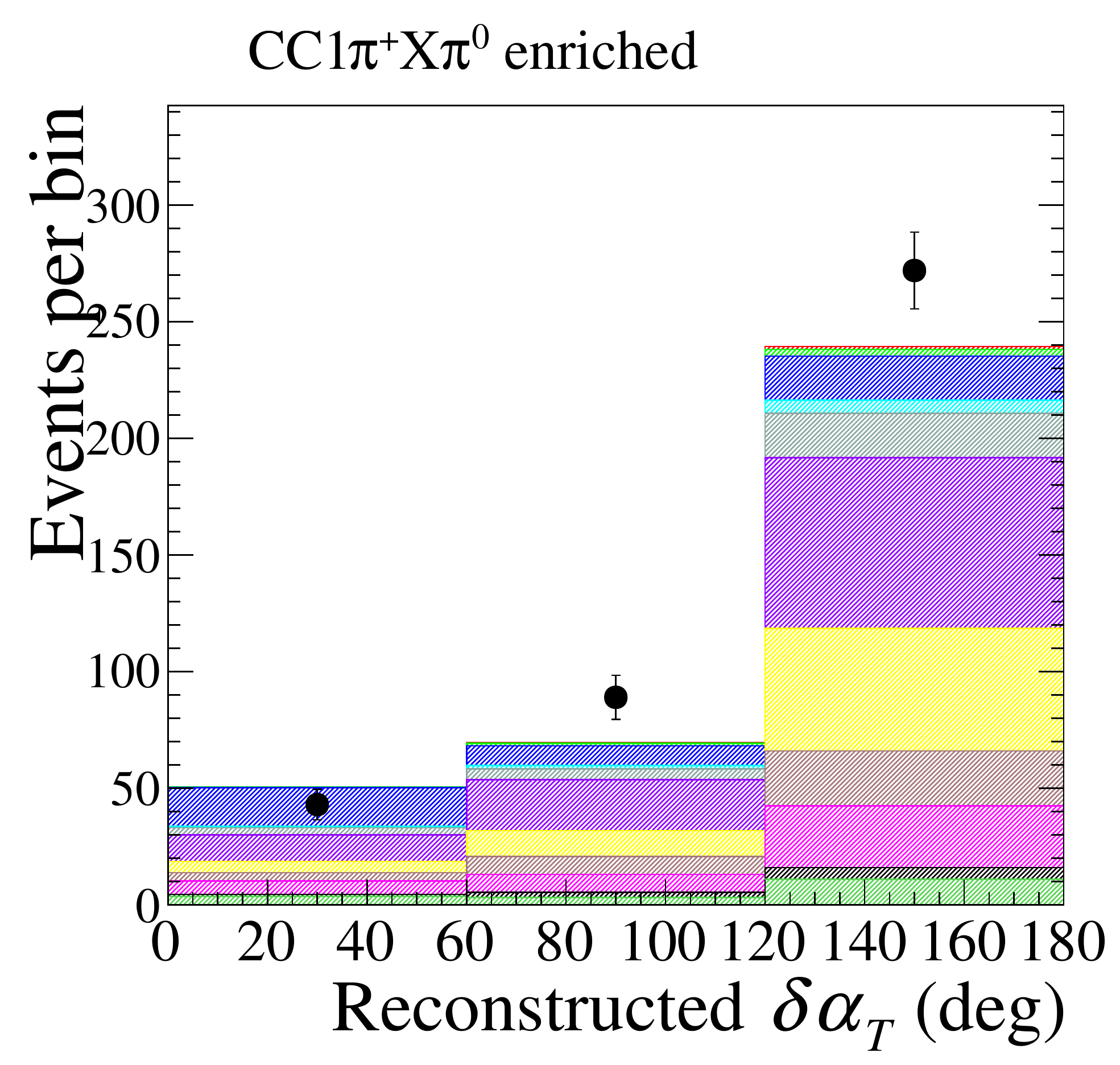}}
    \subfloat{\includegraphics[width=0.19\linewidth]{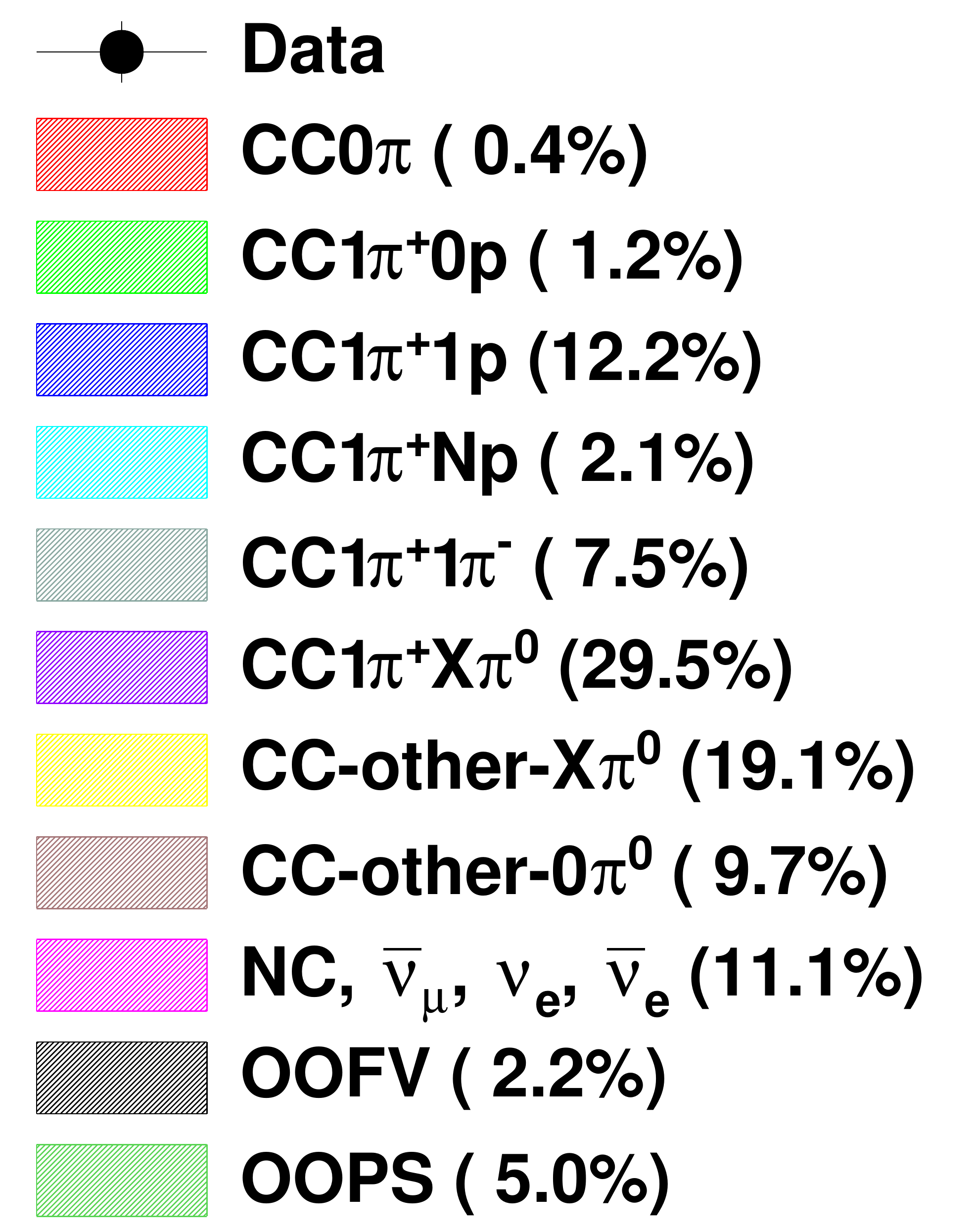}}
    \hfill
    \subfloat{\includegraphics[width=0.27\linewidth]{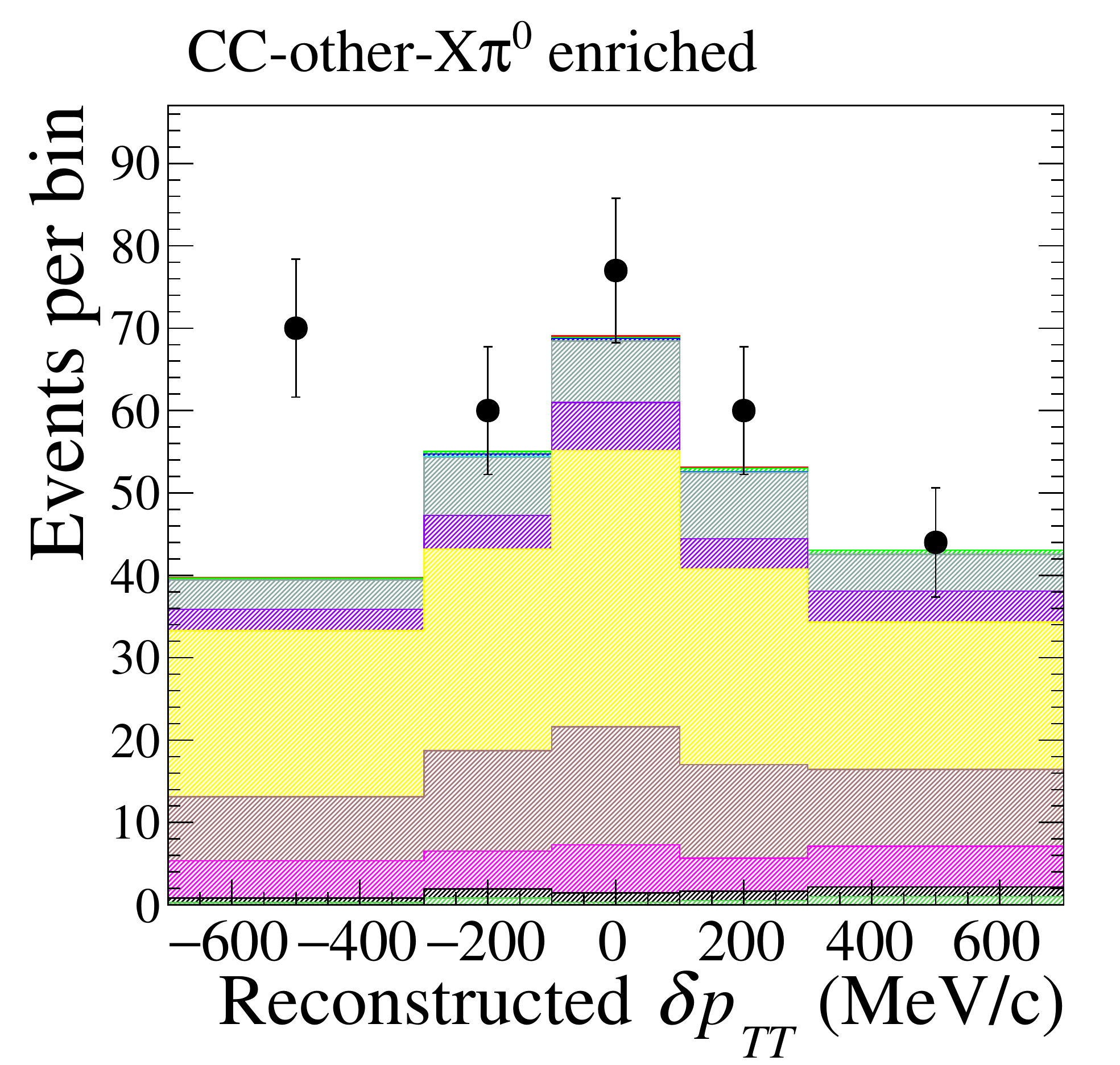}}
    \subfloat{\includegraphics[width=0.27\linewidth]{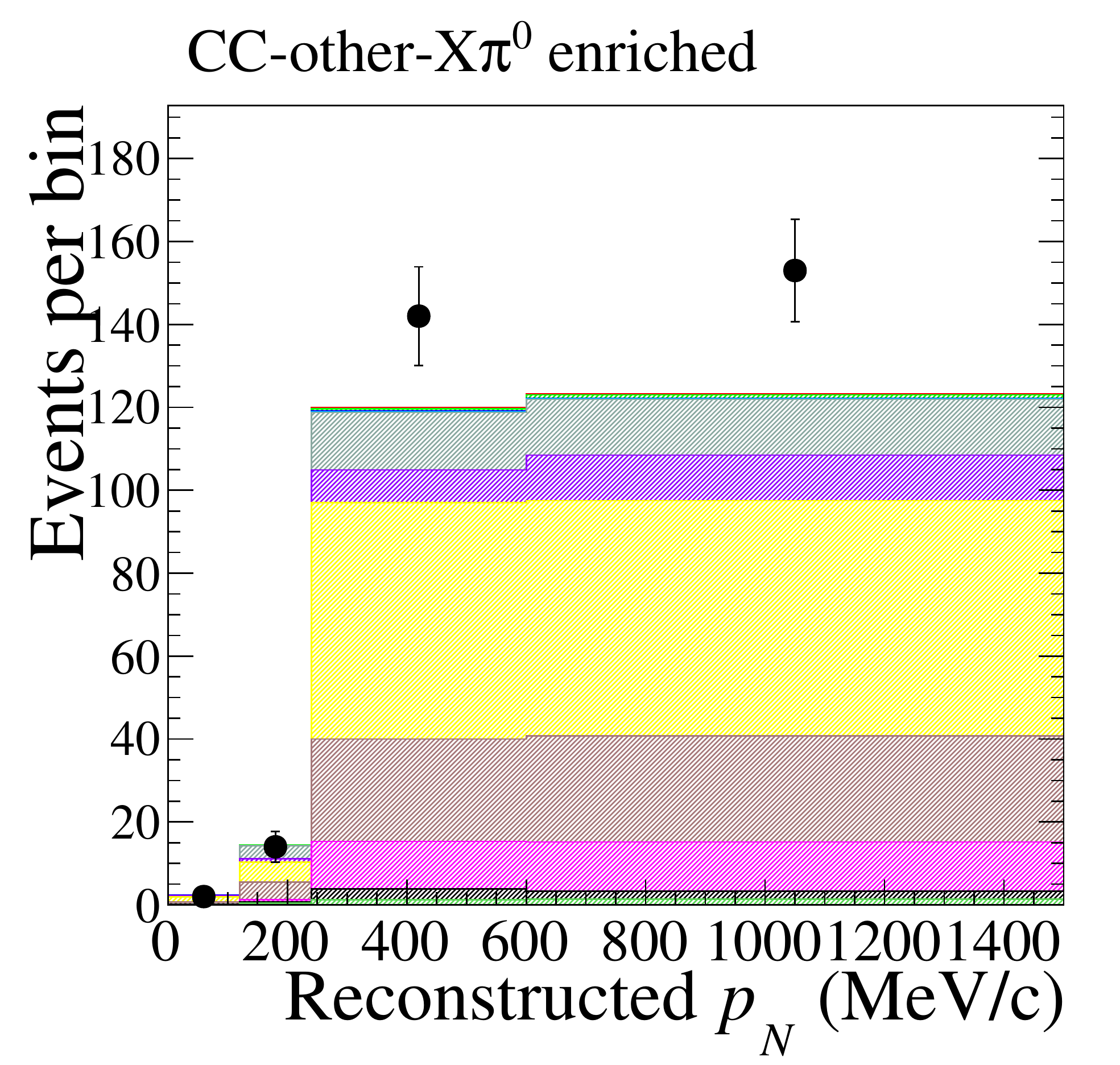}}
    \subfloat{\includegraphics[width=0.27\linewidth]{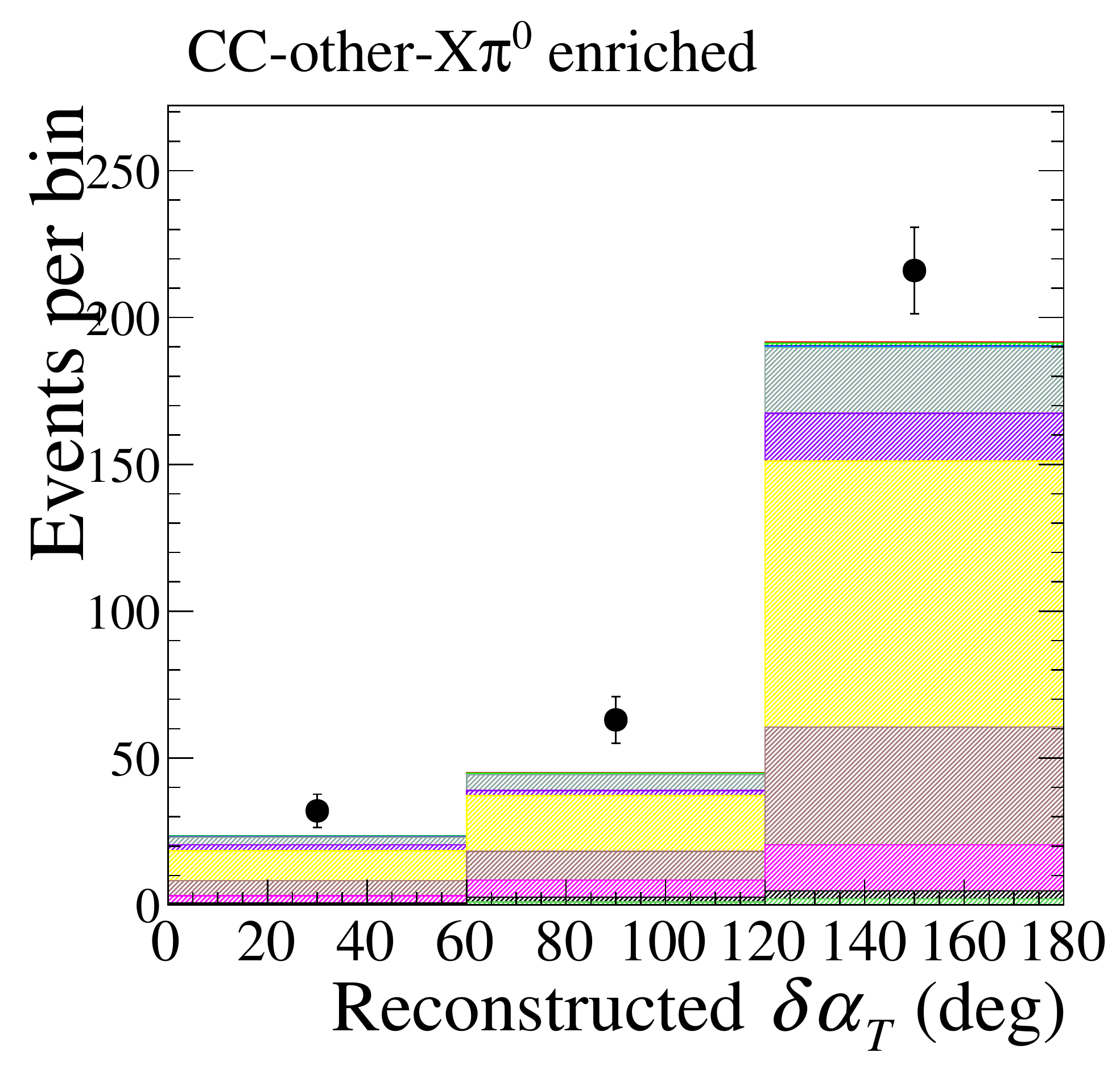}}
    \subfloat{\includegraphics[width=0.19\linewidth]{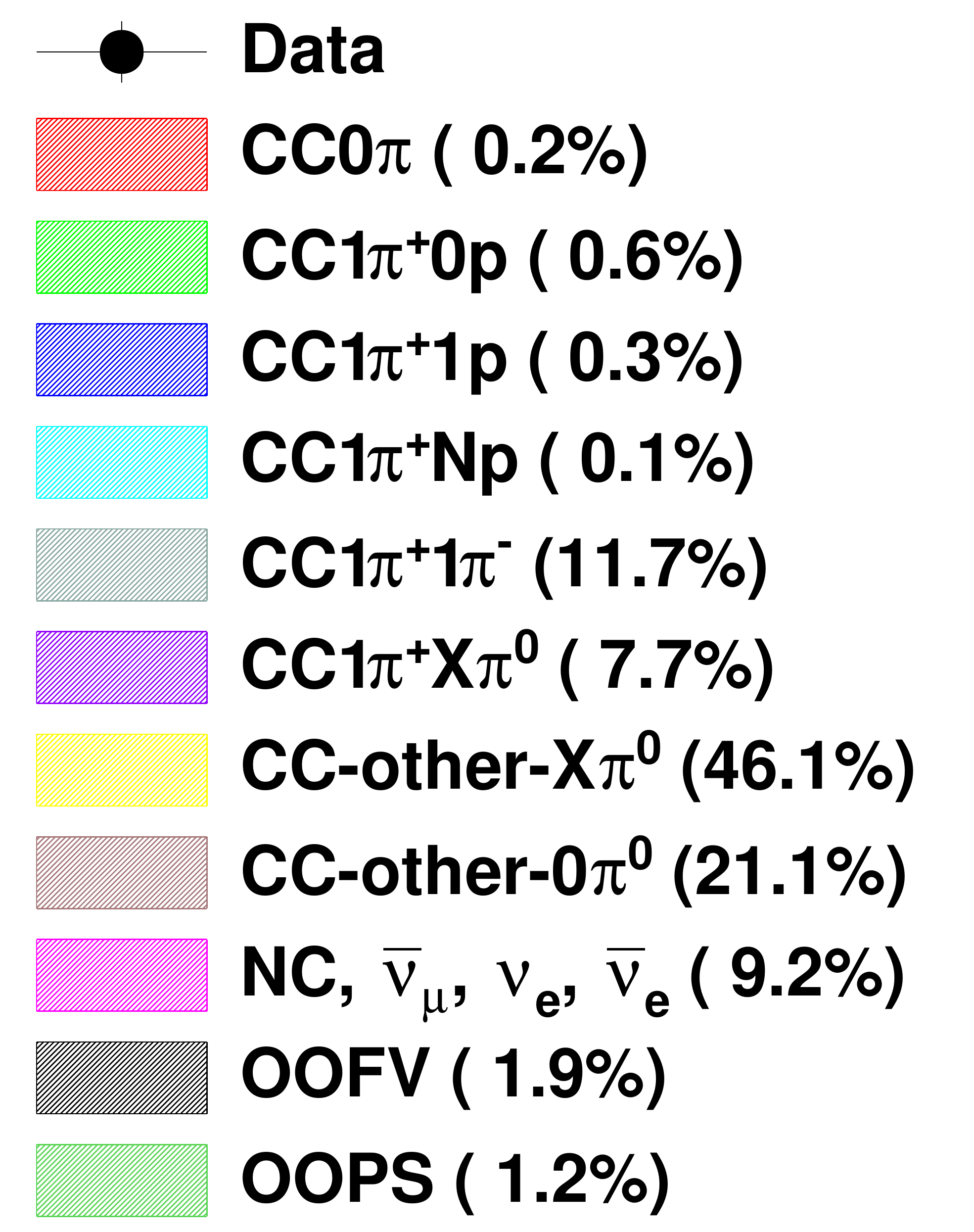}}
    \hfill
    \subfloat{\includegraphics[width=0.27\linewidth]{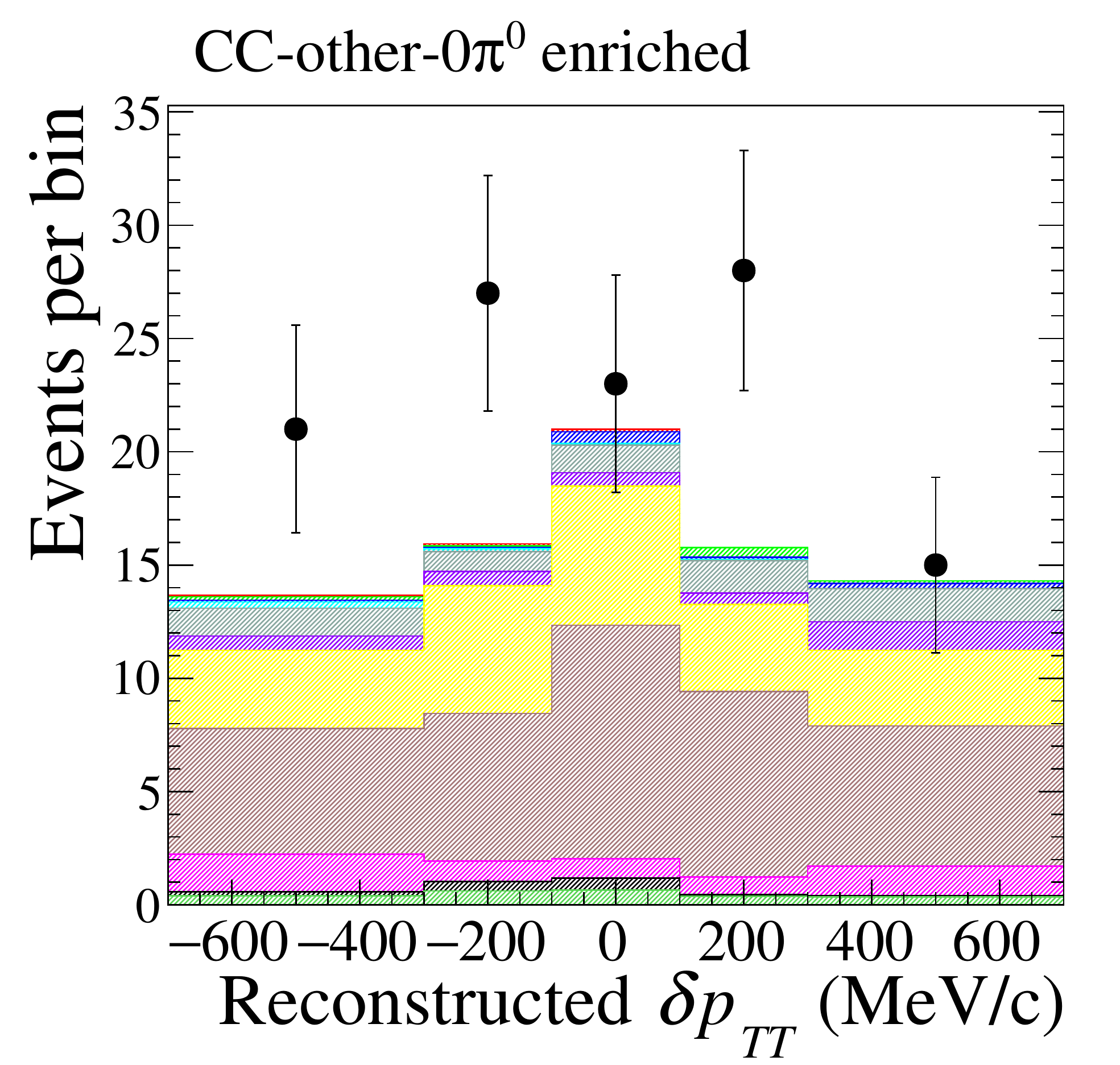}}
    \subfloat{\includegraphics[width=0.27\linewidth]{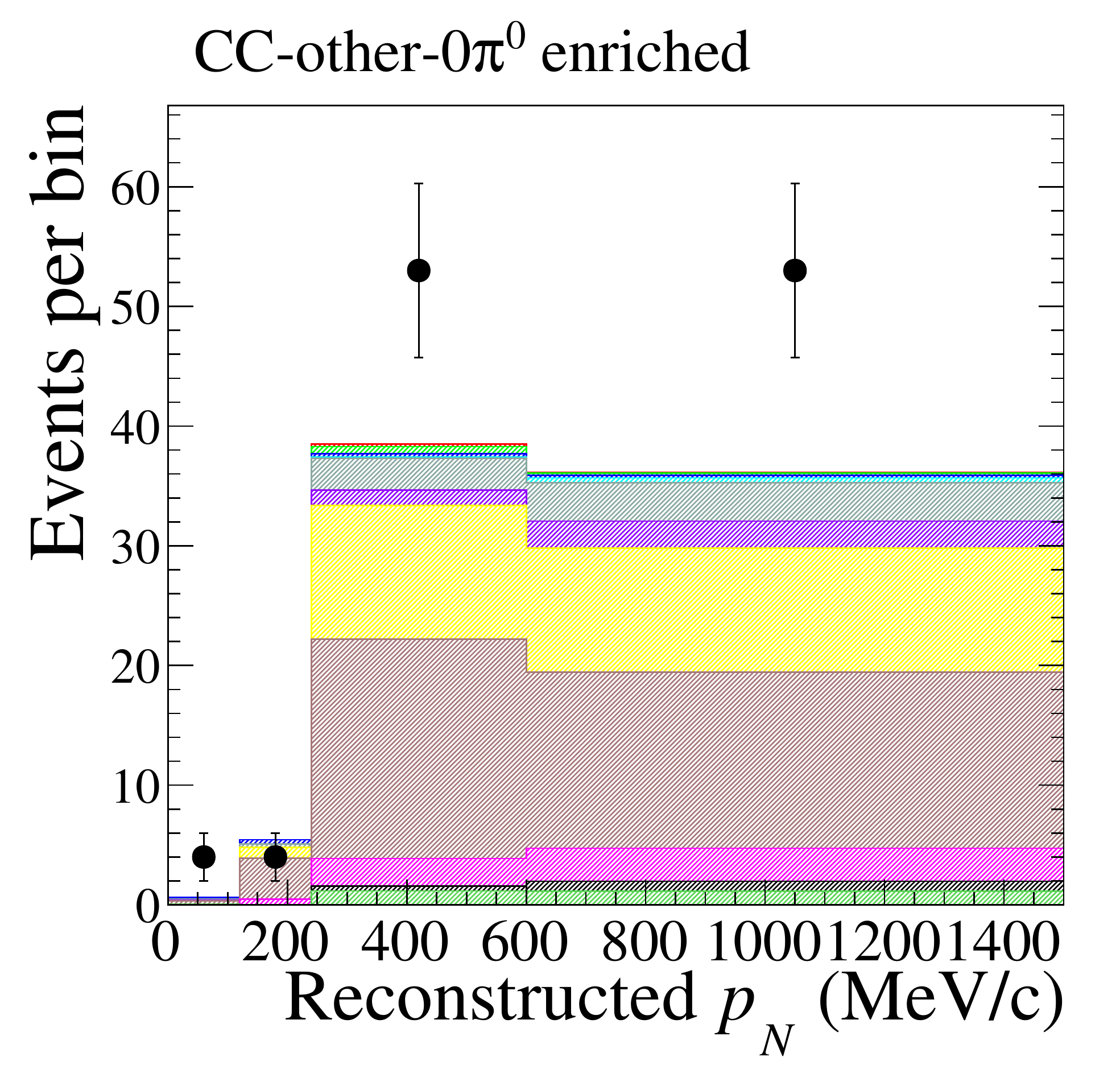}}
    \subfloat{\includegraphics[width=0.27\linewidth]{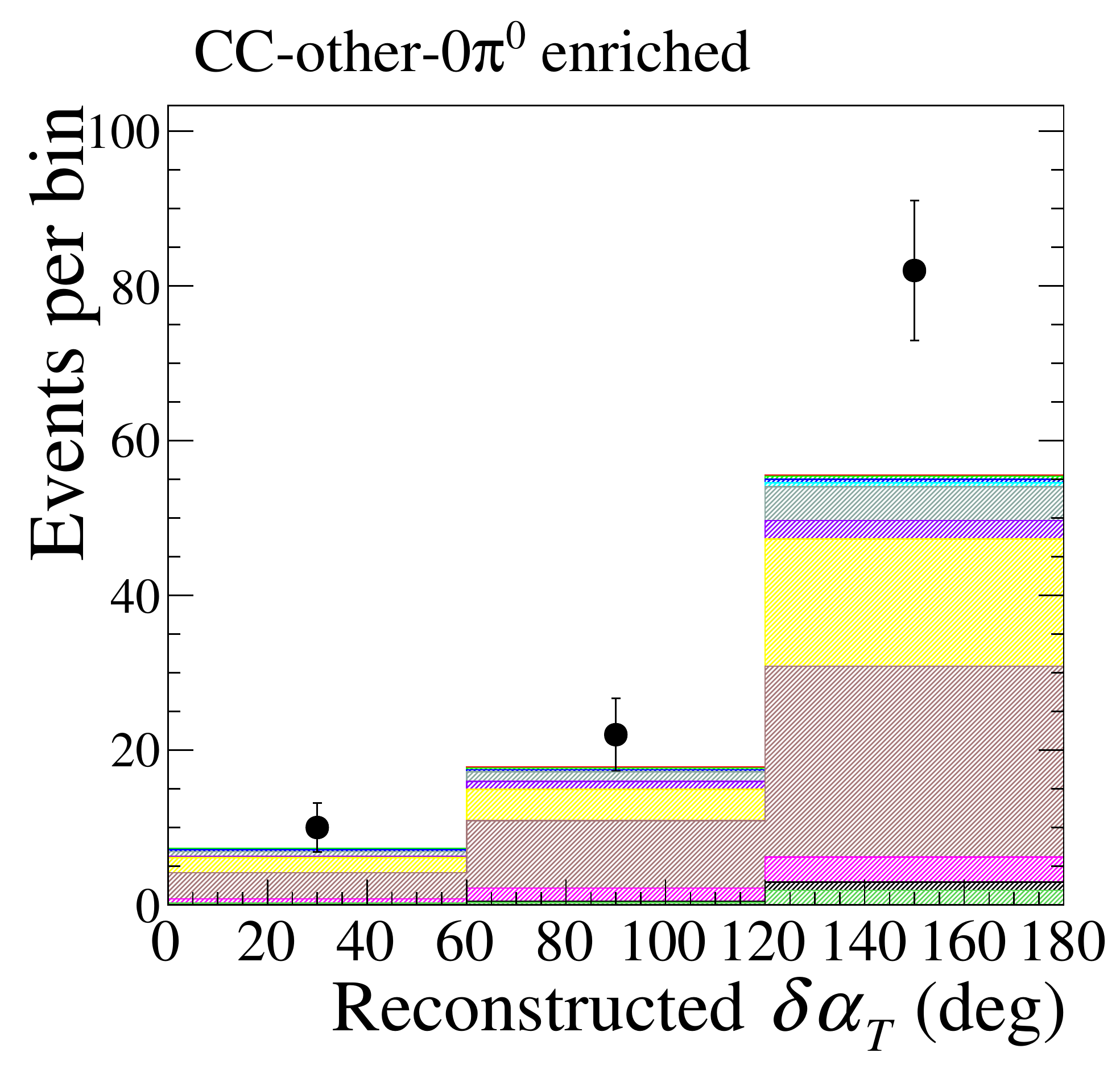}}
    \subfloat{\includegraphics[width=0.19\linewidth]{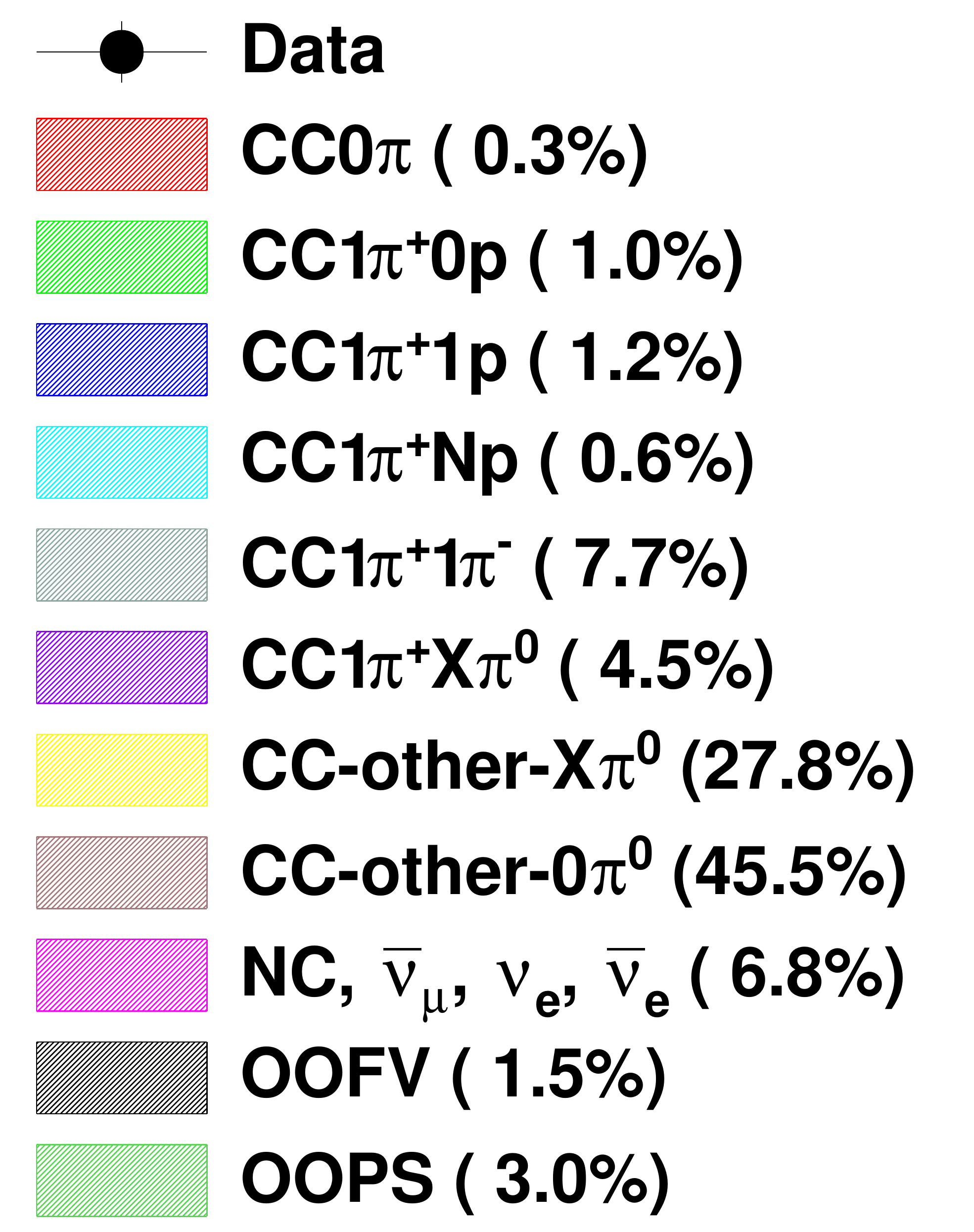}}
    \hfill
\caption{The distribution of events in the four control samples (top to bottom) as a function of reconstructed TKI variables (left to right), broken down into true final-state topology predicted by the nominal MC. The legends show the fraction of events in each control sample.  Histograms are stacked. The MC has been normalized to 11.6$\times 10^{20}$ POT, the equivalent number of POT collected for the data. The error bars show the statistical uncertainty in data.} \label{fig:control_sample}
\end{figure*}

To better constrain the CC-other background in the signal sample, dedicated control samples (on the right of \cref{fig:sample_schematics}) are selected based on the number of charged and neutral pions identified in the events. Following the FGD1-TPC $\mu^-$, $\pi^+$ and p tracks selection described in \cref{sec:signal_sample}, the control samples require the identification of additional $\pi^\pm$ tracks in the FGD/TPC or the identification of a $\pi^0$ in the tracker-ECal. These events are then classified into four samples according to the additional identified pions: 
\begin{enumerate}[(i)]
    \item CC$1\pi^+$1$\pi^-$ enriched sample - events with one $\pi^-$ candidate from FGD1 or the TPC;
    \item CC$1\pi^+$X$\pi^0$ enriched sample - events with $\pi^0$ candidates from the ECal;
    \item CC-other-X$\pi^0$ enriched sample - events with charged pion candidates from FGD1 or the TPC, and $\pi^0$ candidates from the ECal;
    \item CC-other-0$\pi^0$ enriched sample - events with charged pion candidates from FGD1 or the TPC, excluding the case of single $\pi^-$ candidate.
\end{enumerate}
The four separate samples allow for better characterization of the pion emission model and detector responses to different particles compared to a single CC-other sample. The same kinematic cuts in \cref{tab:signal_sample_ps_cut} are applied to the $\mu^-$, highest momentum $\pi^+$ and p tracks, and the TKI variables are calculated using only these tracks. The selected events are binned in the reconstructed TKI variable only, using the same binning in \cref{tab:signal_sample_binning}. \cref{fig:control_sample} shows the reconstructed TKI variable distributions for the four control samples. The nominal MC shows a deficit of events and also some shape discrepancies with respect to data, indicating the need for background correction.

\section{Analysis method}\label{sec:analysis}
\subsection{Binned likelihood fitting}
The analysis is performed using an unregularized binned likelihood fit as in Refs.~\cite{PhysRevD.93.112012,Abe:2018pwo,PhysRevD.98.012004,PhysRevD.101.112001,PhysRevD.101.112004}, with control samples to constrain the background, to unfold the detector smearing and extract the number of selected signal events from the signal sample. Compared to previous cross-section analyses, significant improvements have been achieved in the analysis framework, including the use of principle component analysis to reduce the dimensionality of the fit, and the proper treatment of MC statistical uncertainties. An unregularized fit means that there is no prior constraint on the shape of TKI from the input signal model, thus reducing model bias on the fitted cross sections. 
The numbers of signal events (and thus cross sections) as a function of the three TKI variables are fitted independently in this study.

The input MC is varied by a set of fit parameters, and the set of parameters which best describes the observed data is extracted together with its associated errors. 
The fit parameters of primary interest are the ``signal template parameters", $c_i$, which scale the number of signal events in the truth TKI variable bin $i$ without prior constraints. The remaining parameters are the nuisance parameters which describe plausible systematic variations of the flux, detector response and neutrino interaction model. The effect of these parameters is propagated to the number of selected events in the reconstructed bins.

The best-fit parameters are found by minimizing the following negative log-likelihood ($\chi^2$):
\begin{equation}\label{eq:fitchi2}
    \chi^2=-2\log(L)=-2\log(L_\textrm{stat})-2\log(L_\textrm{syst}),
\end{equation}
where 
\begin{equation}\label{eq:poisson_llh}
\begin{split}
   \chi^2_{\textrm{stat}}&= -2\log(L_{\textrm{stat}})\\
   &=\sum_j^{\textrm{reco. bins}}2\left(\beta_jN_j^{\textrm{MC}}-N_j^{\textrm{obs}}+N_j^{\textrm{obs}}\log\frac{N_j^{\textrm{obs}}}{\beta_jN_j^{\textrm{MC}}}\right.\\
   &\left.\qquad\qquad\qquad\qquad\qquad\qquad+\frac{(\beta_j-1)^2}{2\sigma^2_j}\right),
\end{split}
\end{equation}
and 
\begin{equation}\label{eq:sys_llh}
\begin{split}
    \chi^2_{\textrm{syst}}&=-2\log(L_{\textrm{syst}})\\
    &=(\vec{a}^{\textrm{ syst}}-\vec{a}^{\textrm{ syst}}_{\textrm{prior}})^T(V^{\textrm{syst}}_{\textrm{ cov}})^{-1}(\vec{a}^{\textrm{ syst}}-\vec{a}^{\textrm{ syst}}_{\textrm{prior}}).
\end{split}
\end{equation}
\cref{eq:poisson_llh} is the modified Poisson likelihood ratio which includes the statistical uncertainty of finite MC statistics using the Barlow-Beeston method~\cite{BARLOW1993219,Prosper:1306523}. $N_j^{MC}$ and $N_j^{obs}$ are the number of events in each reconstructed bin~$j$, for MC and data respectively. $\beta_j$ is the Barlow-Beeston scaling parameter given by
\begin{equation}
    \beta_j=\frac{1}{2}\left(-(N_j^{\textrm{MC}}\sigma_j^2-1)+\sqrt{(N_j^{\textrm{MC}}\sigma_j^2-1)^2+4N_j^{\textrm{obs}}\sigma_j^2}\right),
\end{equation}
and $\sigma_j^2$ is the relative variance of $N_j^{MC}$. In the limit of infinite MC statistics, $\sigma_j\to 0$ and $\beta_j\to 1$ which gives the standard Poisson likelihood ratio. \cref{eq:sys_llh} describes how well the nuisance parameters $\vec{a}^{\textrm{ syst}}$ agree with their prior values $\vec{a}^{\textrm{ syst}}_{\textrm{prior}}$, where $V^{\textrm{syst}}_{\textrm{cov}}$ is the covariance matrix describing the confidence in the prior values as well as correlations between parameters. 

The MC prediction $N_j^\textrm{MC}$ in the signal and control samples is composed of both the signal and background events, which can be written as 
\begin{equation}
    N_j^{\textrm{MC}}=\sum_i^{\textrm{true bins}}(c_iw_{i,j}^{\textrm{sig}}N_{i,j}^{\textrm{sig}}+w_{i,j}^{\textrm{bkg}}N_{i,j}^{\textrm{bkg}}),
\end{equation}
where $N_{i,j}^{\textrm{sig}}$ and $N_{i,j}^{\textrm{bkg}}$ are the number of signal and background events in the truth bin~$i$, contributing to the reconstructed bin~$j$, predicted by the T2K MC; $w_{i,j}^{\textrm{sig}}$ and $w_{i,j}^{\textrm{bkg}}$ are the event weights coming from the same set of systematic variations and thus are correlated.

\subsection{Sources of systematic uncertainties}\label{sec:sys_err}
Three sources of systematic uncertainties are considered in this analysis.
\begin{description}
    \item[Neutrino flux uncertainty] This is parametrized as scale factors in bins of true neutrino energy (same binning as in \cref{fig:fluxerror}). Such scale factors are constrained by their prior uncertainty, encoded in a covariance matrix. At the same energy, identical event weights are applied on the signal and background events.
    \item[Detector uncertainty] The detector response (efficiency and resolution) is not perfectly modelled in the simulation. Dedicated and independent control samples are used to evaluate each possible uncertainty based on the data-MC agreement. The overall detector uncertainty is parametrized as a covariance matrix that describes the rate uncertainty and correlation between each reconstructed bin. The uncertainty related to the modelling of the pion secondary interactions, one of the largest detector systematics in previous T2K analyses, has been reduced by around 40\% using external data and the cascade model implemented in NEUT~\cite{PhysRevD.99.052007}. 
    In the signal sample and control samples without reconstructed $\pi^0$, the biggest uncertainty comes from the modelling of proton secondary interactions which causes a 5\% uncertainty on the event rate. On the other hand, $\pi^0$-tagging uncertainty is dominant (around 10\%) in the control samples with reconstructed $\pi^0$.
    \item[Neutrino interaction model uncertainty] This takes care of both the modelling of signal and background interactions, including FSI. 
    In this analysis, the estimation of signal efficiency and background contamination are most significantly affected by the RES and DIS processes. 
    In the RES channel, there are three model parameters: the resonant axial mass $M_\text{A}^\text{RES}$ (1.07$\pm$0.15~GeV/c$^2$), the value of the axial form factor at zero transferred 4-momentum $C^\text{A}_5$ (0.96$\pm$0.15), and the normalization of the isospin non-resonant component $I_{1/2}$ (0.96$\pm$0.40) predicted in the Rein-Sehgal model. Initial central values and uncertainties for these parameters are obtained in a fit to low energy neutrino-deuterium single pion production data from ANL~\cite{PhysRevD.25.1161,PhysRevD.23.569} and BNL~\cite{PhysRevD.34.2554,PhysRevD.23.2495,Furuno:2003ng} (flux-corrected data in Ref.~\cite{PhysRevD.90.112017} is used), and carbon-like data from MiniBooNE~\cite{PhysRevD.83.052007}. One additional parameter varying the $\Delta^{++}$ decay width with 50\% uncertainty, and ad hoc scale parameters binned in signal particle momenta and angles with a 20\% uncertainty, are included to give extra freedom to the efficiency correction. The ad hoc variations are chosen to cover the efficiency's dependency on the initial state nuclear medium effects, which is not otherwise parametrized.
    \\In the DIS channel, a CC-other shape parameter $x^\text{CC-Other}$ with a 40\% uncertainty is used, which scales the cross section by $(1+x^\text{CC-Other}/E_\nu)$ and gives greater flexibility at low $E_\nu$. Four normalization parameters with a 50\% uncertainty, with the same categorization as the four CC-other topologies, are introduced to better parametrize multiple pion production. The neutral current and electron (anti)neutrino interactions, which are not constrained by the control samples, are given a normalization uncertainty of 30\% and 3\% respectively.
    \\Finally, there are parameters varying the pion and proton FSI. The tunable pion interactions in the nucleus are charge exchange, where the charge of the pion changes; absorption, where the pion is absorbed through two- or three-body processes; elastic scattering, where the pion only exchanges momentum and energy; and inelastic scattering, where additional pions are produced. Their prior is given by Ref.~\cite{PhysRevD.99.052007}. For proton FSI, there is a single parameter scaling the overall interaction probability inside the cascade with a 50\% uncertainty, without tuning specific processes. 
    \\It is verified that with such comprehensive list of parameters, the fit can cover the bias in signal efficiency and background subtraction under extreme model variations as discussed in \cref{sec:err_prop}.
\end{description}

\subsection{Cross section extraction, error propagation and validation}\label{sec:err_prop}
After the number of signal events is extracted from the fit, the differential cross section as a function of the true TKI variable is calculated by the following formula:
\begin{equation}\label{eq:xsec_formula}
    \frac{d\sigma}{dx_i}=\frac{N_i^{\textrm{signal}}}{\epsilon_i\Phi N^{\textrm{FV}}_{\textrm{nucleons}}\Delta x_i},
\end{equation}
where $N_i^{\textrm{signal}}$ is the measured number of signal events in the $i$-th bin, for all CC1$\pi^+$Xp events on hydrocarbon satisfying the kinematic phase restrictions in \cref{tab:signal_ps_cut}. Interactions on other elements are estimated by MC and subtracted. Since the fraction of non-hydrocarbon events is small, the potential bias due to cross-section or detector mismodelling is insignificant. 
$\epsilon_i$ is the selection efficiency in the $i$-th bin, contributed by both the signal and control samples. 
$\Phi$ is the overall flux integral, evaluated at the best-fit flux parameter values, and $N^{\textrm{FV}}_{\textrm{nucleons}}$ is the number of target nucleons (only hydrocarbon) in the fiducial volume. $x_i$ is one of the TKI variables and $\Delta x_i$ is the bin width.

\begin{figure}
\centering
    \subfloat{\includegraphics[width=\linewidth]{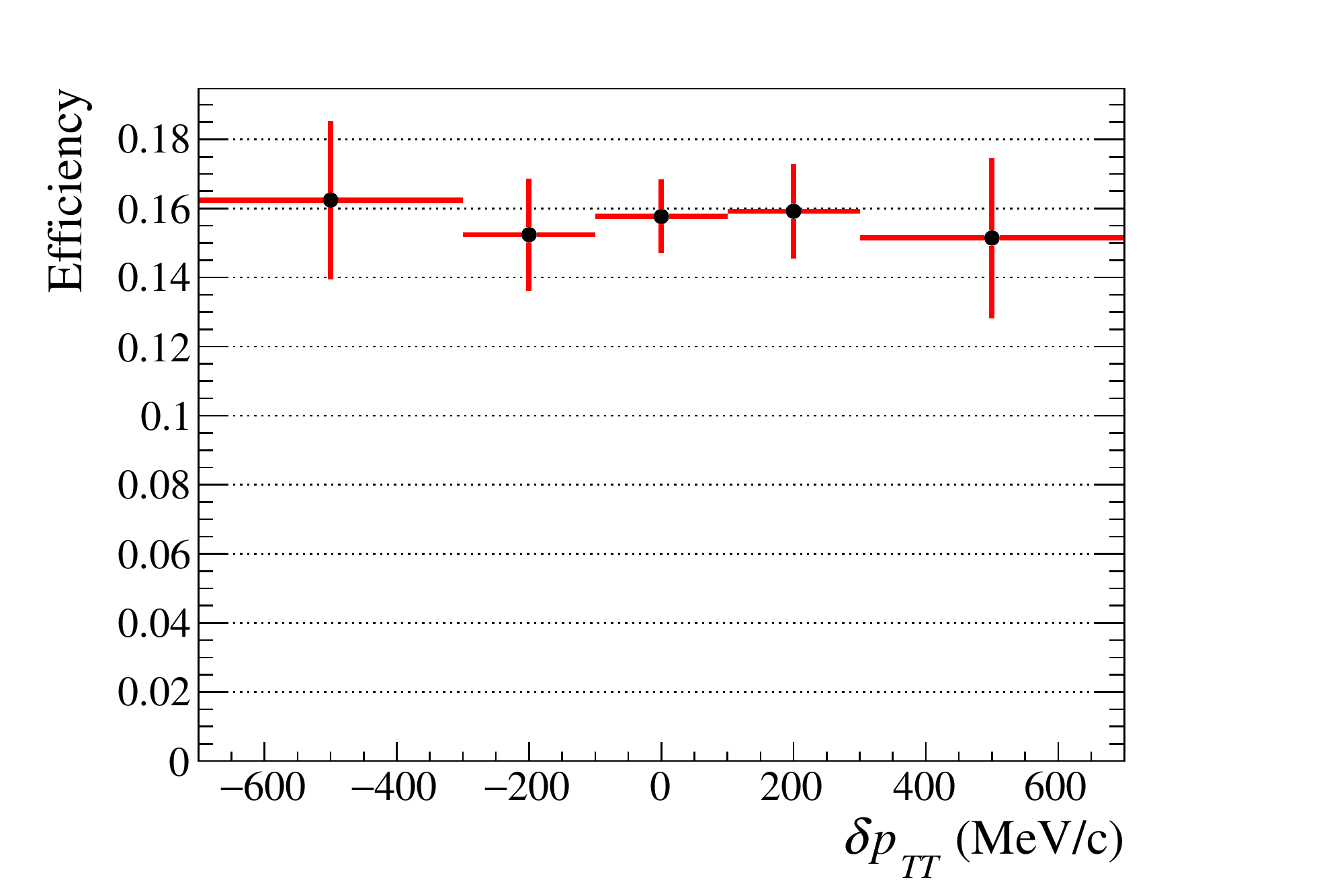}}\hfill
    \subfloat{\includegraphics[width=\linewidth]{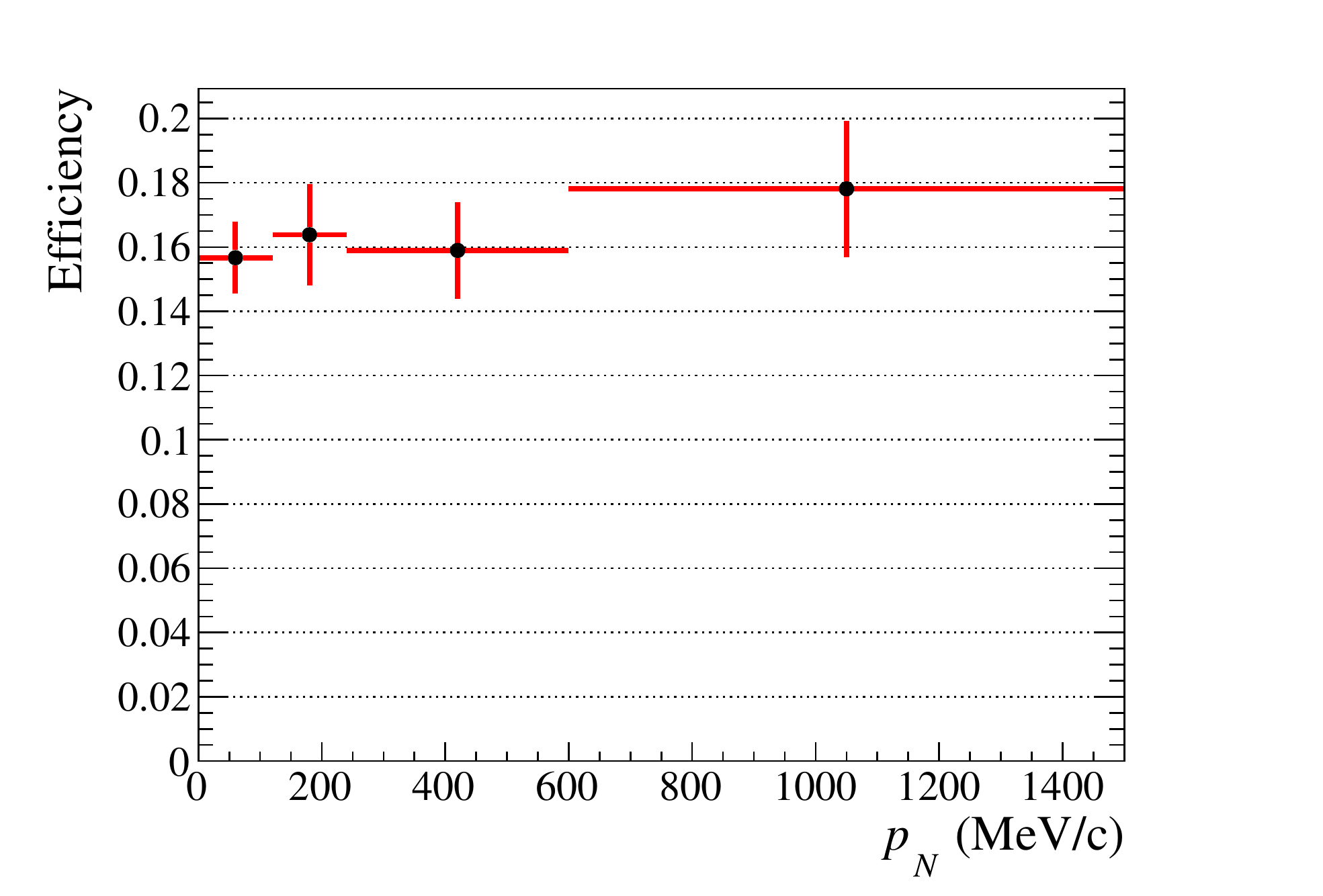}}\hfill
    \subfloat{\includegraphics[width=\linewidth]{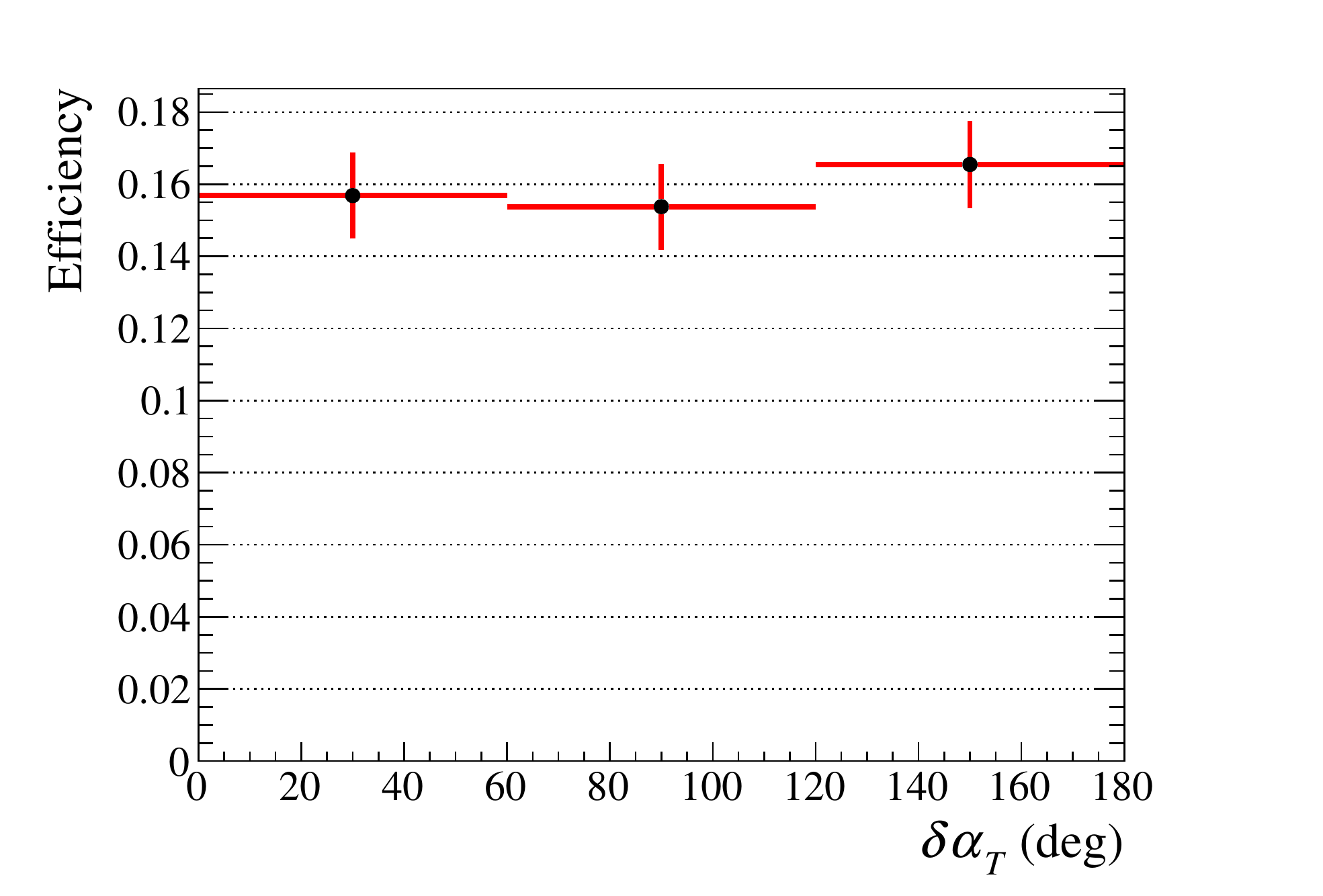}}\hfill
\caption{Mean values and uncertainties of the selection efficiencies as a function of the TKI variables. The error bars include both the statistical and systematic uncertainties propagated from the fit.} \label{fig:postfit_eff}
\end{figure}

We use a similar method as in Refs.~\cite{Abe:2018pwo,PhysRevD.102.012007} to numerically propagate the uncertainty of the fit to the cross section result, assuming the uncertainties of the fit parameters and cross sections are part of a Gaussian distribution. The covariance matrix of the fit parameters is Cholesky decomposed and multiplied by a vector of Gaussian random numbers to generate a set of random parameters. 
These random parameters are added to the best-fit parameters to create 2000 sets of variations (``toys") of parameters. This effectively samples the likelihood space encoded in the covariance matrix, and represents the spread of the plausible parameters according to the statistical and systematic uncertainties from the fit. For each toy, all variables in \cref{eq:xsec_formula} (except $\Delta x_i$), and thus $\frac{d\sigma}{dx_i}$, are re-evaluated with the toy parameters. The flux integral and selection efficiency are changed by the toy parameters. The resultant uncertainty of the flux integral is around 5\%, and \cref{fig:postfit_eff} shows the mean values and uncertainties of the efficiency extracted from toys. The number of target nucleons $N^{\textrm{FV}}_{\textrm{nucleons}}$ is sampled independently with a mean value of $5.5\times 10^{29}$ and an uncertainty of 0.67\%~\cite{AMAUDRUZ20121}. Finally, a covariance matrix $V$ of $\frac{d\sigma}{dx_i}$ is built from such toys. 
This method is different from the one used in previous analyses~\cite{PhysRevD.101.112004,PhysRevD.101.112001}, where the uncertainty was estimated by repeating the fit many times with toys of input MC. 

To ensure our results are not biased, a plethora of mock data studies with alternative neutrino event generators, nuclear ground state models, background models and altered flux models have been performed. It has been verified that even in the case of extreme deviations from the input MC model, such as doubling the signal/background interactions or completely turning off the FSI, the cross section extraction method employed can always recover the truth values to within a 1$\sigma$-uncertainty.  
The fit performance for every mock data study has been quantified by computing the post-fit p-value. 
First, 1000 sets of MC data samples are produced as a result of statistical and systematic variations of the nominal MC, which are then fitted to build the distribution of the post-fit $\chi^2$ (\cref{eq:fitchi2}).  
The p-value for each mock data study has been computed from this distribution and an acceptance threshold of 5\% has been chosen to quantify good fitter performances. 
All the mock data studies performed (without applying statistical fluctuations) have a p-value around 90\%, showing that the model differences are well covered by the conservative systematic uncertainties. On the other hand, the agreement on the measured cross sections is quantified by the $\chi^2_\text{tot}$ statistic:
\begin{equation}\label{eq:chi2_tot}
\begin{split}
    \chi^2_\text{tot}=&\sum_i\sum_j\left(\frac{d\sigma^\text{truth}}{dx_i}- \frac{d\sigma^\text{meas}}{dx_i}\right)
    \\&\cdot(V^{-1})_{ij}\left(\frac{d\sigma^\text{truth}}{dx_j}- \frac{d\sigma^\text{meas}}{dx_j}\right),
\end{split}
\end{equation}
where $\sigma^\text{meas}$ is the measured cross section, and $\sigma^\text{truth}$ is the truth cross section in the mock data. All mock data fits return a $\chi^2_\text{tot}$/ndof less than 0.4, where ndof is the number of degrees of freedom,  and a p-value greater 80\%, showing the robustness of the cross section extraction method employed for this analysis. 

\section{Results}\label{sec:results}
\begin{figure*}
\centering
    \subfloat{\includegraphics[width=0.33\linewidth]{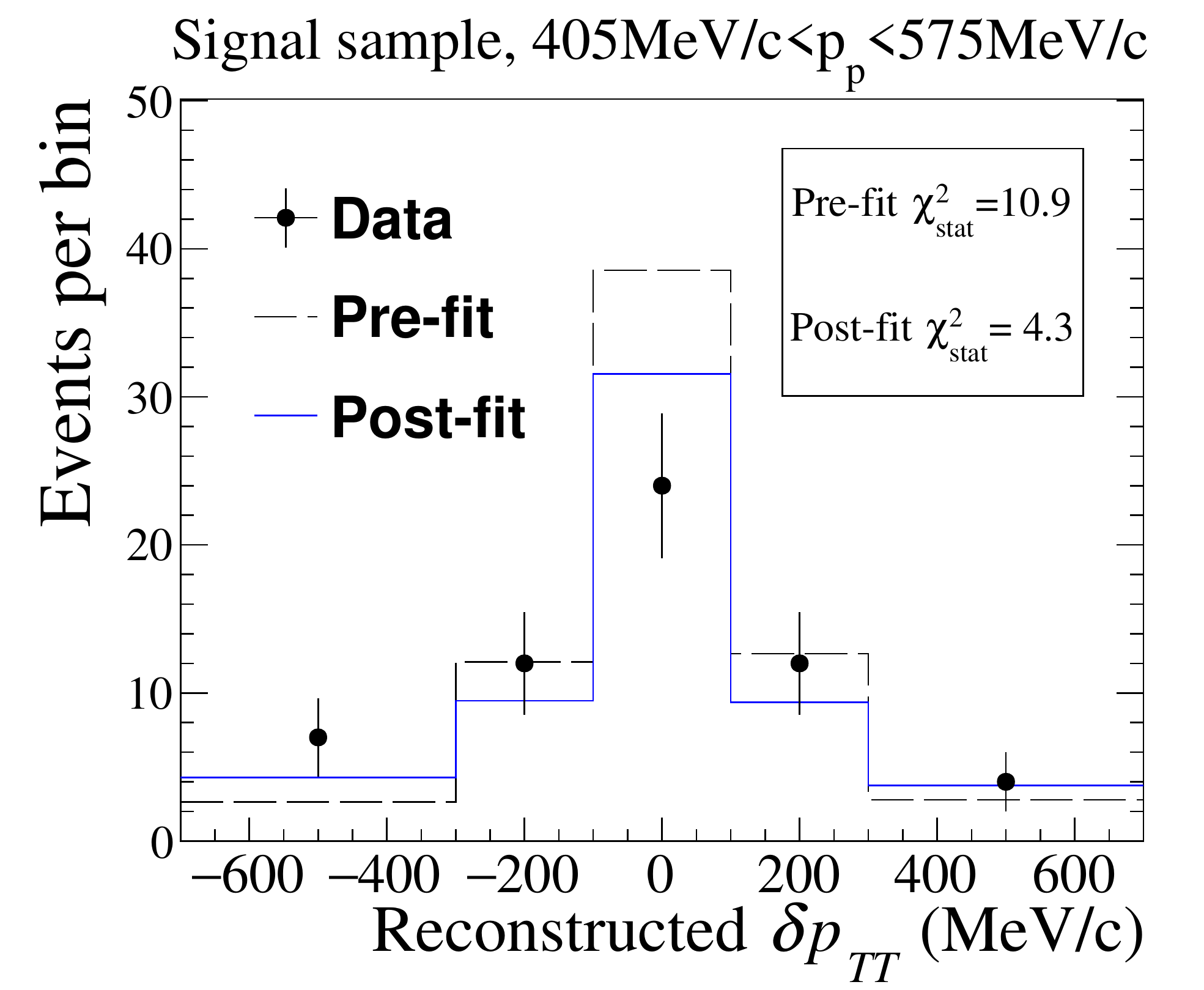}}
    \subfloat{\includegraphics[width=0.33\linewidth]{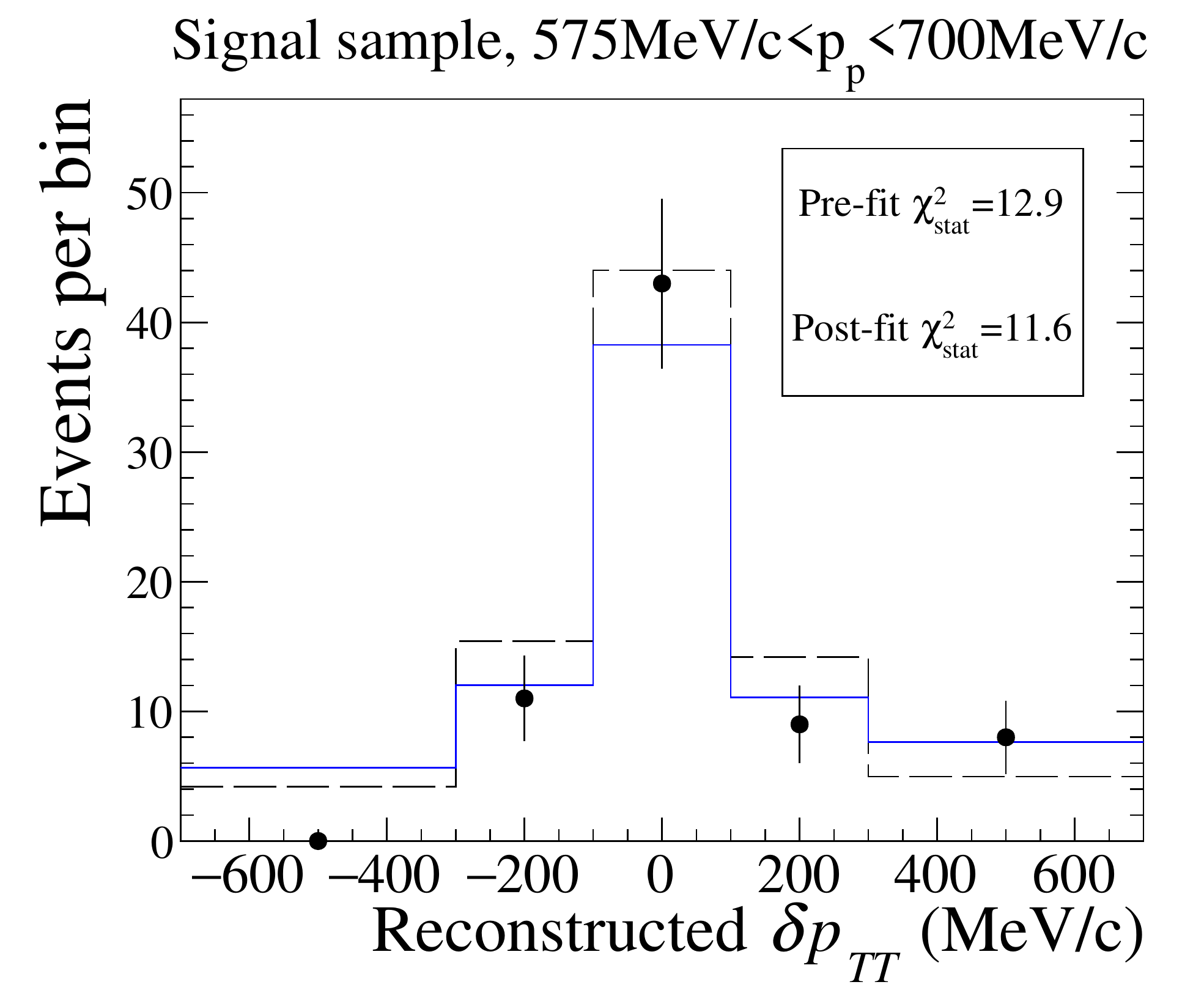}}
    \subfloat{\includegraphics[width=0.33\linewidth]{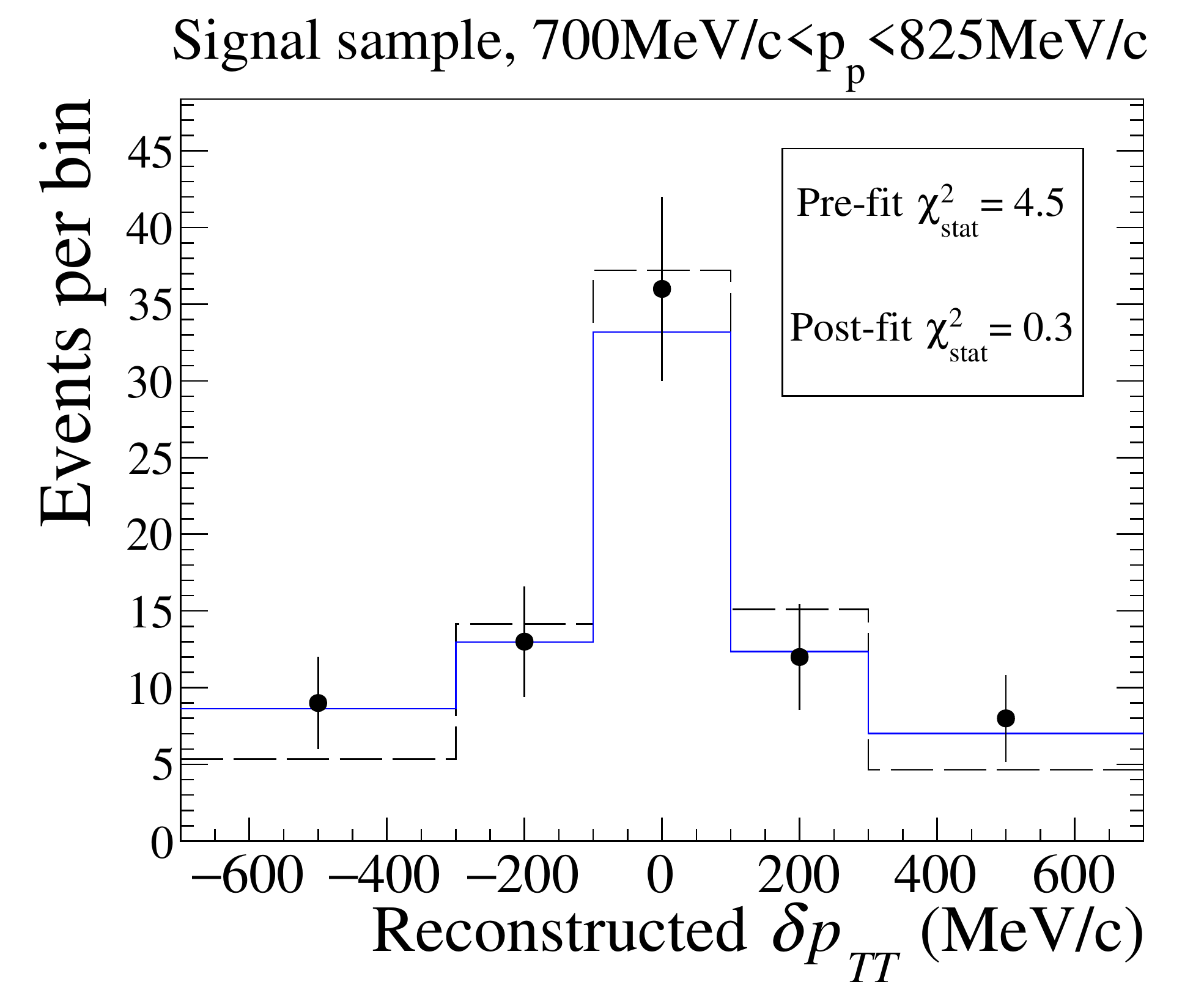}}\hfill
    \subfloat{\includegraphics[width=0.33\linewidth]{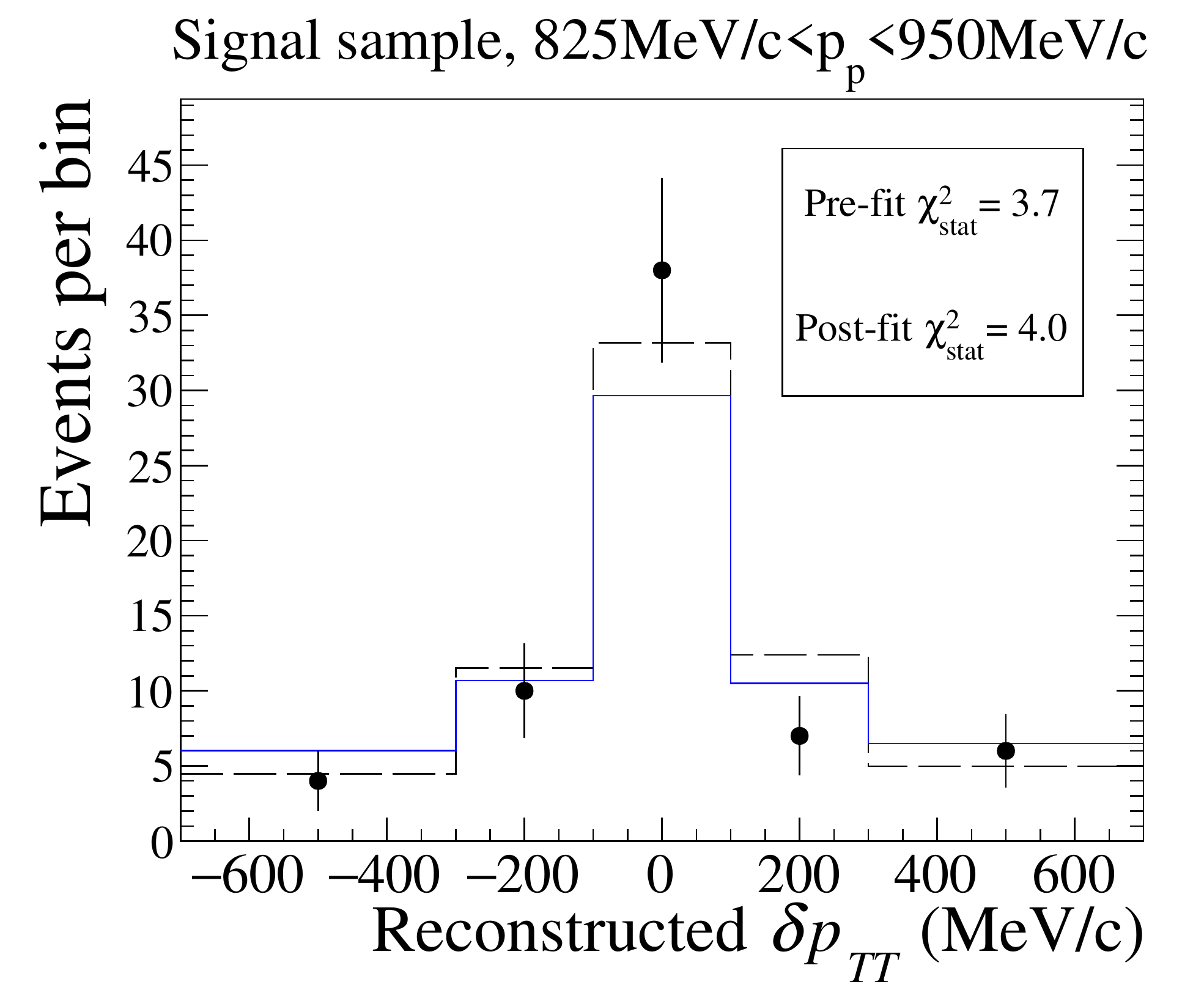}}
    \subfloat{\includegraphics[width=0.33\linewidth]{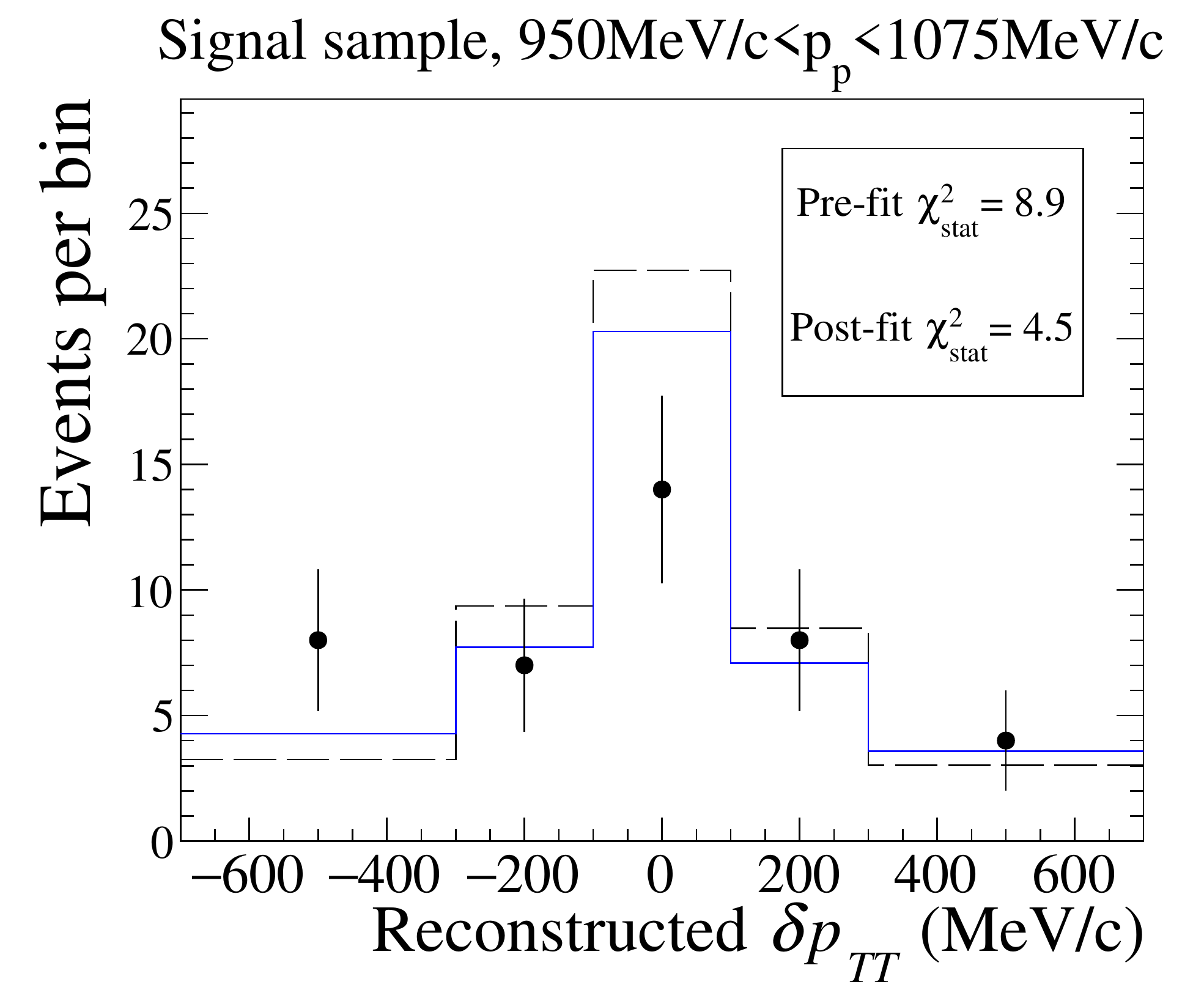}}
    \subfloat{\includegraphics[width=0.33\linewidth]{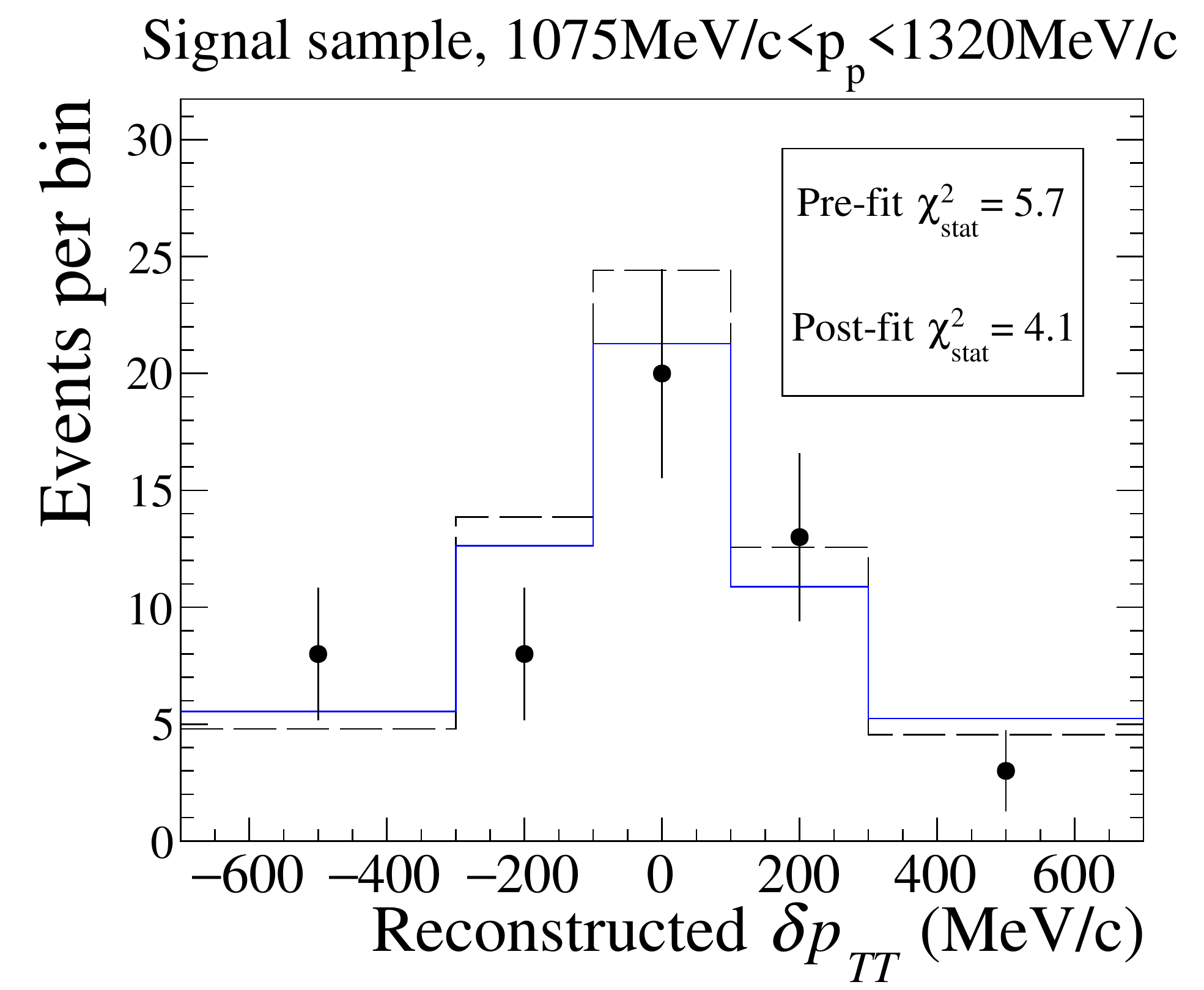}}\hfill
    \subfloat{\includegraphics[width=0.33\linewidth]{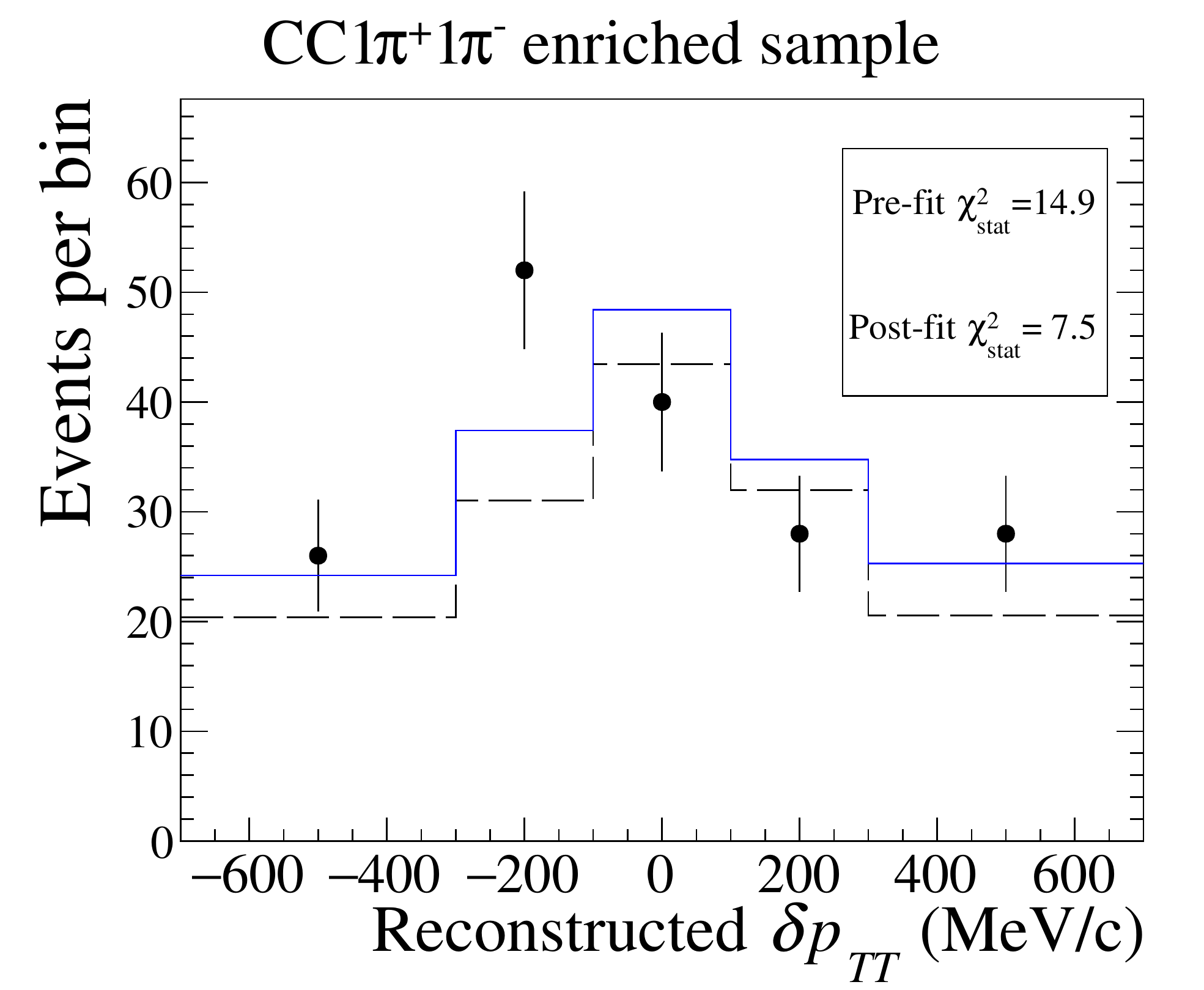}}
    \subfloat{\includegraphics[width=0.33\linewidth]{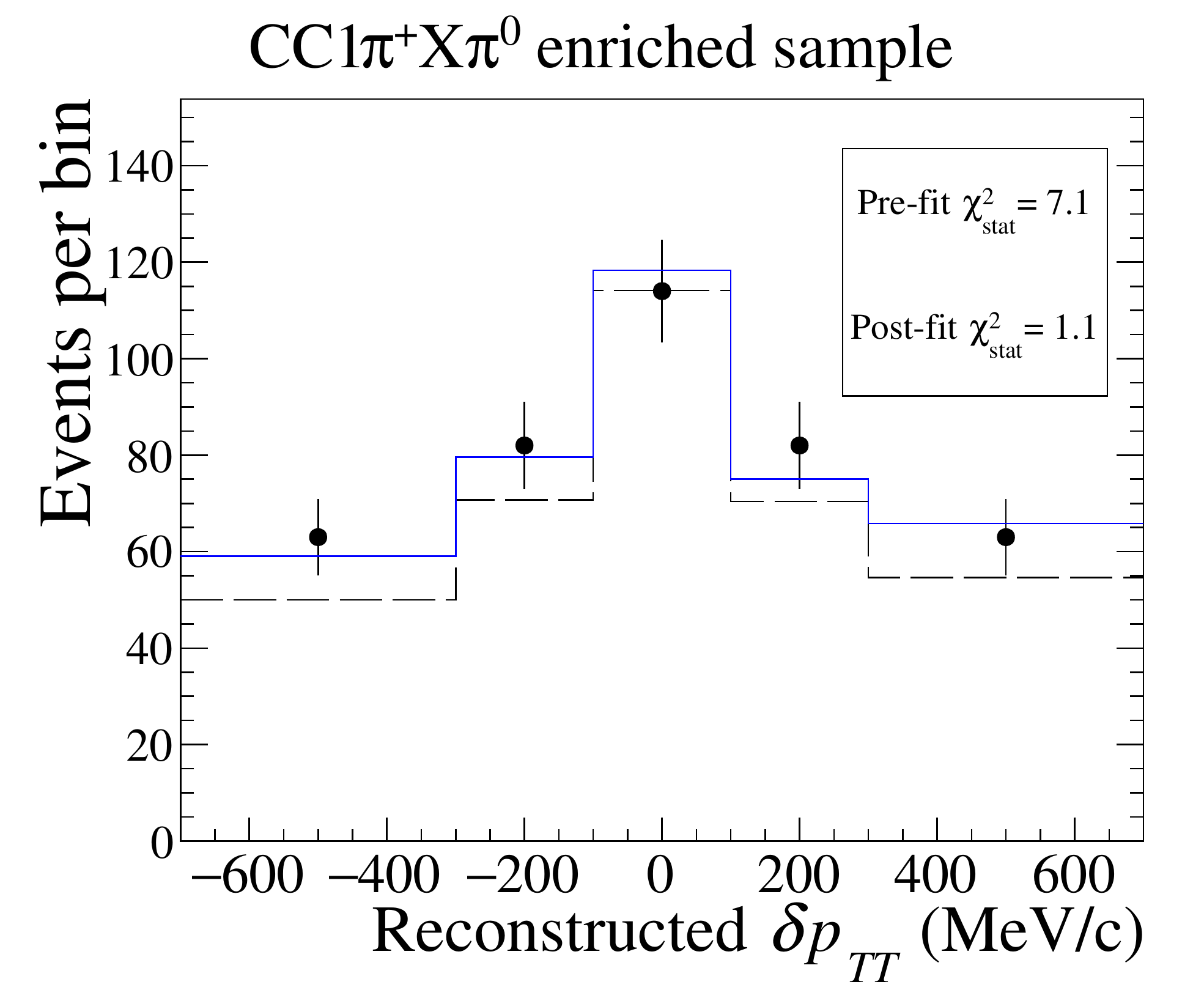}}\hfill
    \subfloat{\includegraphics[width=0.33\linewidth]{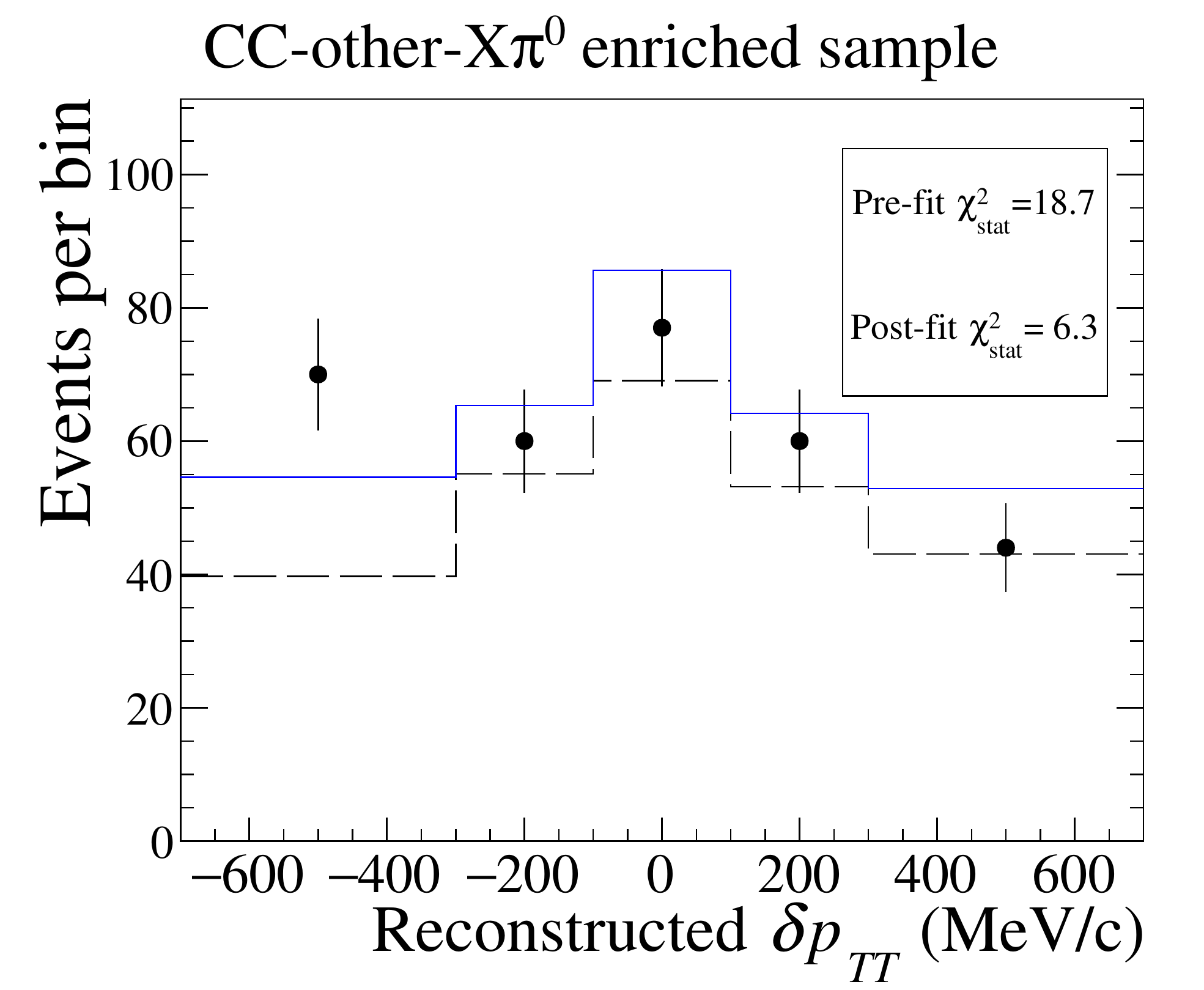}}
    \subfloat{\includegraphics[width=0.33\linewidth]{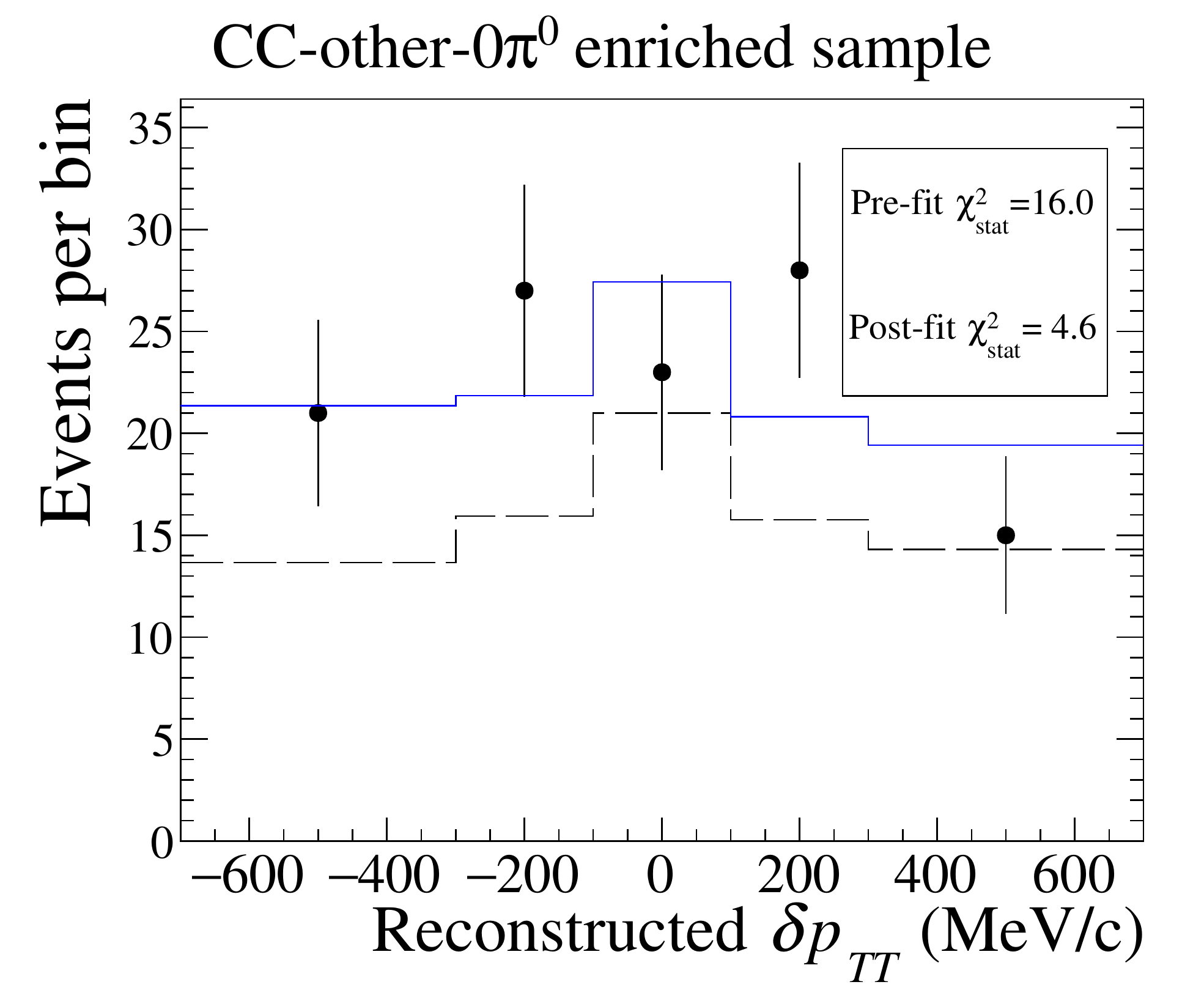}}\hfill
\caption{Distribution of events in the signal and control samples in the $\delta p_{TT}$ fit.  $\chi^2_\text{stat}$ corresponds to the statistical contribution of the fit $\chi^2$ (\cref{eq:poisson_llh}) in that sample. The  MC prediction before (dashed) and after (solid) the fit are also shown. The error bars show the statistical uncertainty in data.} \label{fig:dptt_postfit}
\end{figure*}

\begin{figure*}
\centering
    \subfloat{\includegraphics[width=0.33\linewidth]{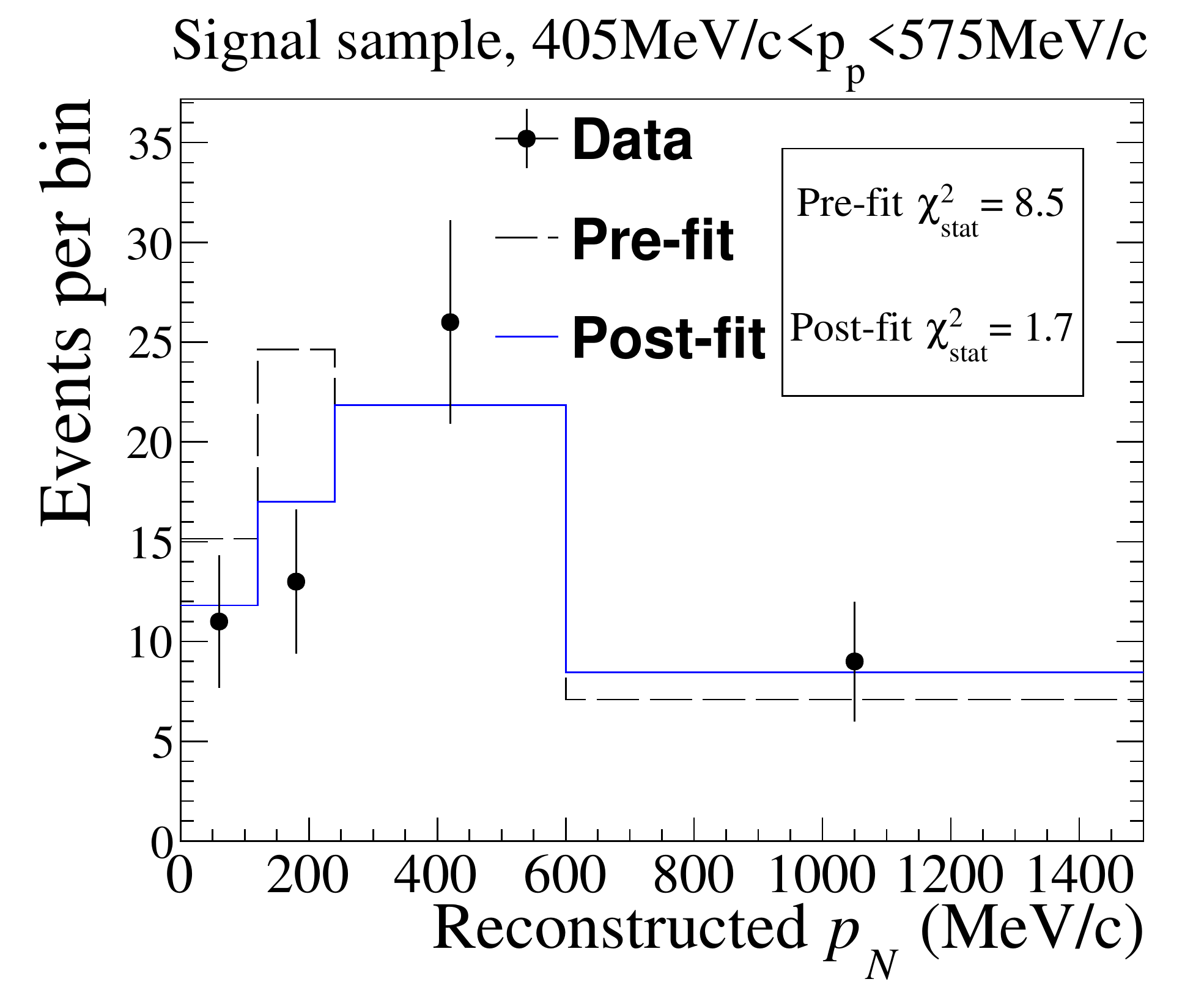}}
    \subfloat{\includegraphics[width=0.33\linewidth]{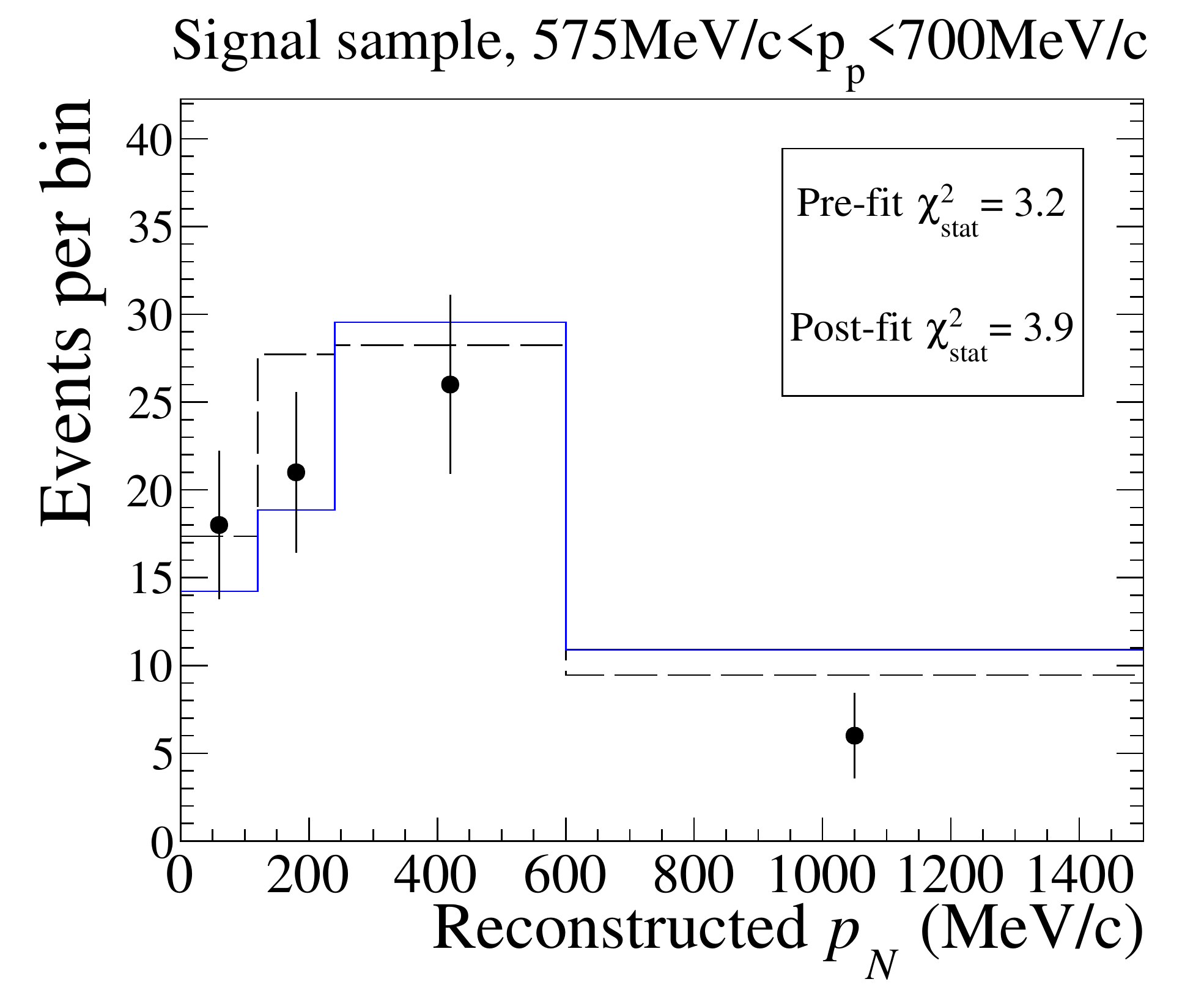}}
    \subfloat{\includegraphics[width=0.33\linewidth]{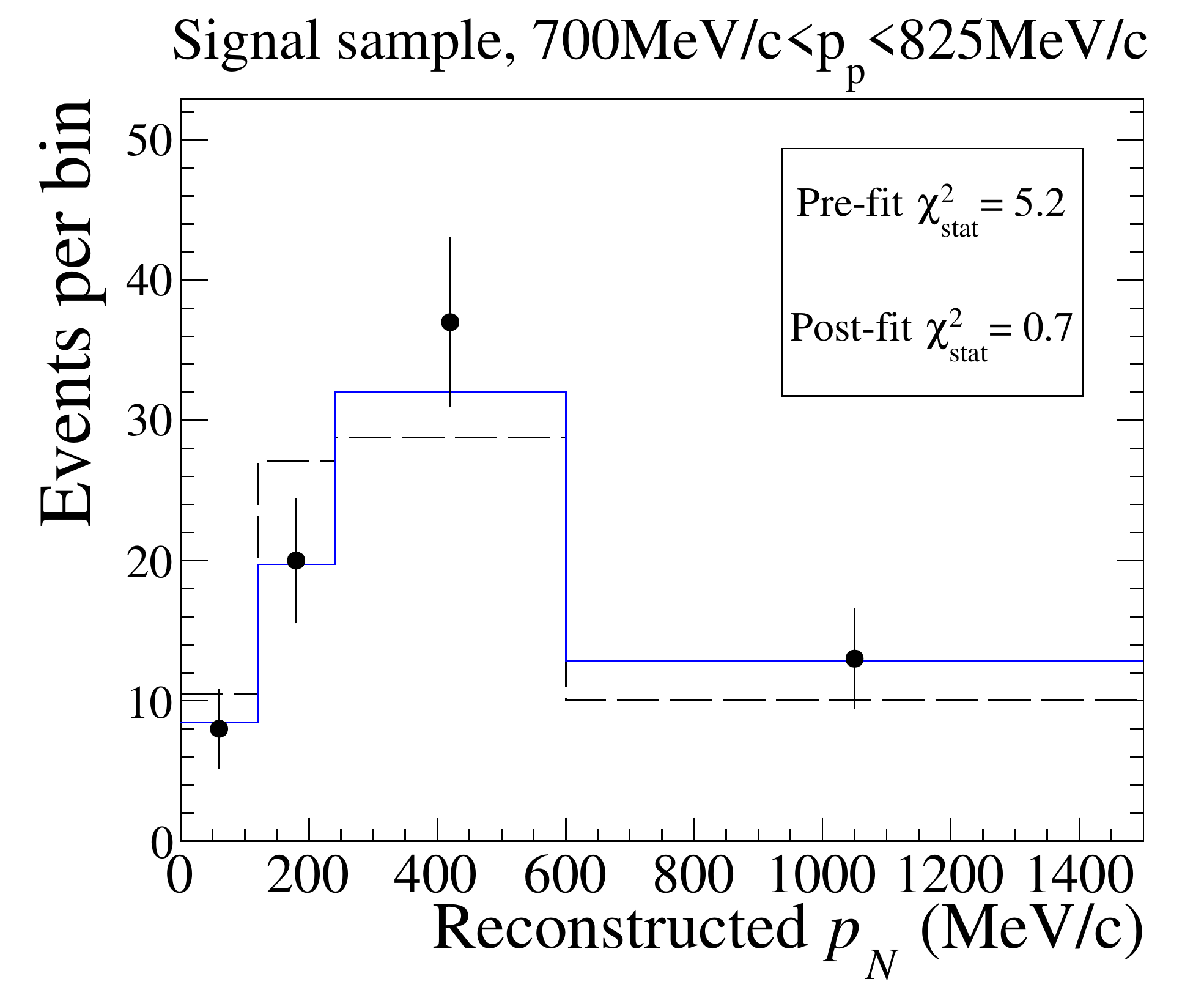}}\hfill
    \subfloat{\includegraphics[width=0.33\linewidth]{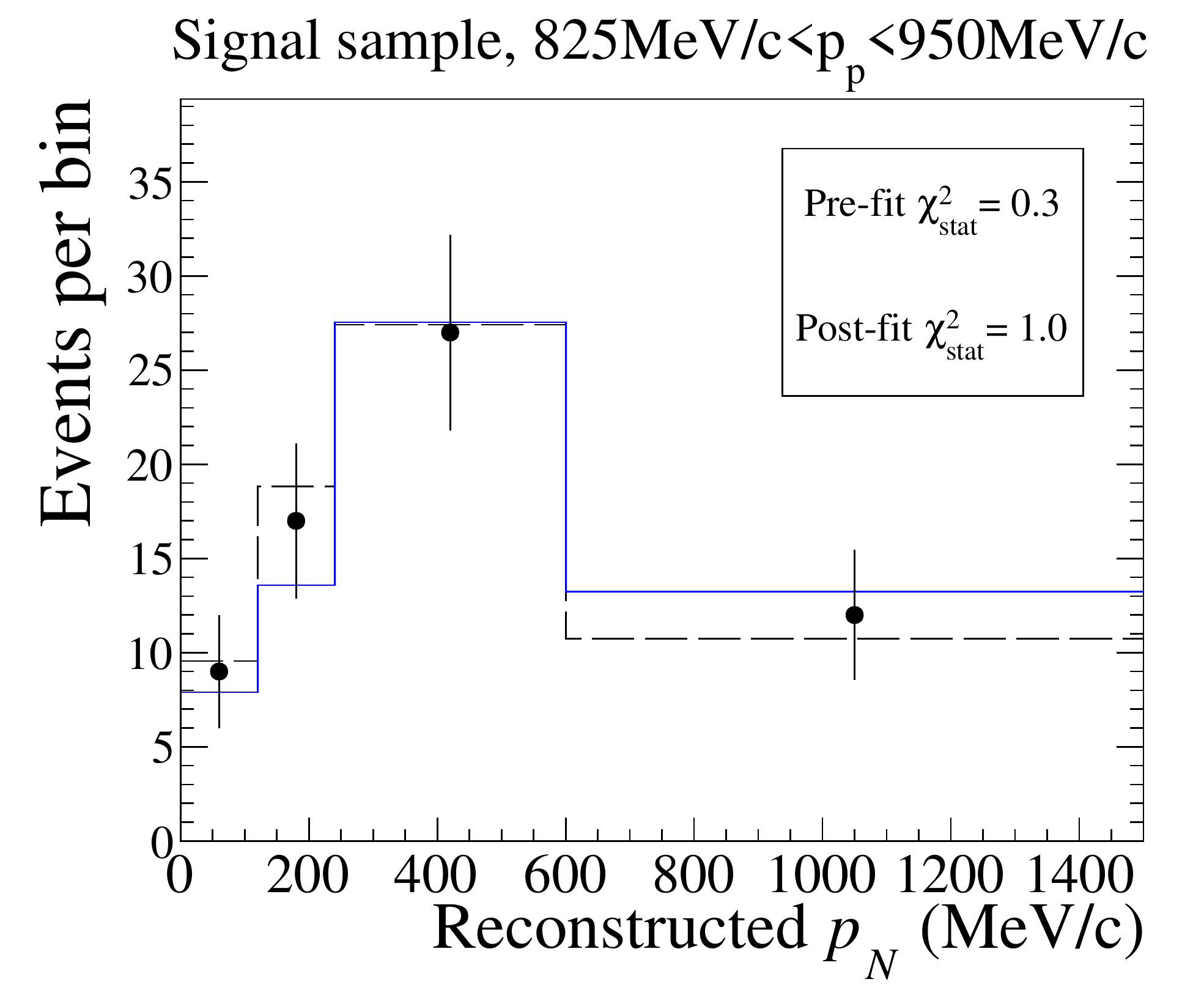}}
    \subfloat{\includegraphics[width=0.33\linewidth]{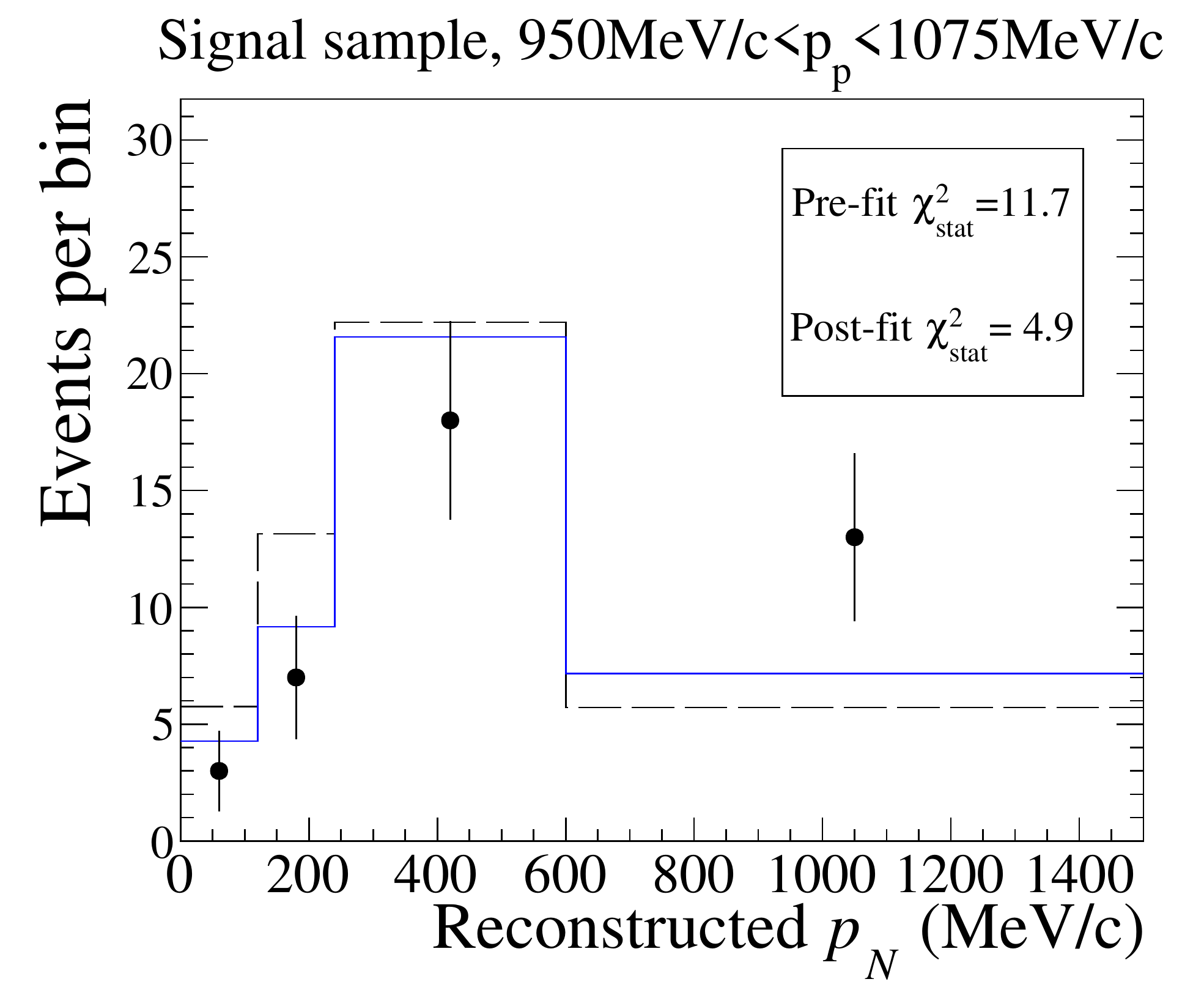}}
    \subfloat{\includegraphics[width=0.33\linewidth]{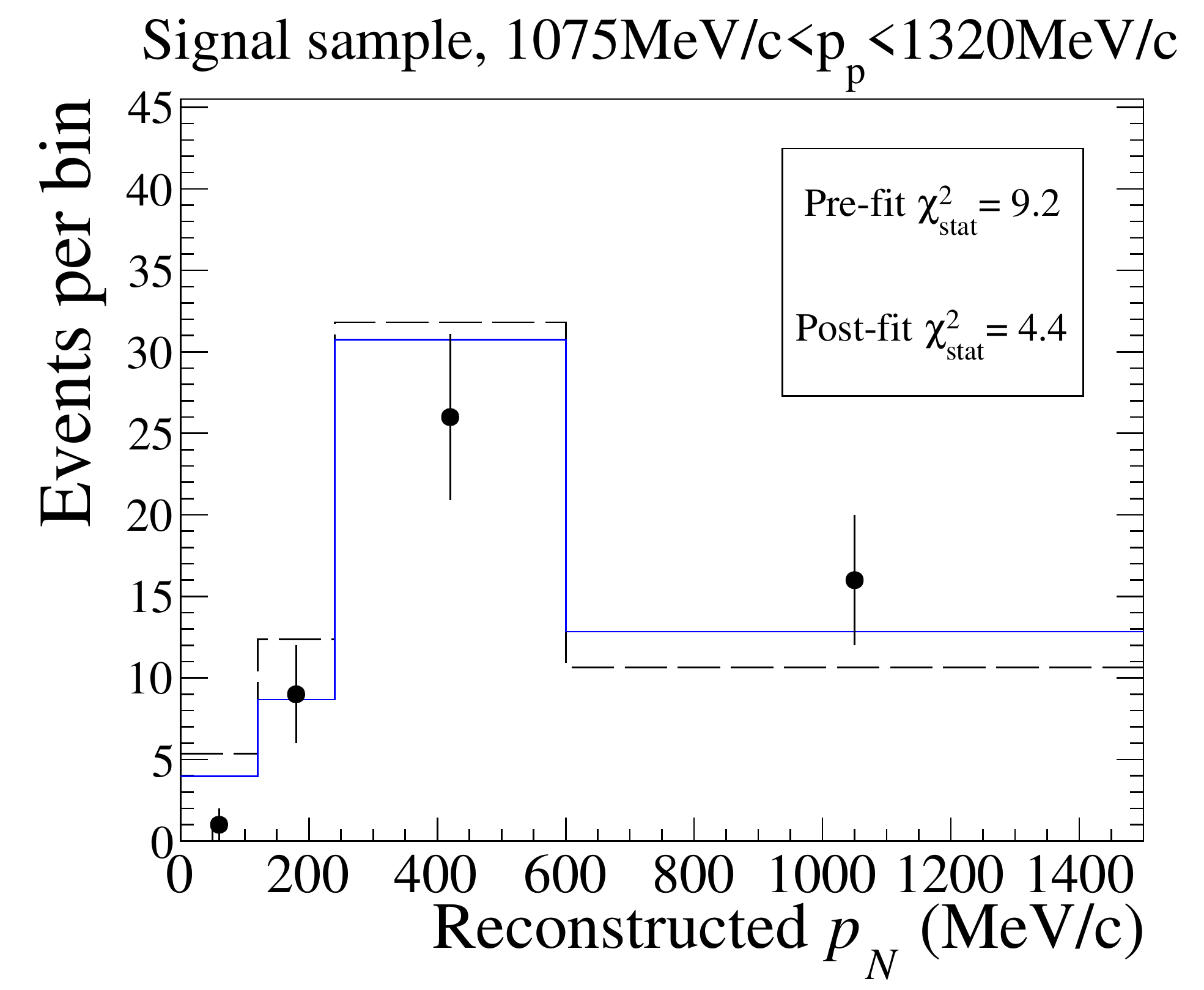}}\hfill
    \subfloat{\includegraphics[width=0.33\linewidth]{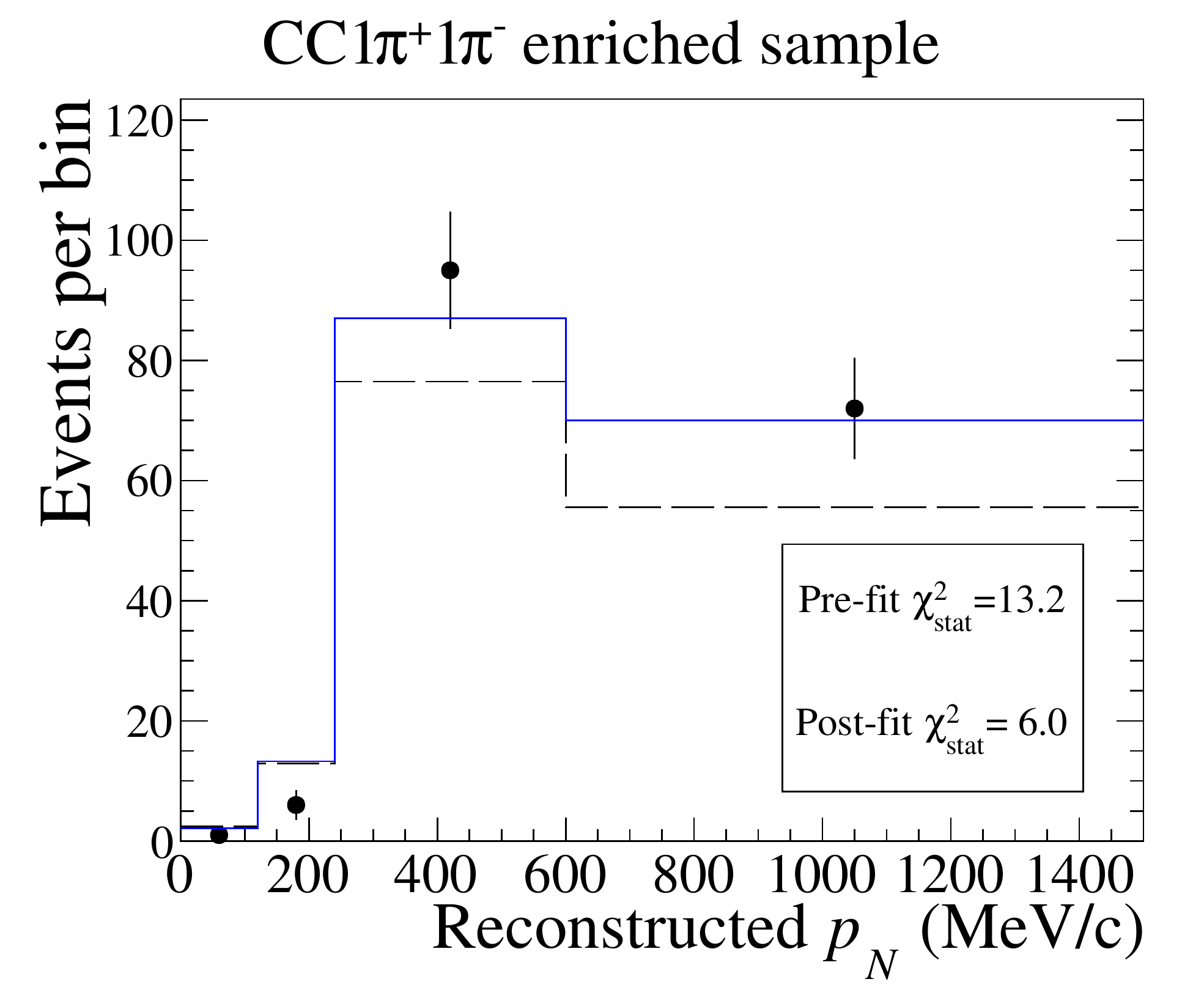}}
    \subfloat{\includegraphics[width=0.33\linewidth]{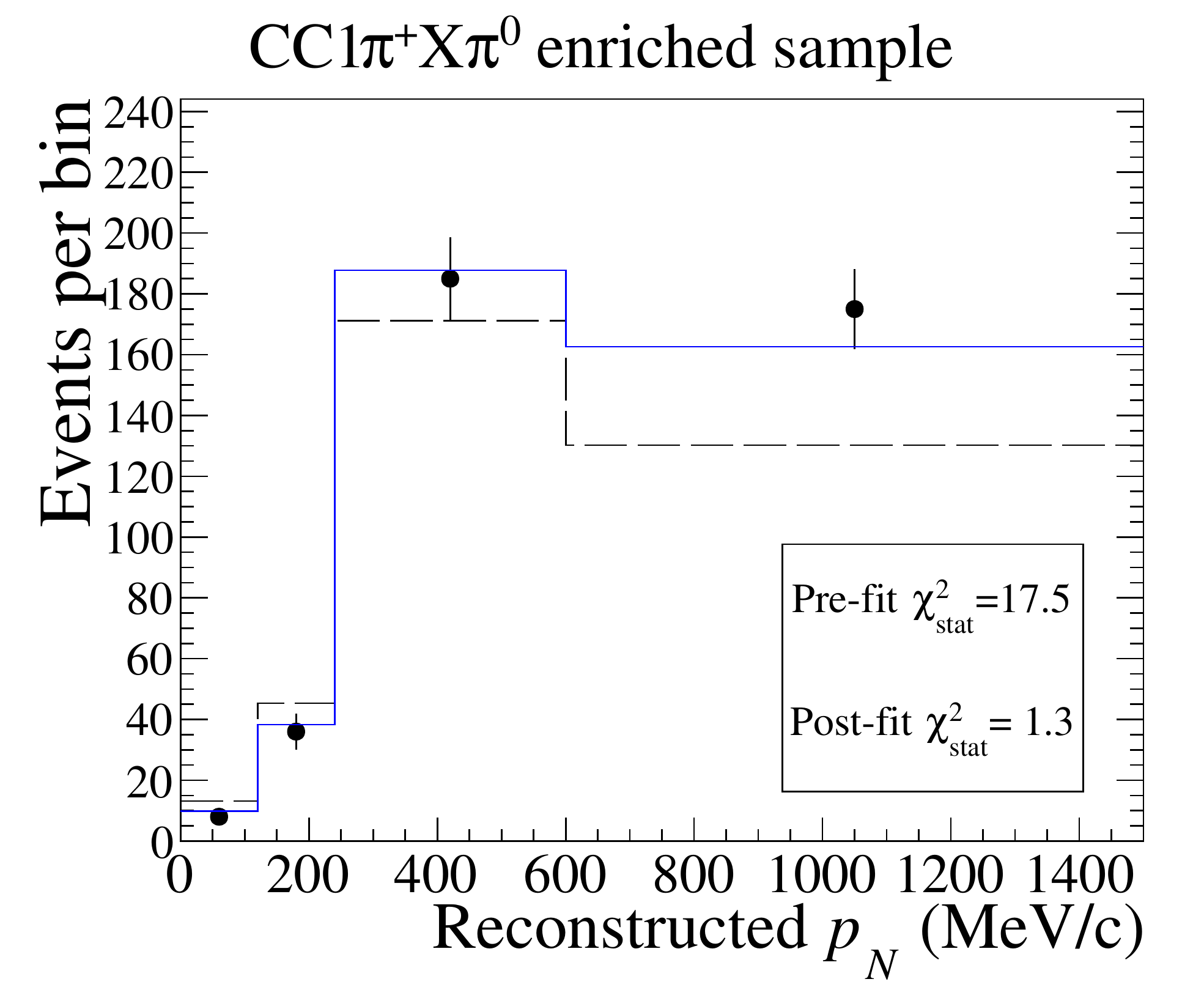}}\hfill
    \subfloat{\includegraphics[width=0.33\linewidth]{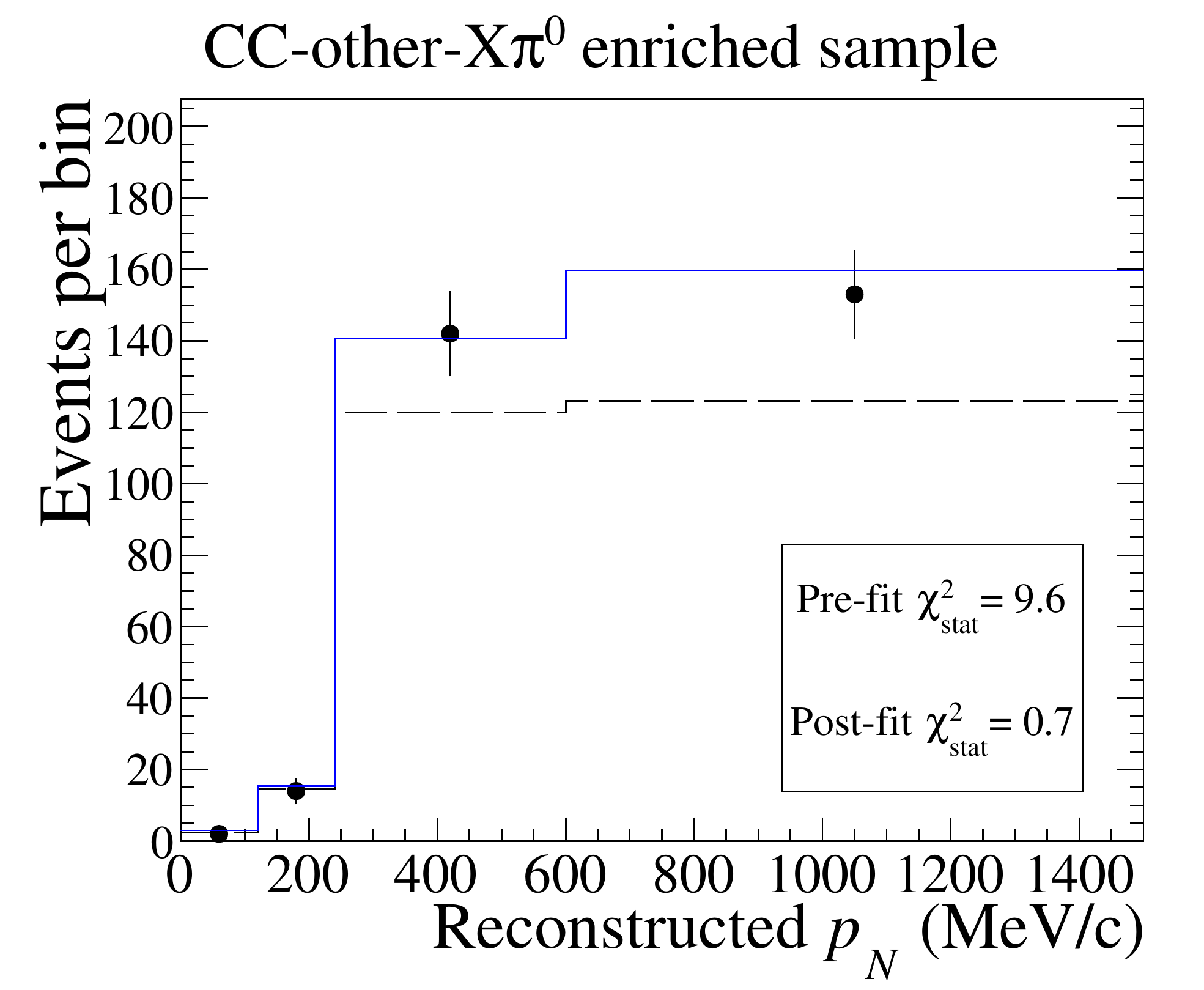}}
    \subfloat{\includegraphics[width=0.33\linewidth]{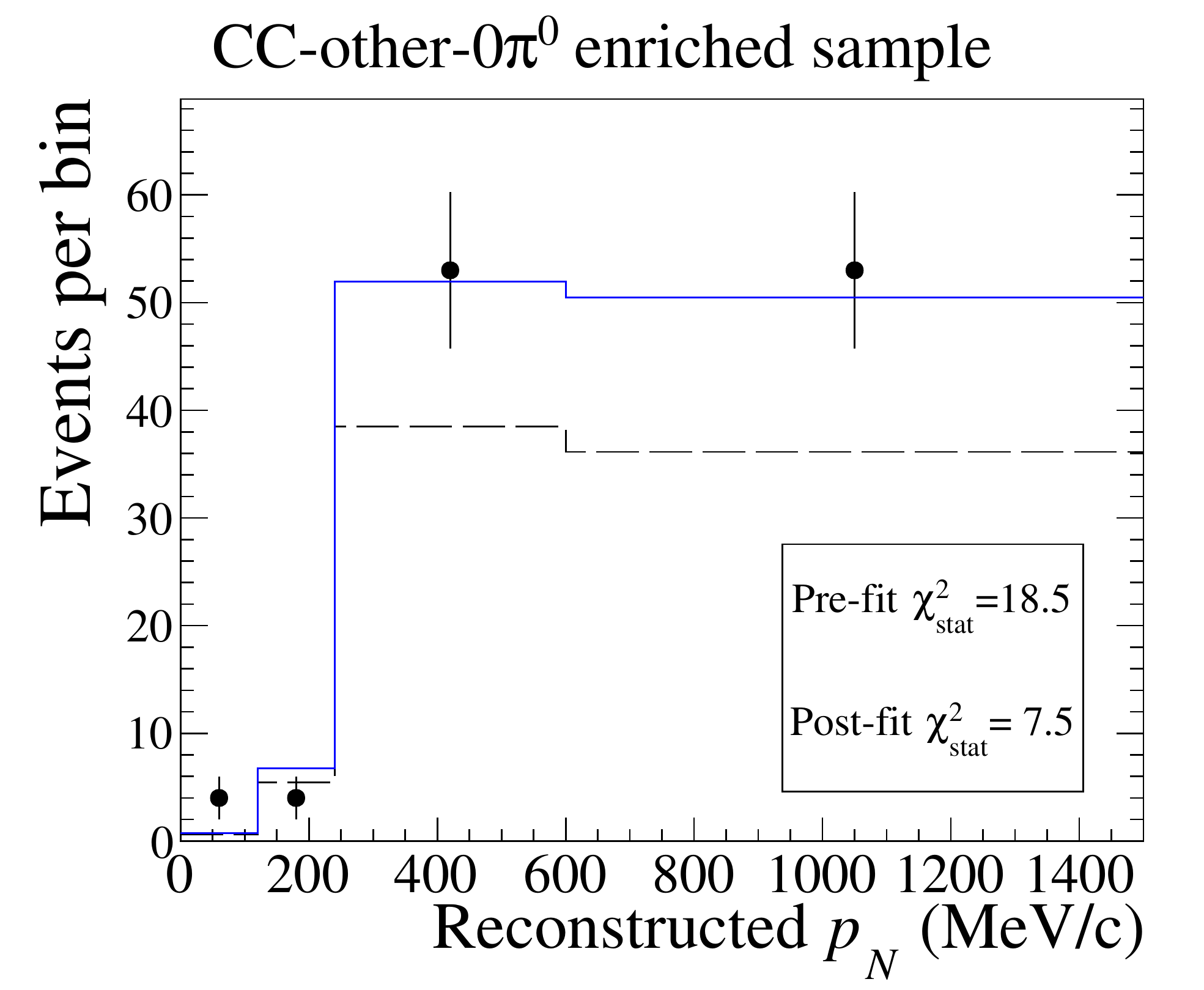}}\hfill
\caption{Distribution of events in the signal and control samples in the $p_{N}$ fit. $\chi^2_\text{stat}$ corresponds to the statistical contribution of the fit $\chi^2$ (\cref{eq:poisson_llh}) in that sample. The  MC prediction before (dashed) and after (solid) the fit are also shown. The error bars show the statistical uncertainty in data.} \label{fig:pN_postfit}
\end{figure*}

\begin{figure*}
\centering
    \subfloat{\includegraphics[width=0.33\linewidth]{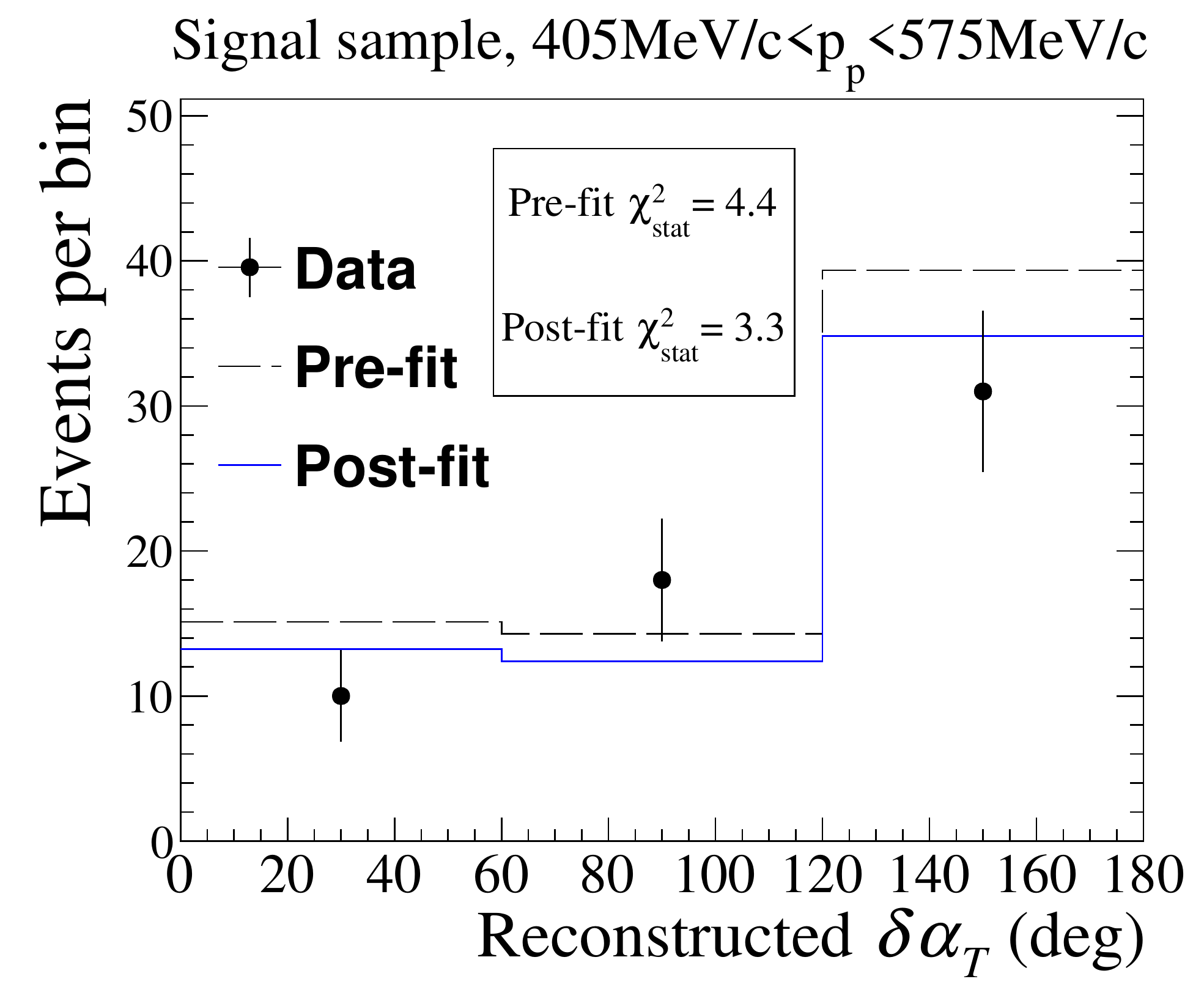}}
    \subfloat{\includegraphics[width=0.33\linewidth]{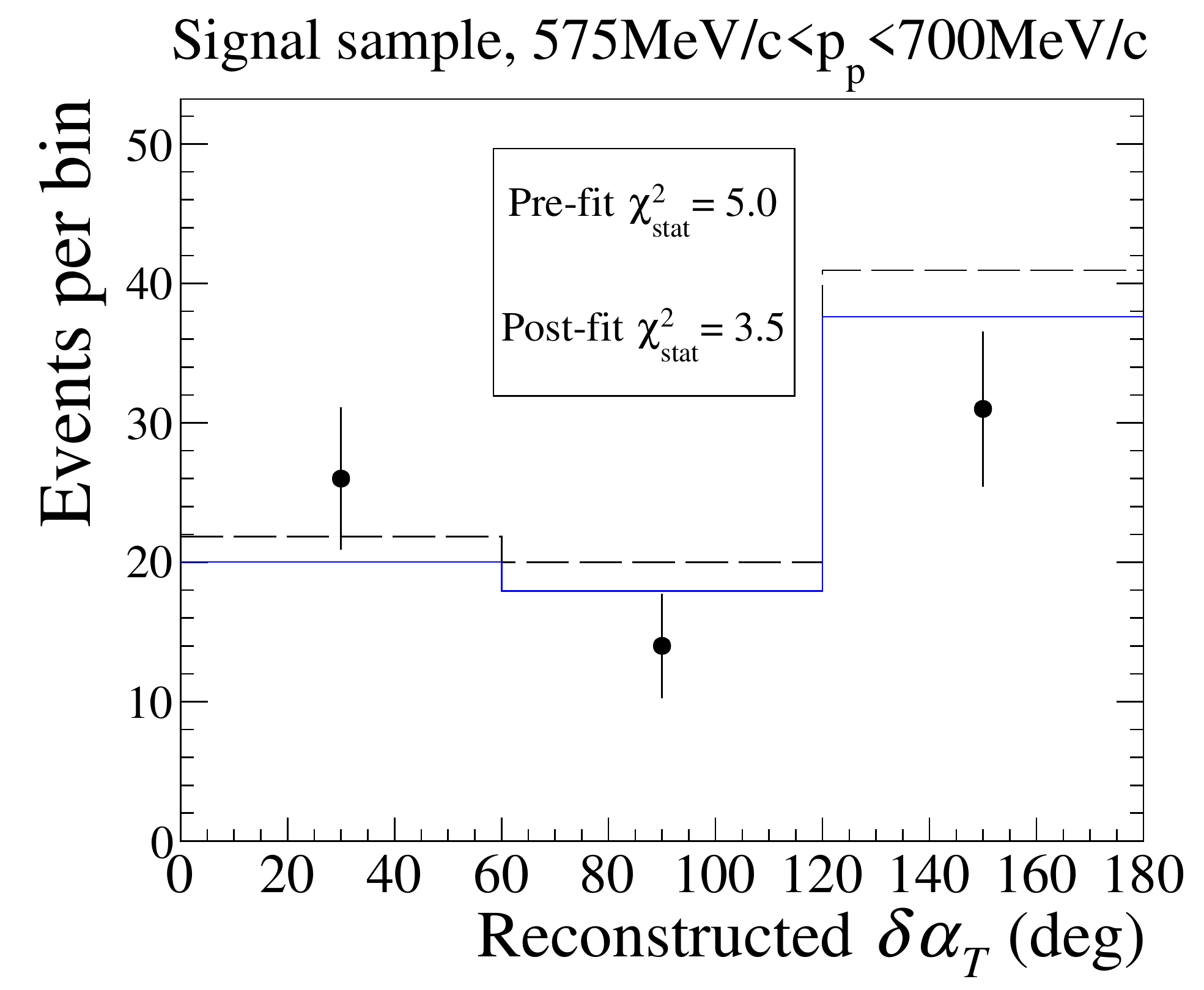}}
    \subfloat{\includegraphics[width=0.33\linewidth]{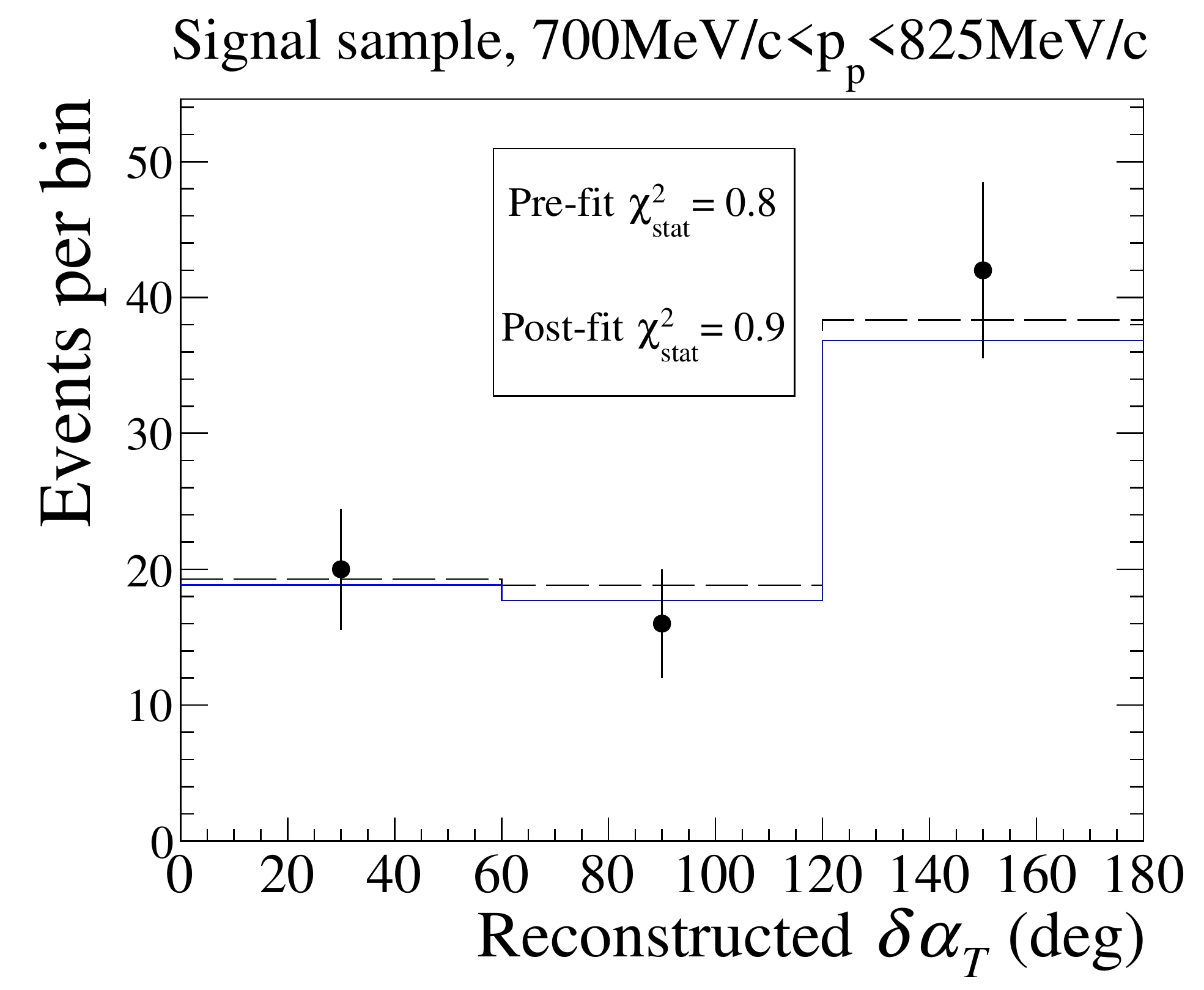}}\hfill
    \subfloat{\includegraphics[width=0.33\linewidth]{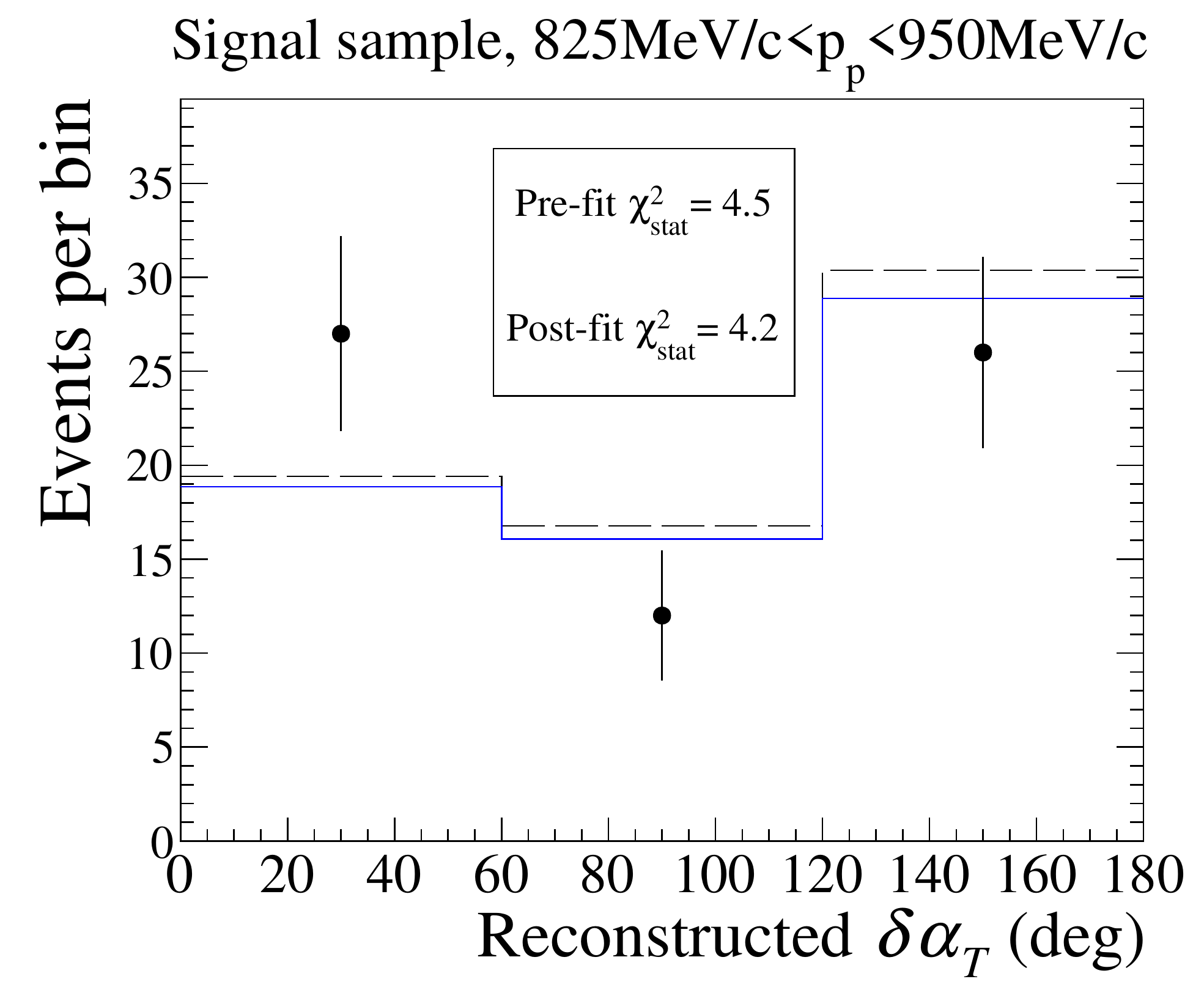}}
    \subfloat{\includegraphics[width=0.33\linewidth]{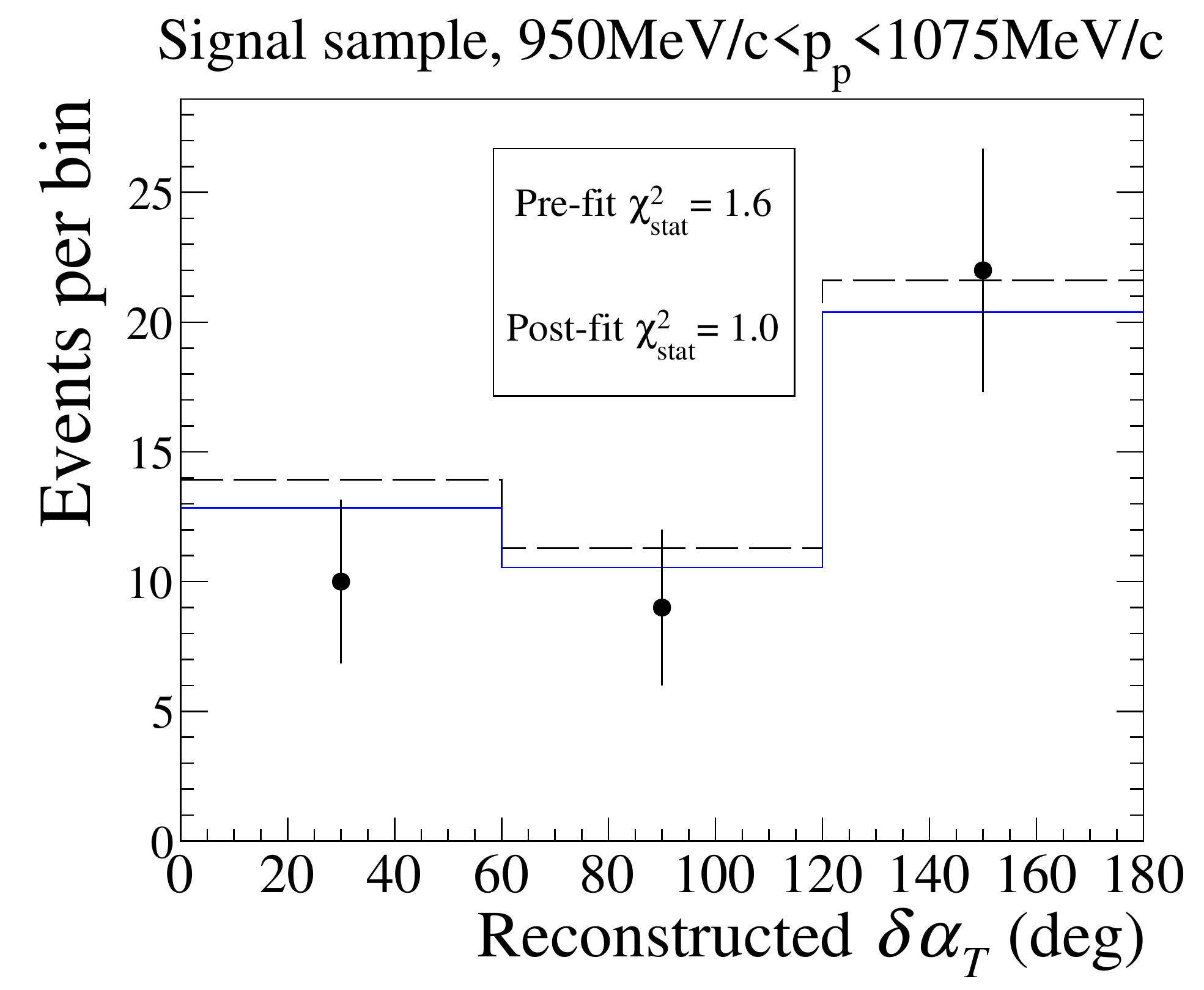}}
    \subfloat{\includegraphics[width=0.33\linewidth]{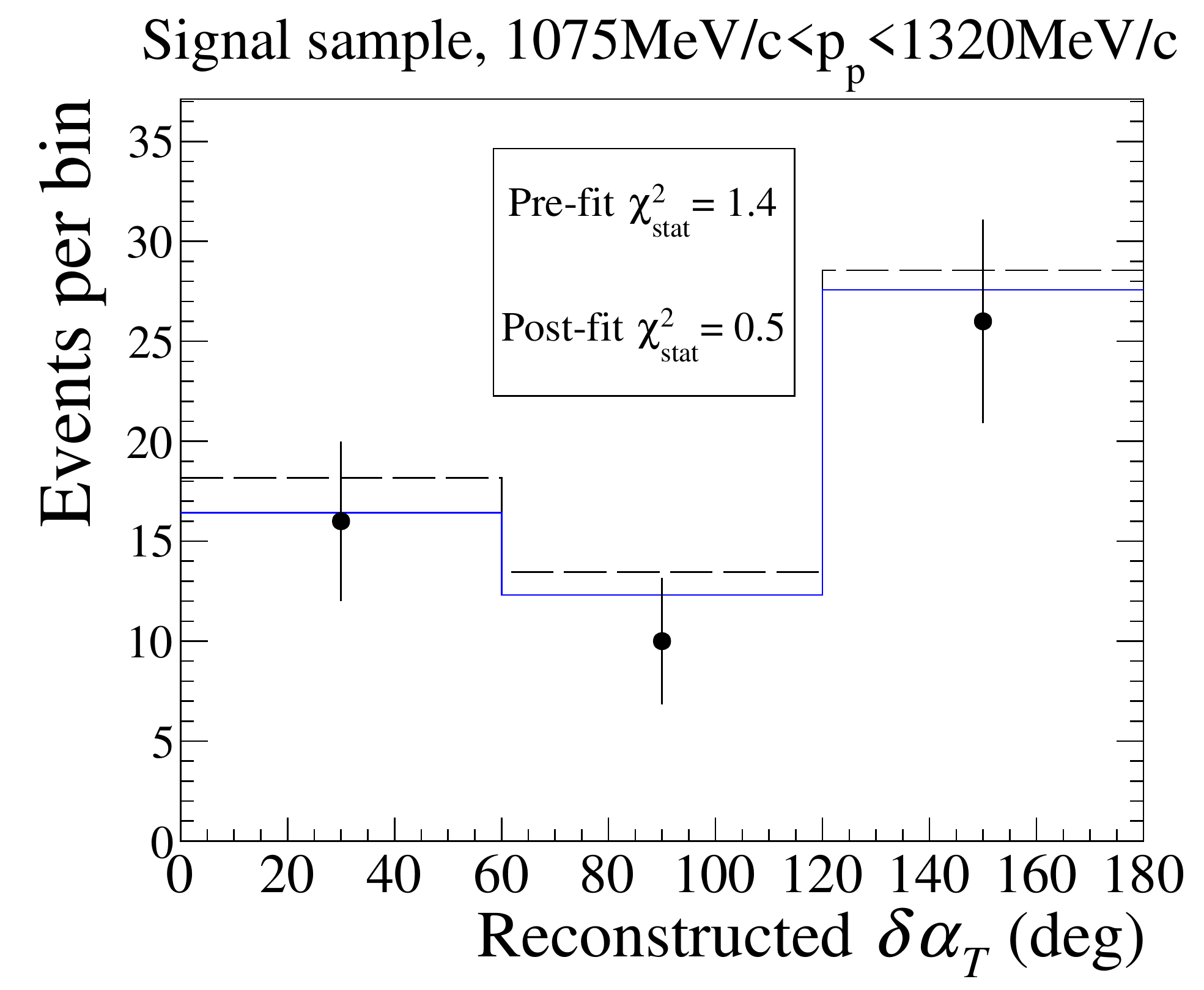}}\hfill
    \subfloat{\includegraphics[width=0.33\linewidth]{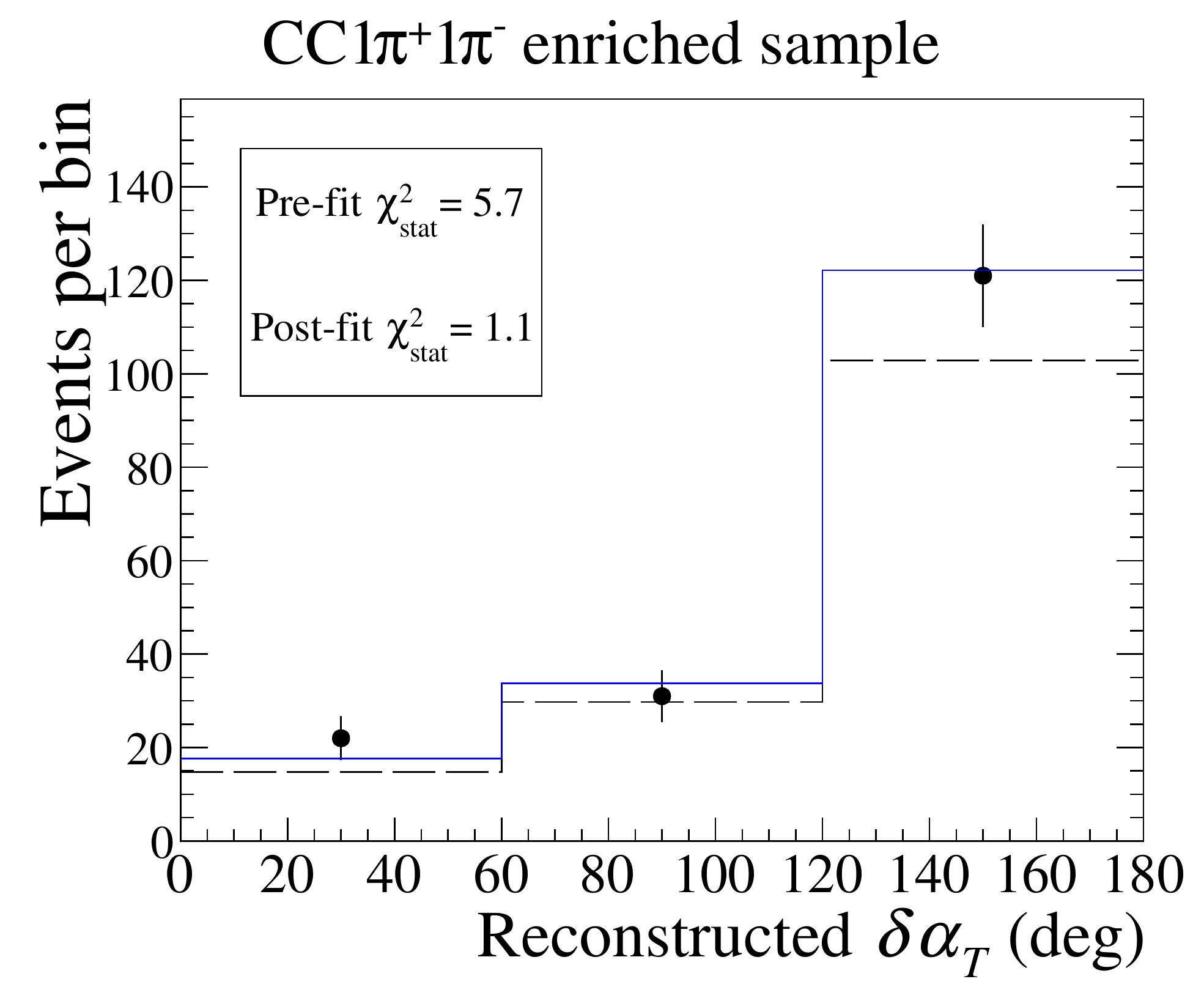}}
    \subfloat{\includegraphics[width=0.33\linewidth]{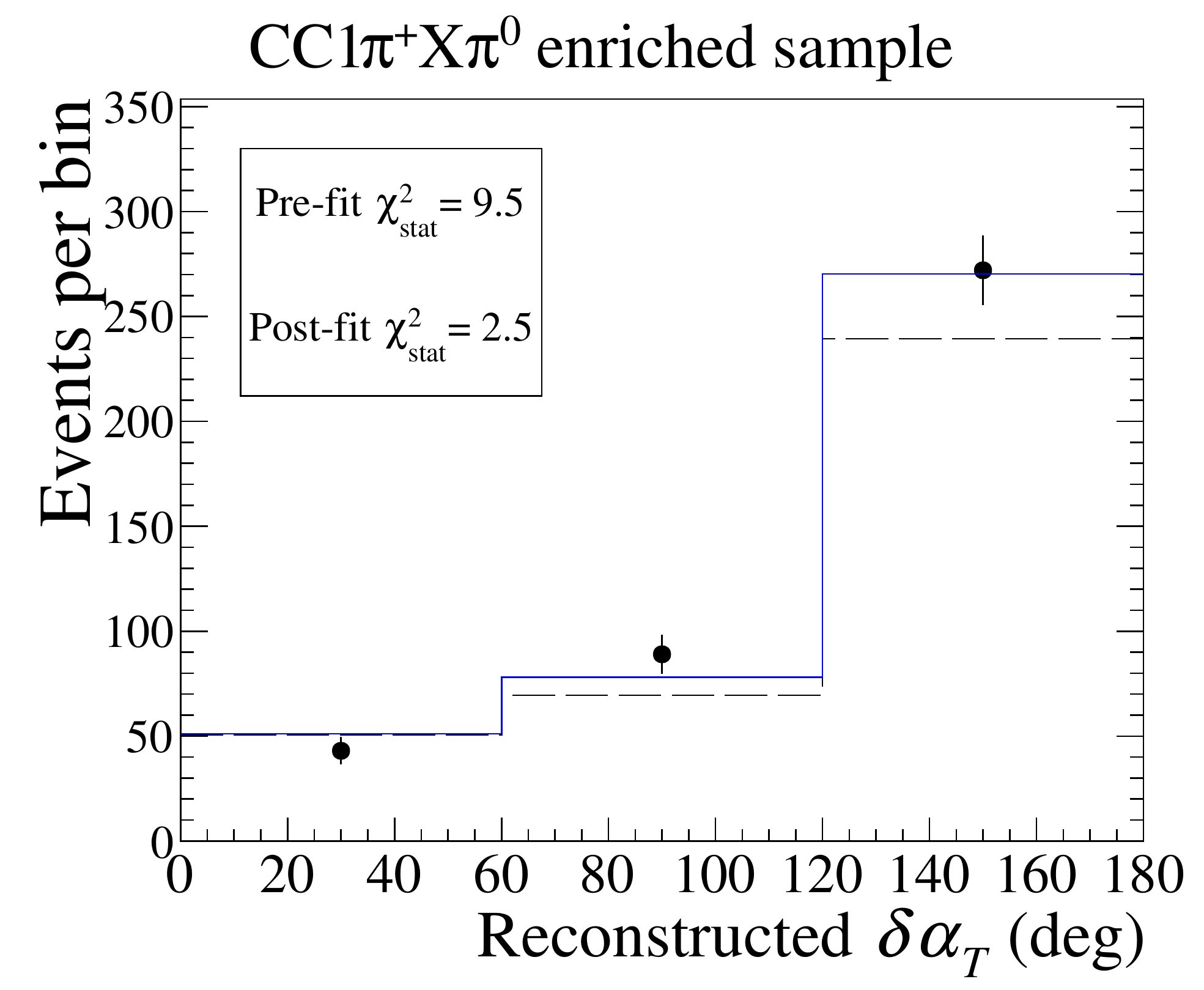}}\hfill
    \subfloat{\includegraphics[width=0.33\linewidth]{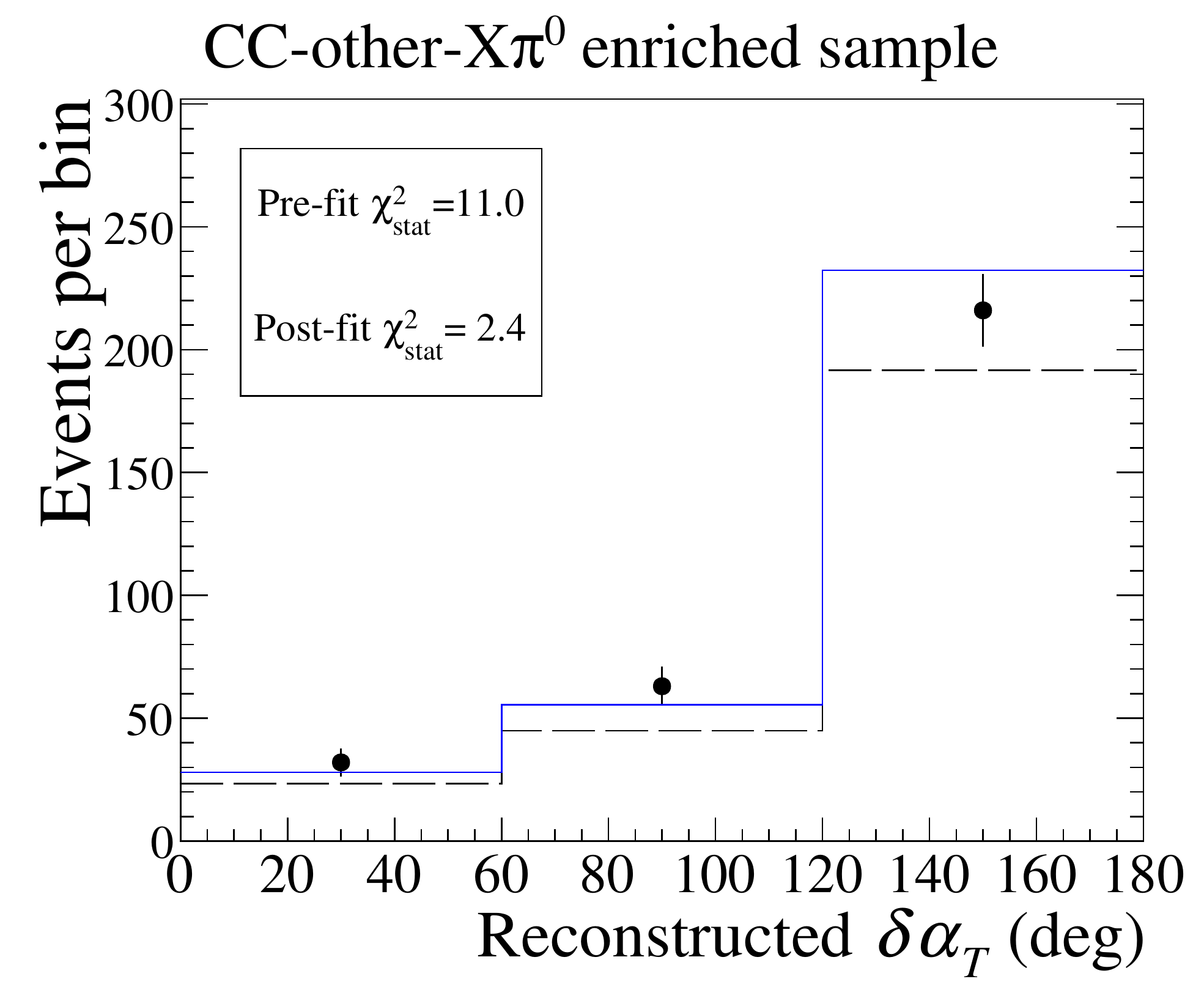}}
    \subfloat{\includegraphics[width=0.33\linewidth]{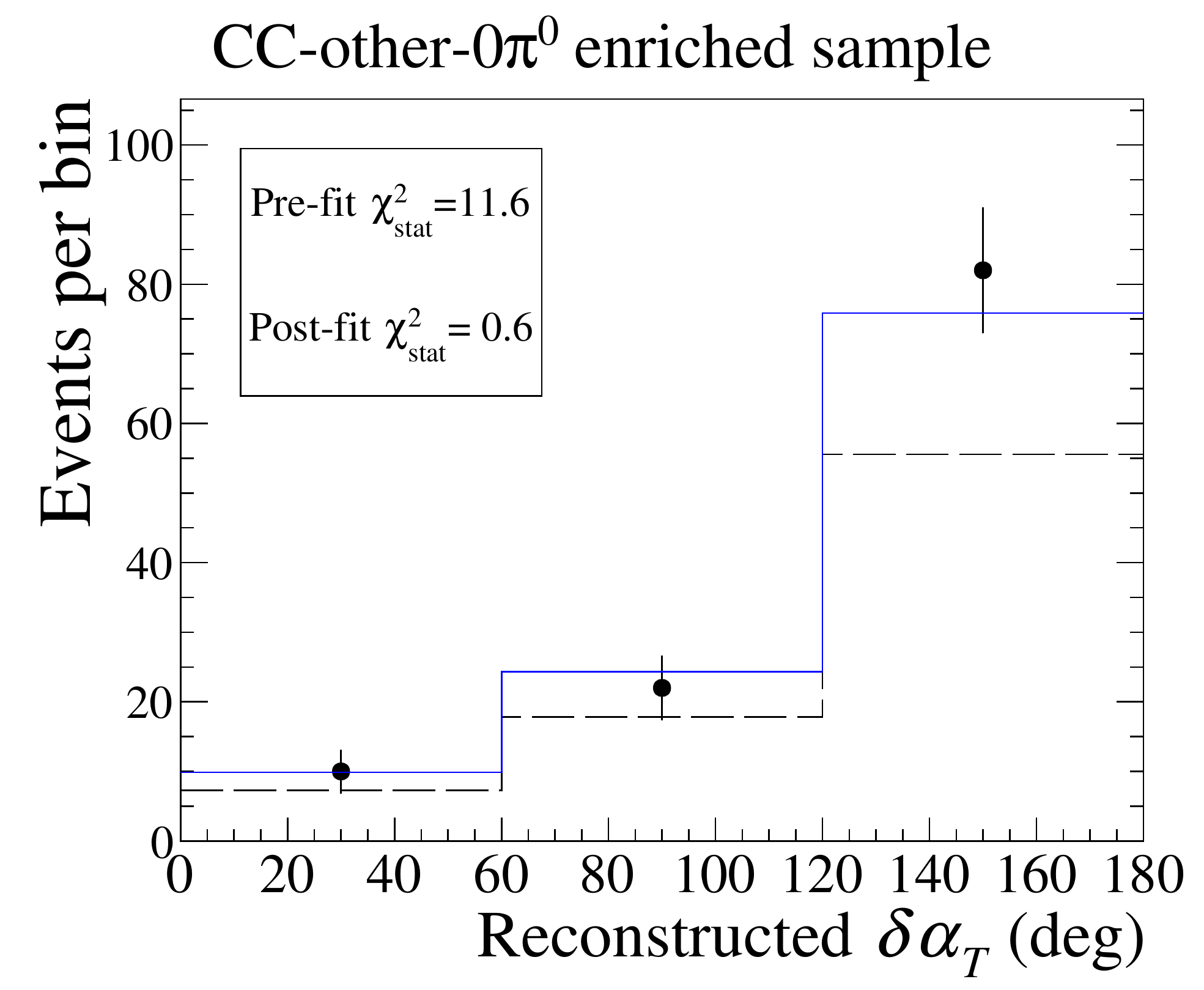}}\hfill
\caption{Distribution of events in the signal and control samples in the $\delta\alpha_{T}$ fit. $\chi^2_\text{stat}$ corresponds to the statistical contribution of the fit $\chi^2$ (\cref{eq:poisson_llh}) in that sample. The  MC prediction before (dashed) and after (solid) the fit are also shown. The error bars show the statistical uncertainty in data.} \label{fig:dat_postfit}
\end{figure*}

\cref{fig:dptt_postfit,fig:pN_postfit,fig:dat_postfit} show the distributions of the reconstructed events in the signal and control samples, together with the prediction from the pre-fit and post-fit MC. Overall, the fit is able to reproduce the observed distributions, with a p-value greater than 10\% for all the TKI variable fits, and is qualified to have a good data-MC agreement in the presence of statistical fluctuations. All nuisance parameters are fitted within their prior uncertainties. The normalization difference in control samples before the fit is well covered by the nuisance parameters, mostly through the CC-other normalization parameters. 
In the signal sample, there are few bins of reconstructed $p_\textrm{p}$ where the post-fit $\chi^2_\text{stat}$ is worse than the pre-fit one. This indicates there might not be enough freedom in the shape of the signal particle kinematics. However, from the mock data studies, it is concluded that the potential bias is much smaller than the statistical uncertainty and has little impact on this analysis.

\begin{figure*}
\centering
    \subfloat{\includegraphics[width=0.49\linewidth]{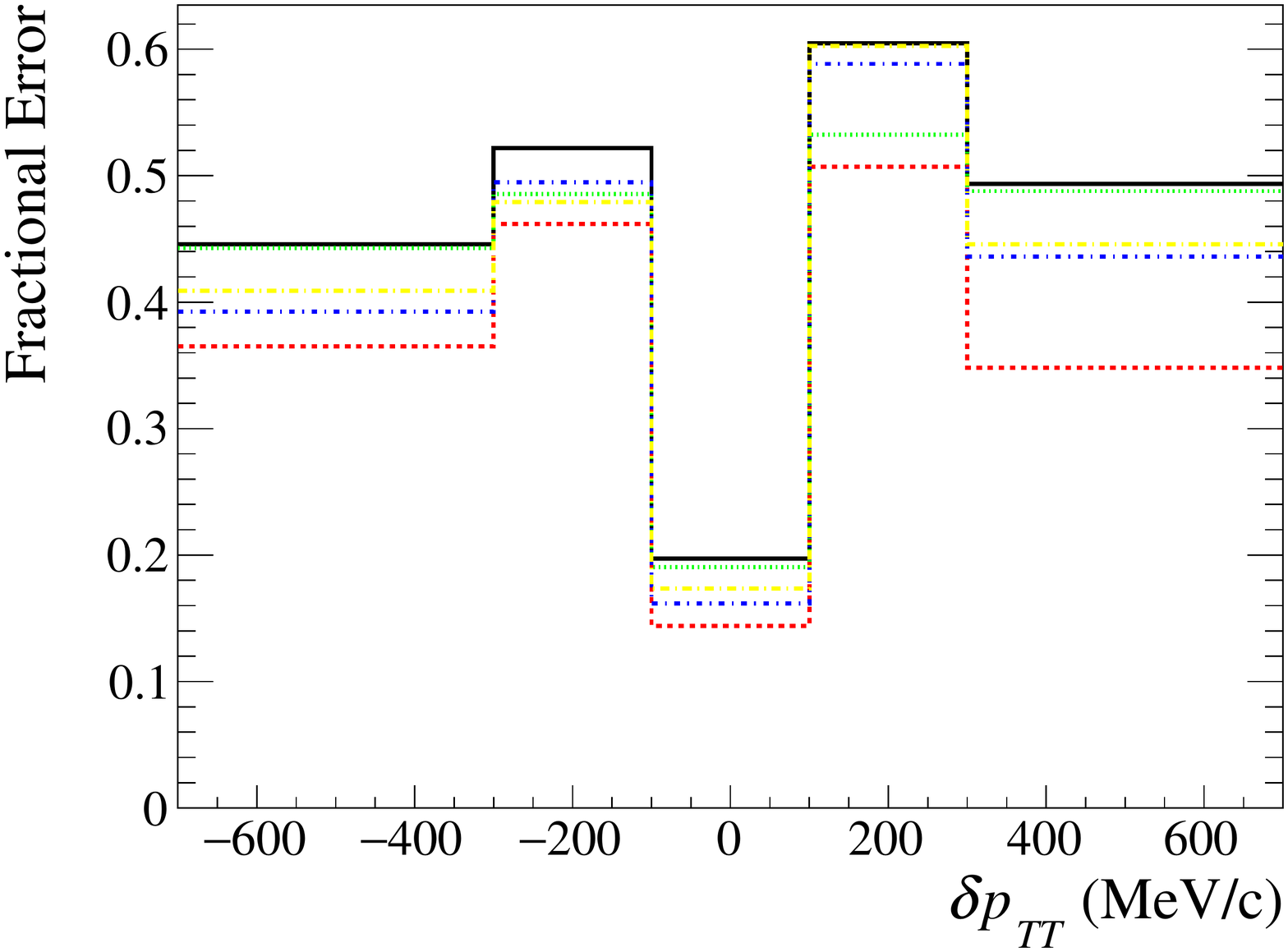}}
    \subfloat{\includegraphics[width=0.49\linewidth]{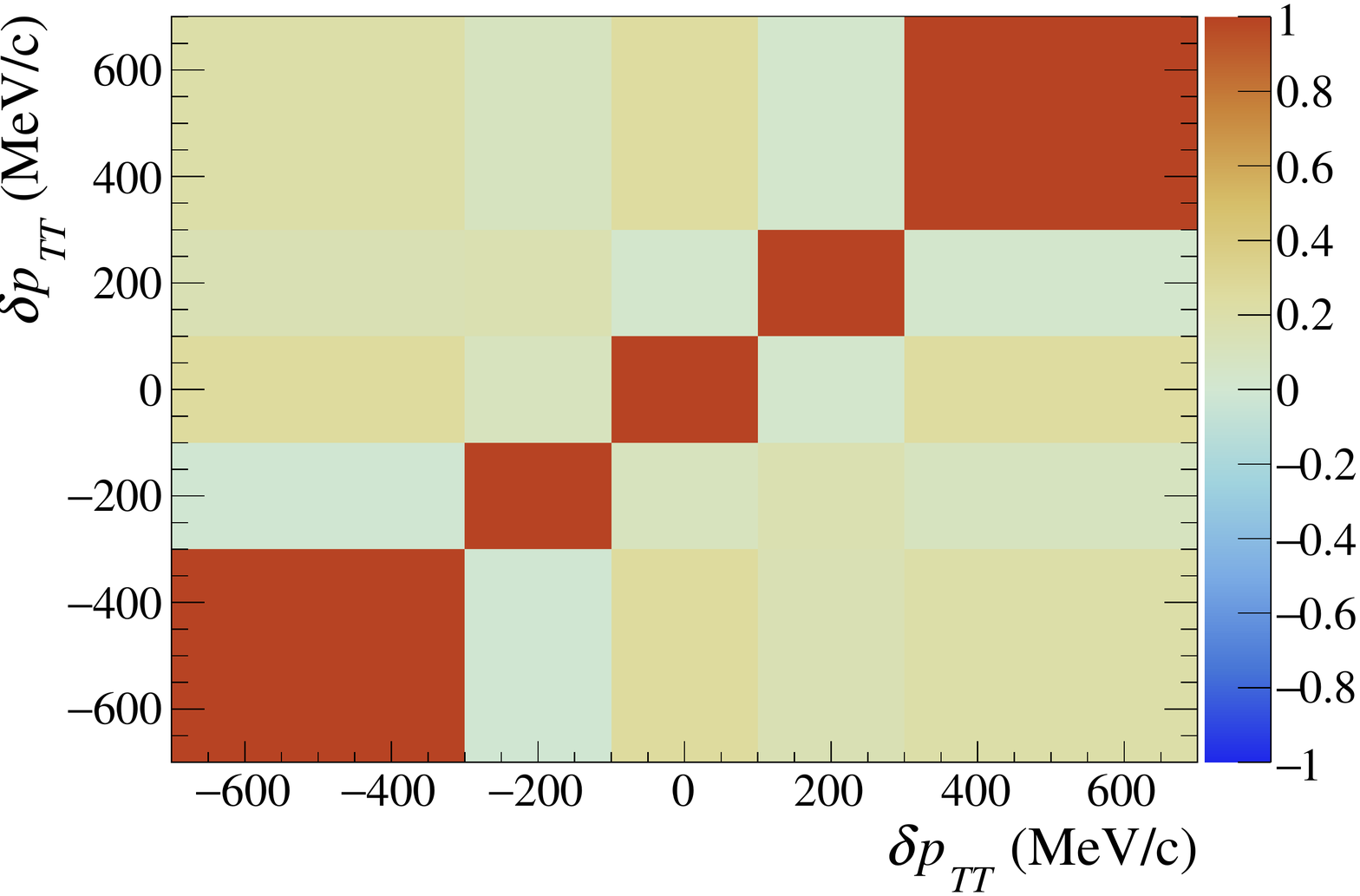}}\hfill
    \subfloat{\includegraphics[width=0.49\linewidth]{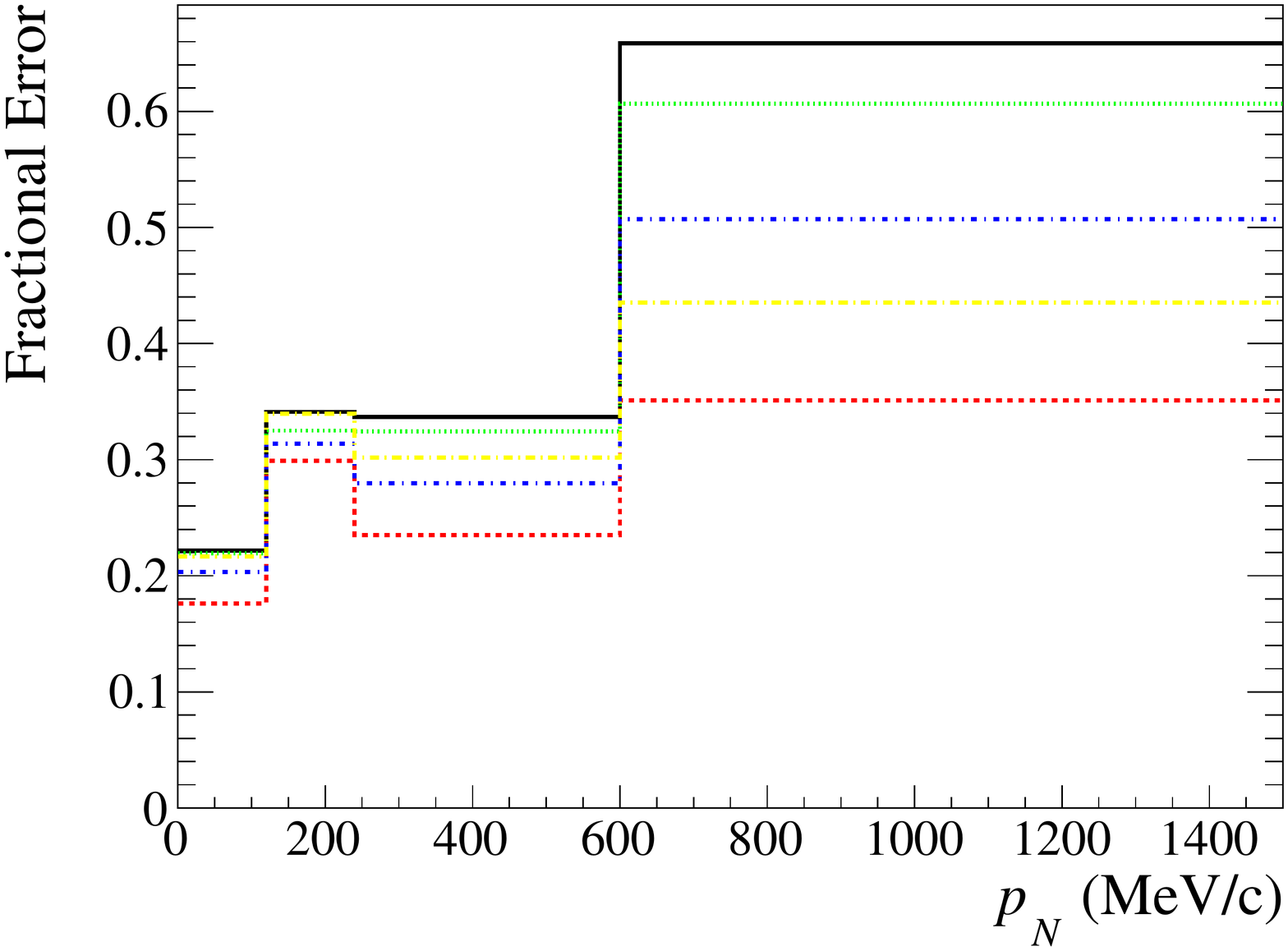}}
    \subfloat{\includegraphics[width=0.49\linewidth]{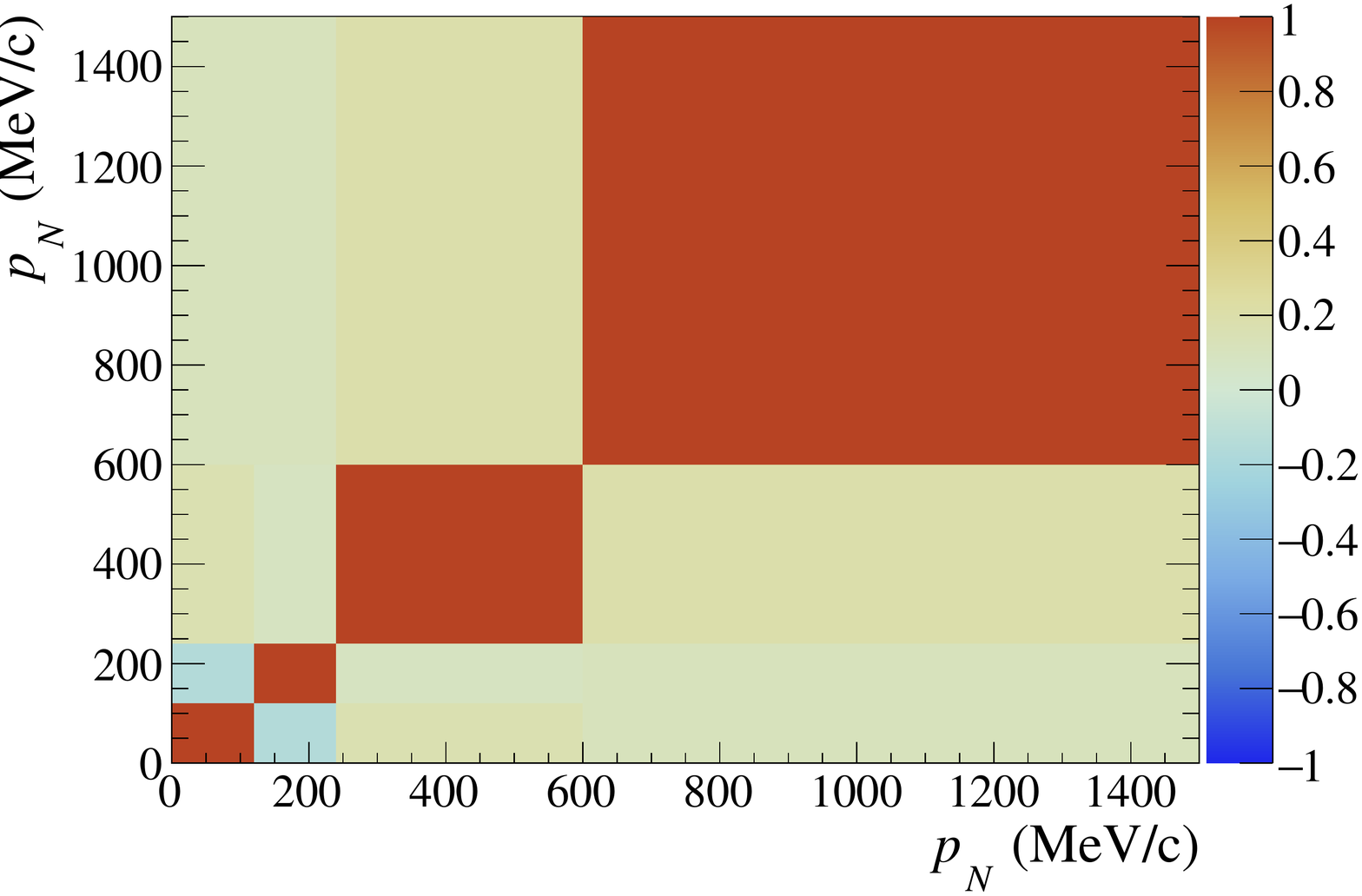}}\hfill
    \subfloat{\includegraphics[width=0.49\linewidth]{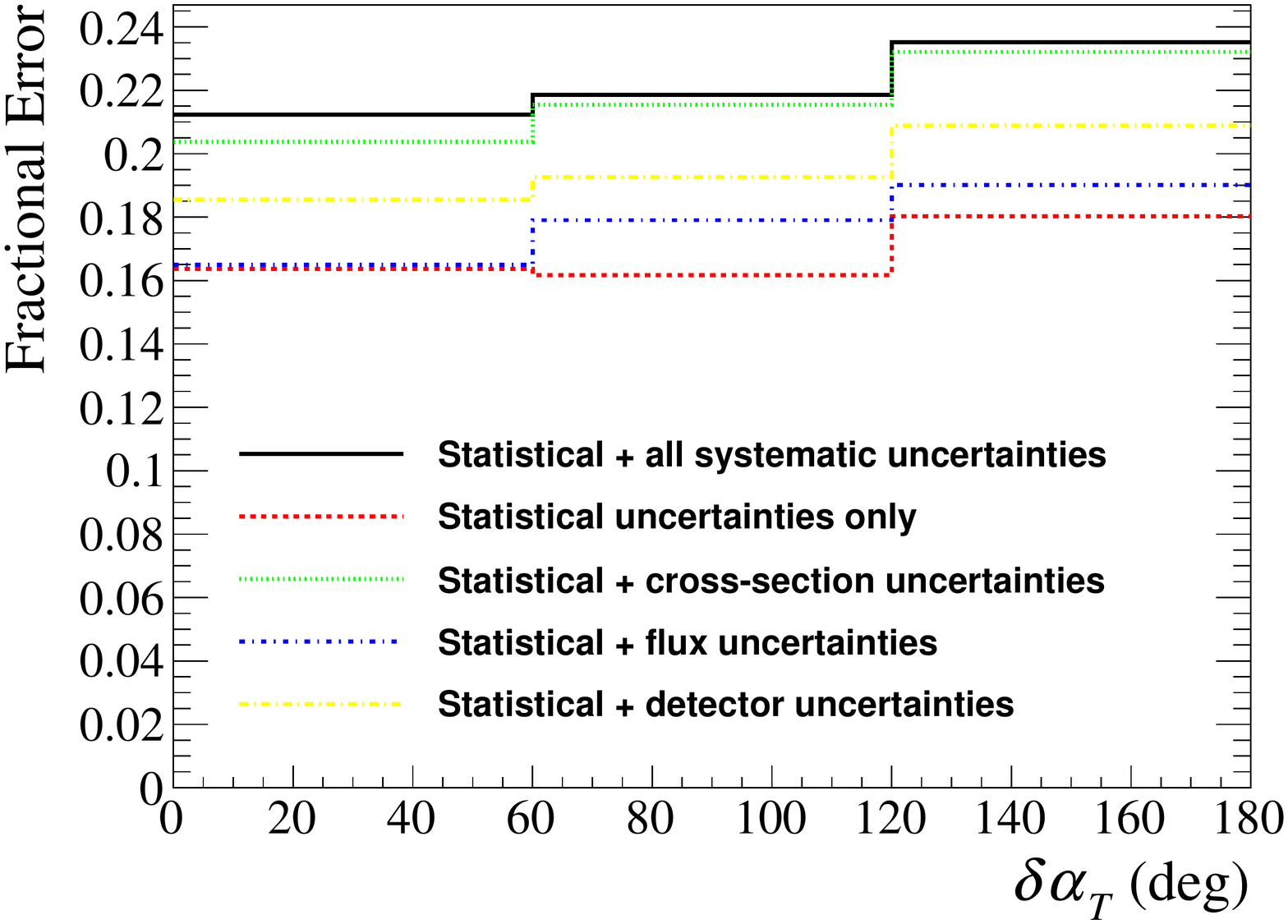}}
    \subfloat{\includegraphics[width=0.49\linewidth]{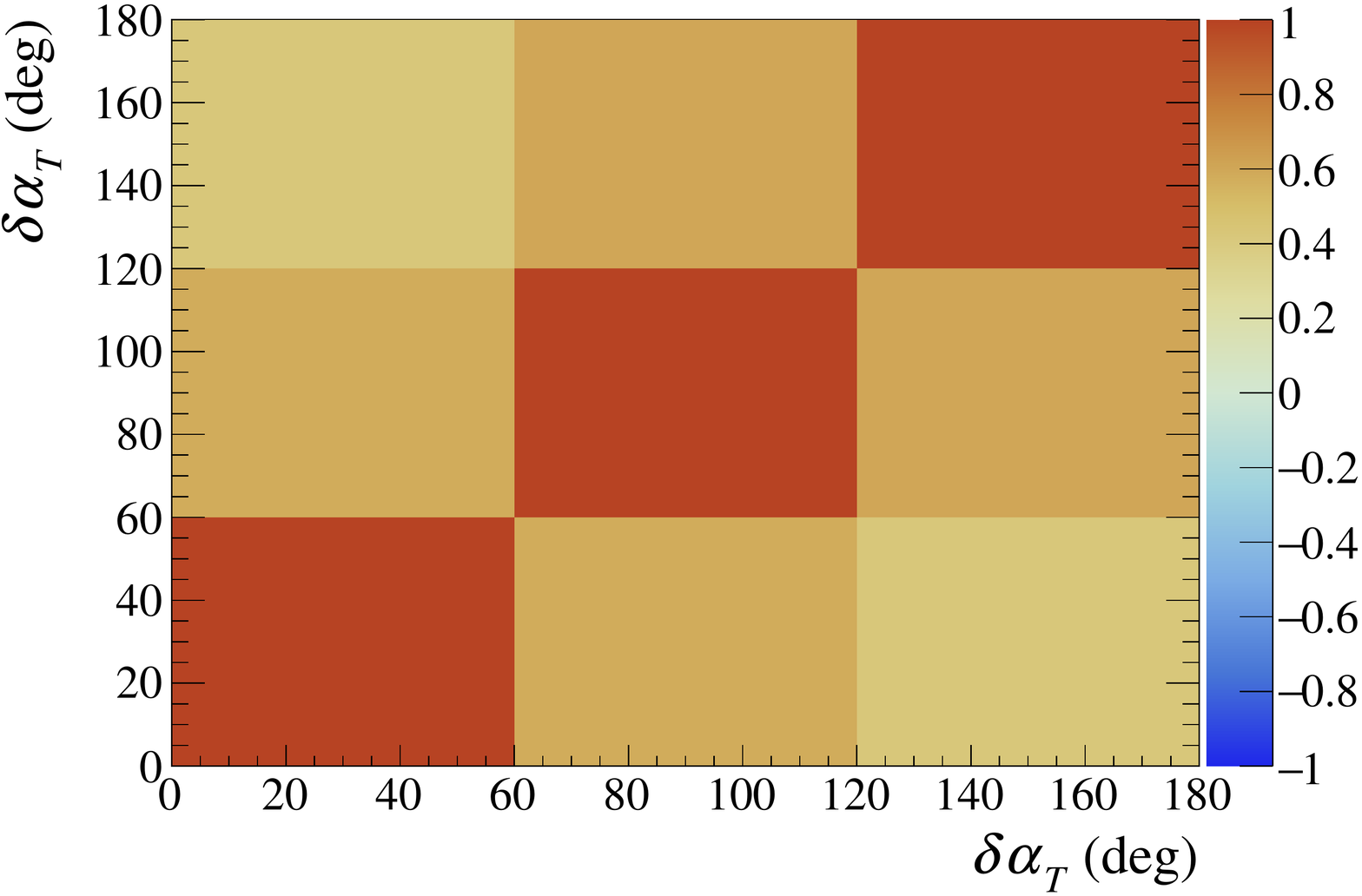}}\hfill
\caption{Error of the measured differential cross sections in each bin (left) and the correlation between bins (right), for the $\delta p_{TT}$ (top), $p_N$ (middle) and $\delta\alpha_{T}$ (bottom) fit respectively. The statistical error is shown in red, and the total statistical and systematic error in black. Contributions from each source of systematic uncertainties are shown one by one: neutrino flux in blue, detector in yellow, and neutrino interaction model in green. } \label{fig:xsec_err}
\end{figure*}

\cref{fig:xsec_err} estimates the uncertainties of the cross sections as a function of the TKI variables, together with the correlation between bins. Contributions from each kind of systematic uncertainties are estimated by running the fit with only the relevant nuisance parameters.
As expected, the statistical error is much larger than the individual or combined systematic uncertainties. The largest systematic uncertainties are those related to the neutrino interaction model, which affect both the signal selection efficiency and background estimation. The bin-by-bin correlation in $\delta\alpha_{T}$ is larger than that in $\delta p_{TT}$ and $p_N$ because the cross section on hydrogen is uniform across all bins of $\delta\alpha_{T}$.

\subsection{Comparisons to models}\label{sec:nugen}
In the following, the measured cross sections are compared to different neutrino interaction models. The agreement is quantified by the $\chi^2_\text{tot}$ statistic in \cref{eq:chi2_tot}, with $\sigma^\text{truth}$ replaced by the model prediction $\sigma^\text{model}$.

On the other hand, the overall normalization uncertainty, which is fully correlated between bins, may constitute a relatively large fraction of the uncertainty. Therefore, the $\chi^2_\text{tot}$ statistics may suffer from ``Peelle's pertinent puzzle"~\cite{Peelle,doi:10.1063/1.1945011}, in which the assumption in \cref{eq:chi2_tot} that the variance is distributed as a multivariate Gaussian may not be valid for highly correlated results. To mitigate this problem, the shape only $\chi^2_\text{shape}$ is also provided:
\begin{equation}\label{eq:chi2_shape}
\begin{split}
    \chi^2_\text{shape}=&\sum_i\sum_j\left(\frac{d\sigma^\text{model}}{dx_i}\frac{1}{\sigma^\text{model}_\text{int}}- \frac{d\sigma^\text{meas}}{dx_i}\frac{1}{\sigma^\text{meas}_\text{int}}\right)
    \\&\cdot(W^{-1})_{ij}\left(\frac{d\sigma^\text{model}}{dx_j}\frac{1}{\sigma^\text{model}_\text{int}}- \frac{d\sigma^\text{meas}}{dx_j}\frac{1}{\sigma^\text{model}_\text{meas}}\right),
\end{split}
\end{equation}
where $\sigma^\text{model}_\text{int}$ and $\sigma^\text{meas}_\text{int}$ are the total integrated cross sections per nucleon estimated from the model and data respectively. The shape only covariance matrix $W$ is built by the same method as described in \cref{sec:err_prop} but on the shape variable $\frac{d\sigma^\text{meas}}{dx_i}\frac{1}{\sigma^\text{meas}_\text{int}}$ instead. It is important to notice that the ndof is one less for $\chi^2_\text{shape}$ compared to $\chi^2_\text{tot}$ since the sum of the shape variables is equal to one by construction, reducing the number of independent dimensions. 

To compare the measured cross sections with model predictions, a sufficiently large number of events are generated on hydrocarbon from each model using the T2K flux. Events satisfying the CC1$\pi^+$Xp signal definition in \cref{tab:signal_ps_cut} are selected to calculate the cross sections per target nucleon.  The number of target nucleons for each CH is equal to 13 which includes all seven protons and six neutrons. The following models are considered.
\begin{enumerate}[(i)]
\item NEUT version 5.4.0: models implemented in this event generator are described in \cref{sec:simulation}. RFG is used as the nuclear ground state for pion production.
\item GENIE~\cite{Andreopoulos:2009rq,Andreopoulos:2015wxa} version 3.0.6: two model configurations are compared: the ``BRRFG+hA" model uses the G18\_01a physics configuration, with the Rein-Sehgal (RS) model for pion production,  Bodek-Ritchie empirical corrections of RFG (BRRFG~\cite{PhysRevD.23.1070,PhysRevD.24.1400}) as the nuclear ground state model and the hA (``empirical") FSI model; the ``LFG+hN" model uses the G18\_10b physics configuration, with the Berger-Sehgal (BS) model~\cite{PhysRevD.76.113004} for pion production, local Fermi gas (LFG) as nuclear ground state and the hN (``cascade") FSI model. For both models, the 2018a free nucleon cross section model re-tune~\cite{Andreopoulos:2019gvw} is used. Specific to pion production, CC1$\pi$ and CC2$\pi$  cross section data on deuterium targets from ANL~\cite{PhysRevD.90.112017,PhysRevD.28.2714,PhysRevLett.30.335,PhysRevD.25.1161}, BNL~\cite{PhysRevD.90.112017,PhysRevD.34.2554}, BEBC~\cite{ALLEN1980269,ALLASIA1990285,ALLEN1986221} and FNAL~\cite{PhysRevLett.41.1008} bubble chamber experiments were used in the re-tune. This mostly affects the cross-section normalization ($\sim$15\% reduction in total cross section) and gives much better $\chi^2_\textrm{tot}$ agreement.
\item GiBUU~\cite{BUSS20121} version 2019: it uses an LFG-based nuclear ground state to describe all neutrino interaction modes and FSI consistently. In the RES channel, 13 resonances are included and the non-resonant contribution is described by a phenomenological model. Rather than a simple cascade model, GiBUU models FSI by solving the dynamical evolution of the particle phase space density in the nuclear mean field potential. 
\item NuWro~\cite{GOLAN2012499} version 19.02: four different nuclear ground state models are implemented. These include three Fermi gas models:  
LFG, RFG, and BRRFG; and an effective approximation of a spectral function (ESF)~\cite{Ankowski:2005wi}. The Adler-Rarita-Schwinger single $\Delta$ model~\cite{PhysRevD.80.093001,Juszczak:2005zs} is used for RES, and the FSI cascade model is based on the Metropolis algorithm~\cite{Niewczas:2019fro}.
\end{enumerate}

\begin{figure*}
\centering
    \subfloat{\includegraphics[width=0.49\linewidth]{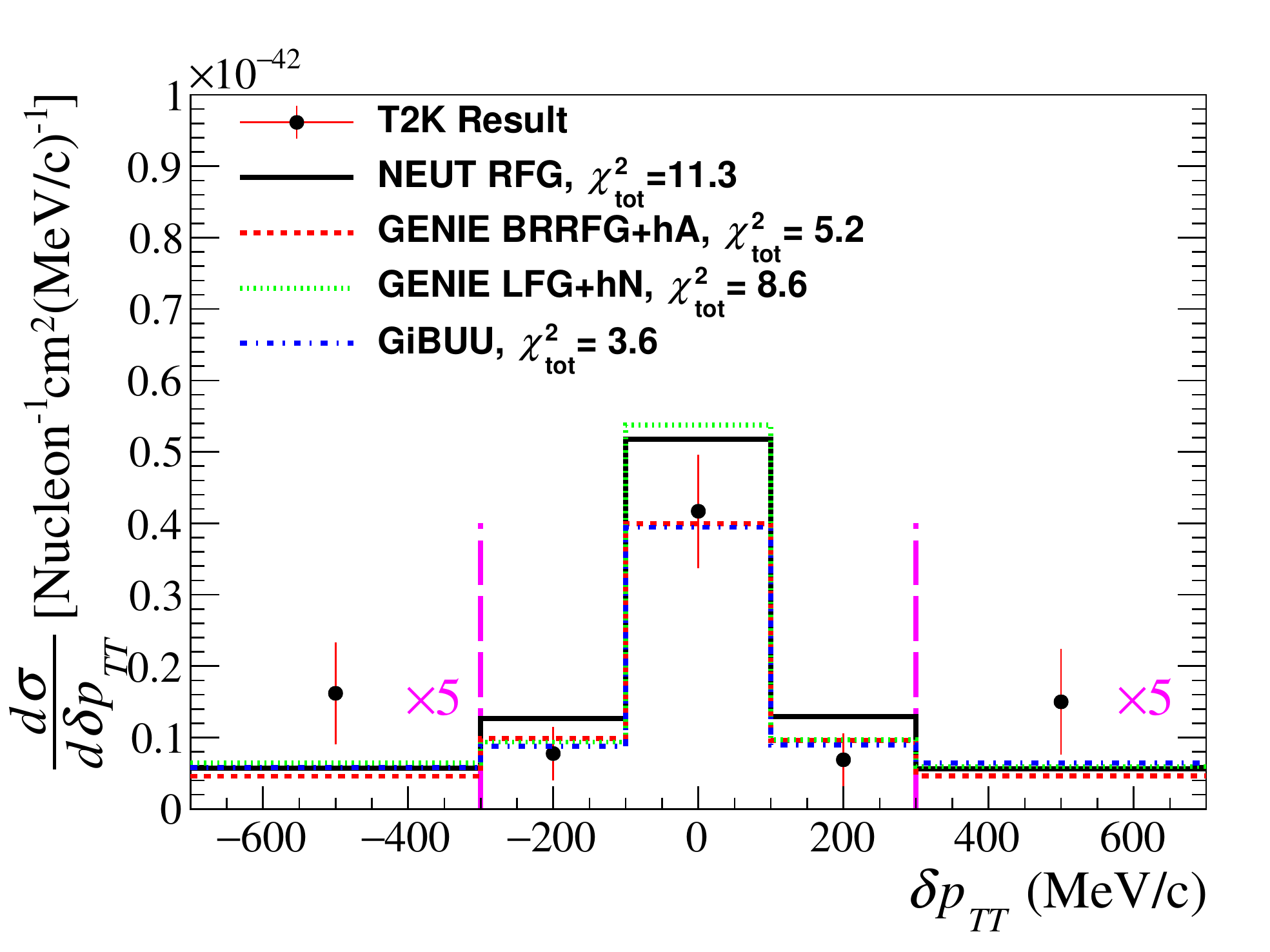}}
    \subfloat{\includegraphics[width=0.49\linewidth]{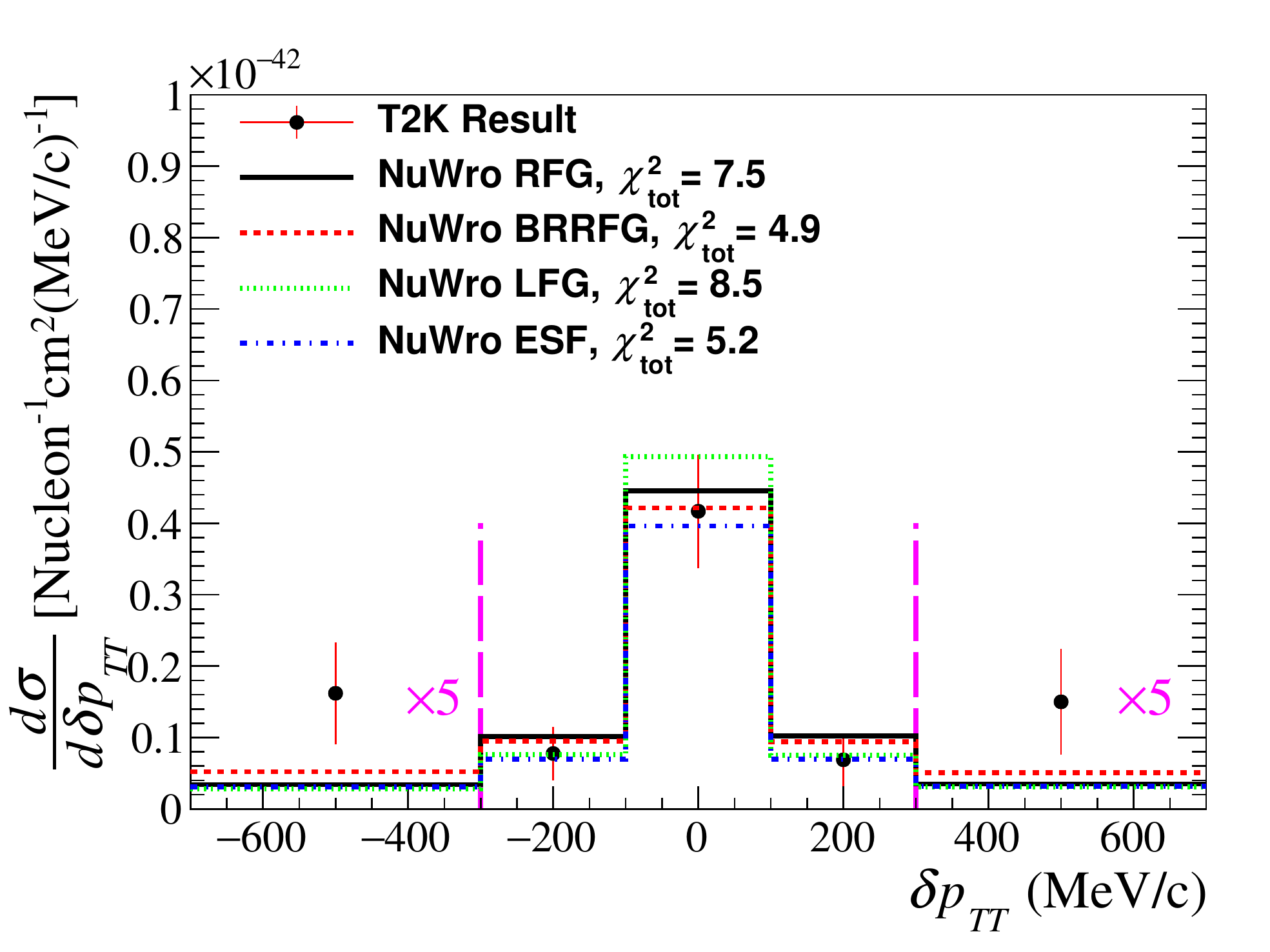}}\hfill
    \subfloat{\includegraphics[width=0.49\linewidth]{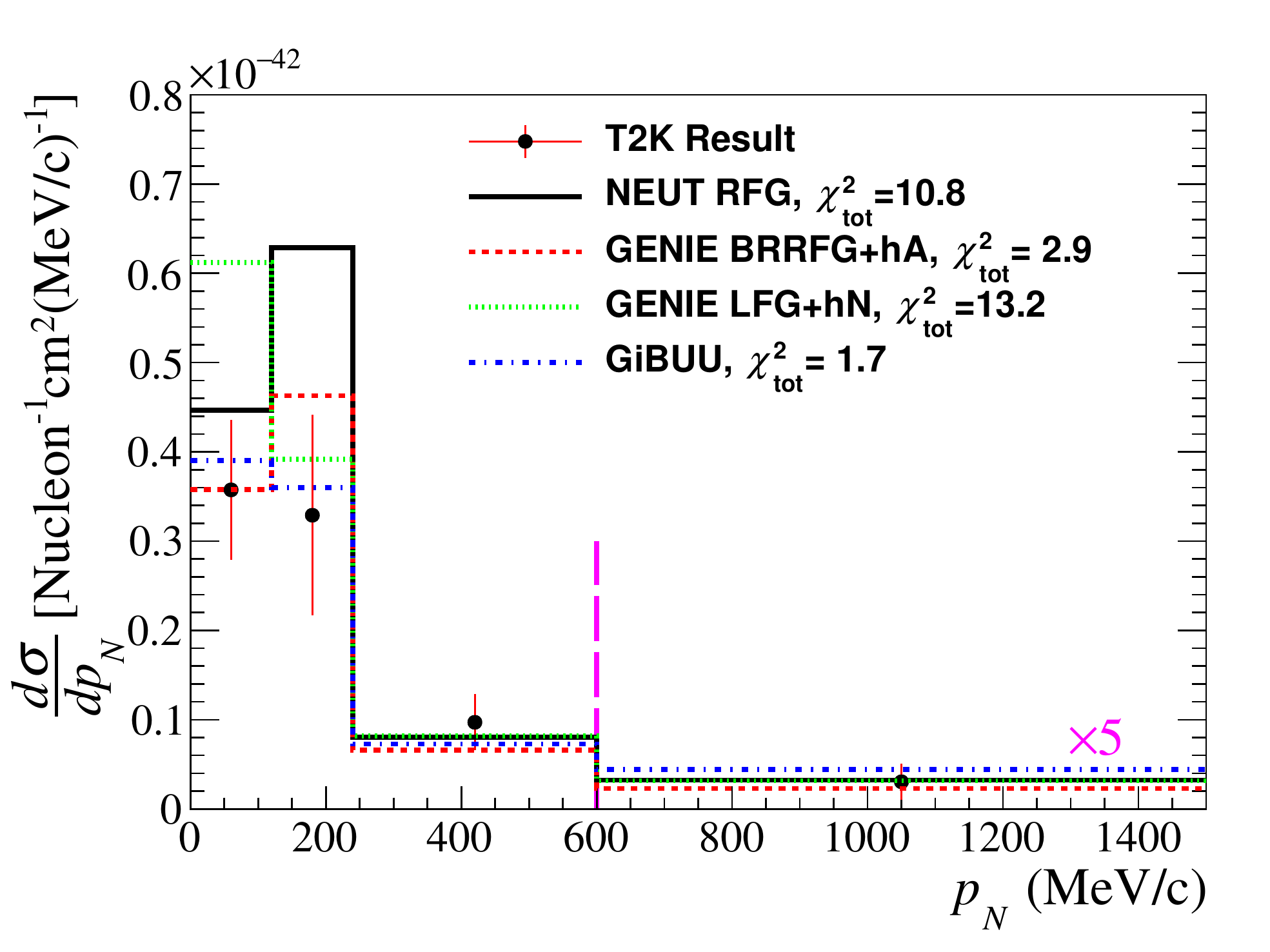}}
    \subfloat{\includegraphics[width=0.49\linewidth]{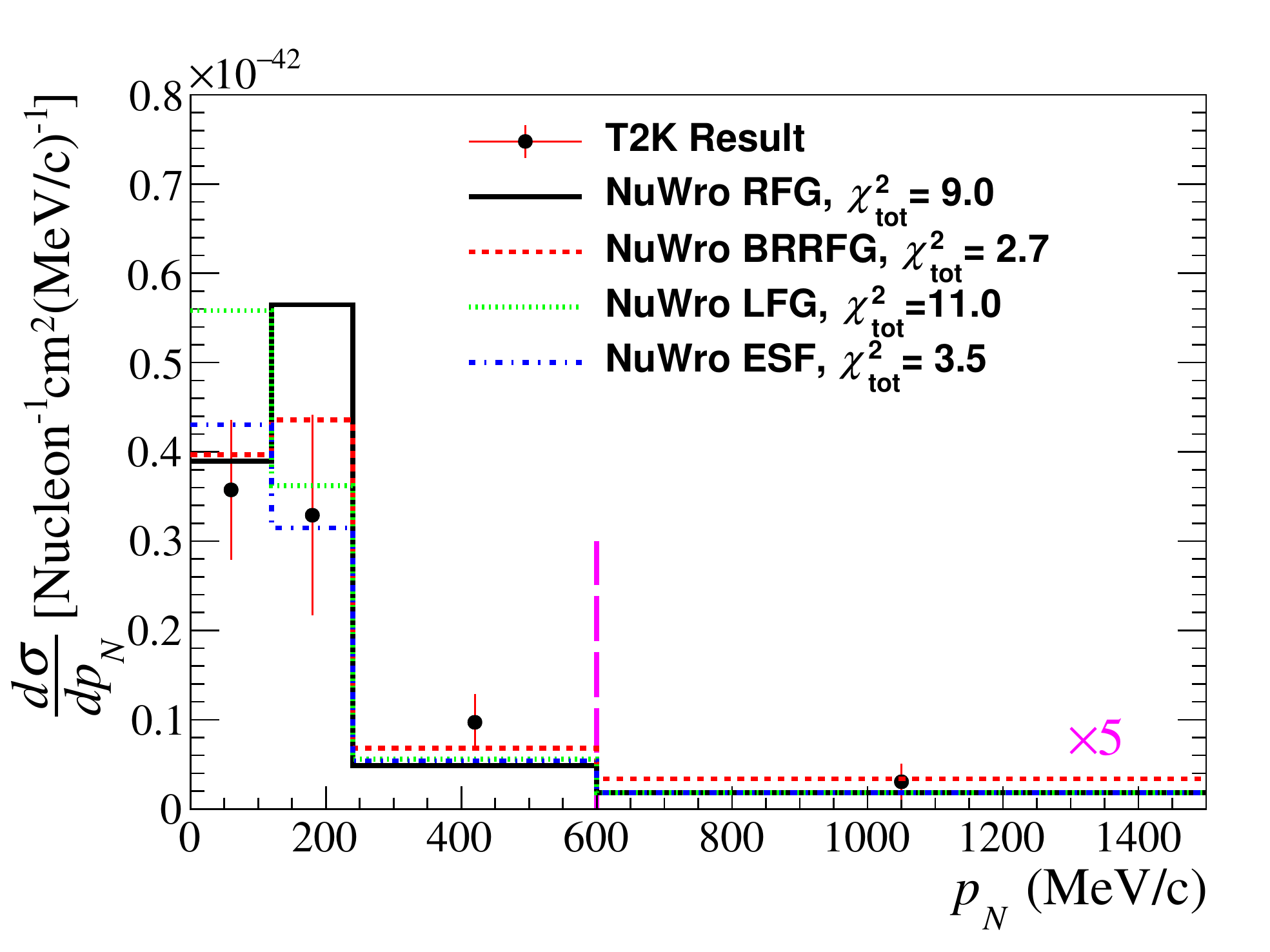}}\hfill
    \subfloat{\includegraphics[width=0.49\linewidth]{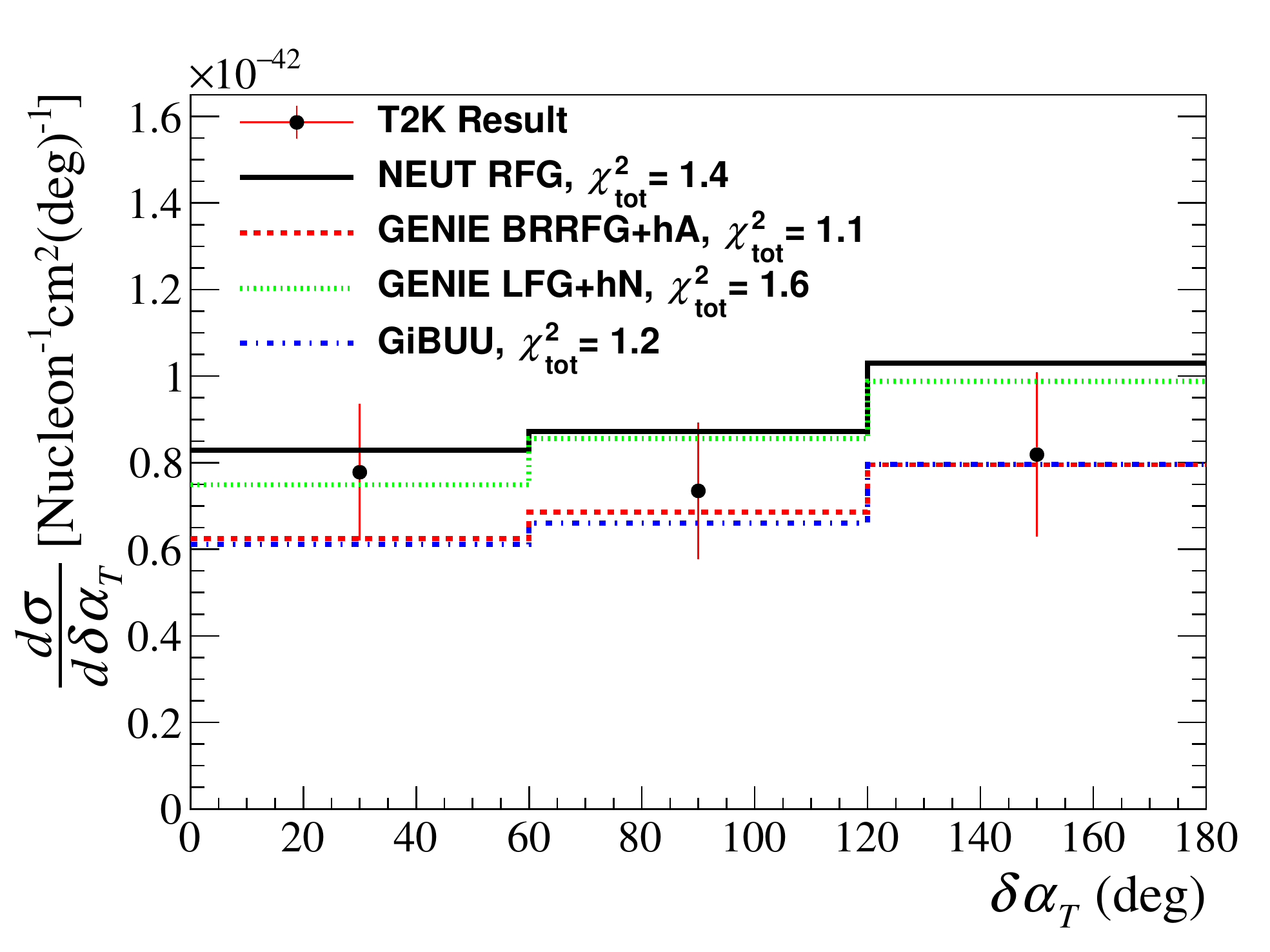}}
    \subfloat{\includegraphics[width=0.49\linewidth]{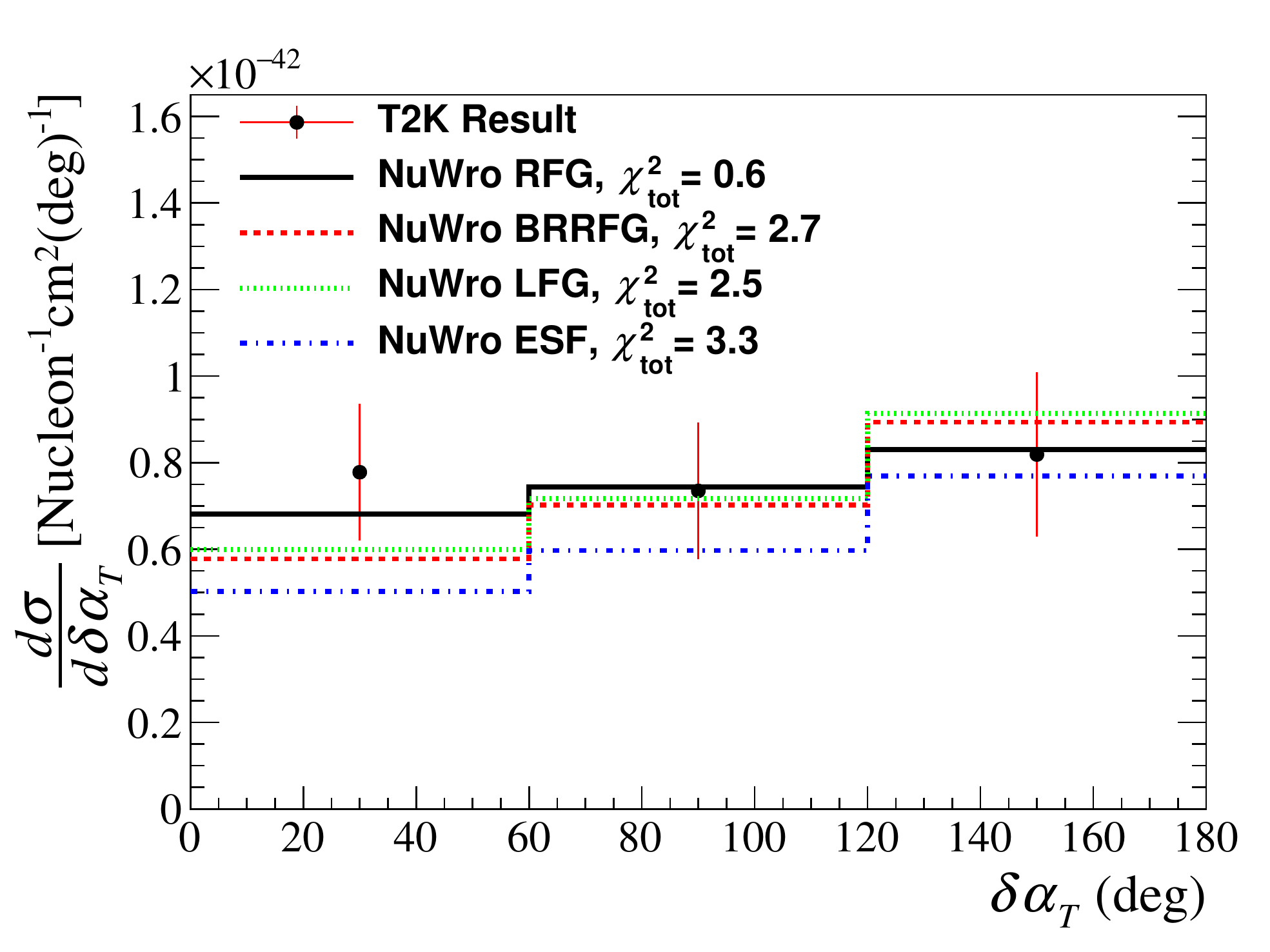}}\hfill
\caption{Measured differential cross sections per nucleon as a function of $\delta p_{TT}$ (top), $p_N$ (middle) and $\delta\alpha_{T}$ (bottom), together with predictions from NEUT, GENIE, GiBUU (left) and NuWro (right). In the tails of $\delta p_{TT}$ and $p_N$ (beyond the magenta lines), the cross sections are scaled by a factor of 5 for better visualization. The legend also shows the $\chi^2_\text{tot}$ from \cref{eq:chi2_tot}.} \label{fig:xsec_model}
\end{figure*}

\cref{fig:xsec_model} shows the comparisons between the measured cross sections and model predictions. The full $\chi^2_\text{tot}$ and shape only $\chi^2_\text{shape}$ are summarized in \cref{tab:chi2}. 
It is observed that $\chi^2_\text{shape}$ is usually much smaller than $\chi^2_\text{tot}$, implying a large part of the model separation power in this analysis comes from normalization differences.
\begin{table}[h!]
    \centering
	\caption{\label{tab:chi2}%
		$\chi^2_\text{tot}$ and $\chi^2_\text{shape}$ for the three TKI variable measurements. The ndof of $\chi^2_\text{tot}$ is equal to 5, 4, and 3 for $\delta p_{TT}$, $p_N$ and $\delta\alpha_{T}$ respectively. The ndof of $\chi^2_\text{shape}$ is one less than that of $\chi^2_\text{tot}$. 
	}
		\begin{tabular}{l@{\hskip 8pt}c@{\hskip 8pt}c@{\hskip 8pt}c}
		\hline\hline
		   & \multicolumn{3}{c}{$\chi^2_\text{tot}$
		   ($\chi^2_\text{shape}$)}\\
		   \hline
			Generator & $\delta p_{TT}$ & $p_N$ & $\delta\alpha_{T}$ \\
			\hline
			NEUT RFG & 11.3 (5.1) & 10.8 (2.7) & 1.4 (0.4) \\
			GENIE BRRFG+hA & 5.2 (4.8) & 2.9 (2.2) & 1.1 (0.5)\\
			GENIE LFG+hN & 8.6 (4.2) & 13.2 (2.7) & 1.6 (0.8)\\
			GiBUU & 3.6 (3.3) & 1.7 (1.3) & 1.2 (0.6)\\
			NuWro RFG & 7.5 (5.9) & 9.0 (5.4) & 0.6 (0.4) \\
			NuWro BRRFG & 4.9 (4.2) & 2.7 (1.6) & 2.7 (1.8) \\
			NuWro LFG & 8.5 (6.7) & 11.0 (4.9) & 2.5 (1.7) \\
			NuWro ESF & 5.2 (5.6) & 3.5 (3.0) & 3.3 (1.7) \\
			\hline\hline
		\end{tabular}
\end{table}	

\subsection{Discussion}
Amongst all the models compared, GiBUU shows marginally better agreement with data. Both $\chi^2_\text{tot}$ and $\chi^2_\text{shape}$ are smaller than the corresponding ndof for each TKI variable. 
It is explained in Ref. \cite{Mosel_2019} that GiBUU uses a density- and momentum-dependent mean field to model the nucleon-nucleus potential and prepare the nuclear ground state. Since the same potential is used in all interaction channels, 
it may provide a more accurate prediction than other generators which often treat QE and pion production processes differently. 
Also GiBUU's modelling of FSI in the transport theory is a more complete approach than the commonly used cascade models, which might be a contributing factor to the overall agreement. The nice shape agreement at low $p_N$ suggests that the nuclear ground state modelling in GiBUU is better than other generators. In the tail all models have similar predictions, meaning that we are not sensitive to the FSI differences there.

Within the NuWro models, ESF and BRRFG have better agreement than LFG and RFG. In the pion channel, these nuclear models affect properties like the removal energy and nucleon momentum distribution. This suggests that ESF and BRRFG may provide a more realistic nuclear ground state description. From the $p_N$ result, the characteristic nucleon momentum peak at the Fermi surface ($\sim$220 MeV/c) in RFG is strongly disfavored. On the contrary, LFG predicts a large number of events with $p_N<120$~MeV/c which is also incompatible with data. 

NEUT RFG, GENIE BRRFG and GENIE LFG use the same types of Fermi gas nuclear ground state models as in NuWro. The choices of pion production and FSI models make a difference in their predictions, but in general the same nuclear ground state model shows similar features across generators in the small imbalance regions of $\delta p_{TT}$ and $p_N$. This indicates these observables are a good probe of the nuclear ground state models. 

In general the model separation in $\delta p_{TT}$ and $p_N$ is better than that in $\delta\alpha_T$, with most of the sensitivity coming from the central bin of $\delta p_{TT}$ and the first two bins in $p_N$ where the imbalance is small. While $\delta\alpha_T$ is rather insensitive to the initial nuclear state, the hardening of $\delta\alpha_T$ towards 180$^\circ$ is strongly affected by FSI which usually slow down the final-state hadrons but not the lepton. 
However, with the present signal phase space restrictions, in particular the high proton momentum threshold of 450~MeV/c, many of the CC1$\pi^+$Xp events that undergo FSI are lost, making the measurement less effective. 
With improved detector acceptance in the coming ND280 upgrade~\cite{Abe:2019whr}, $\delta\alpha_T$ will be an extremely useful and independent probe of FSI. 

If only $\chi^2_\text{shape}$ is considered, most models have a $\chi^2_\text{shape}$ less than or roughly equal to the ndof. The worst case is the $p_N$ prediction from NuWro RFG which has a p-value of 15\%. Nevertheless, the large normalization discrepancy exhibited by the RFG and LFG nuclear ground state models cannot be simply explained by flux or other normalization uncertainties. Thus one should be careful in interpreting the model agreement when the difference between $\chi^2_\text{shape}$ and $\chi^2_\text{tot}$ is large.

\begin{figure}[h!]
\centering
    \subfloat{\includegraphics[width=\linewidth]{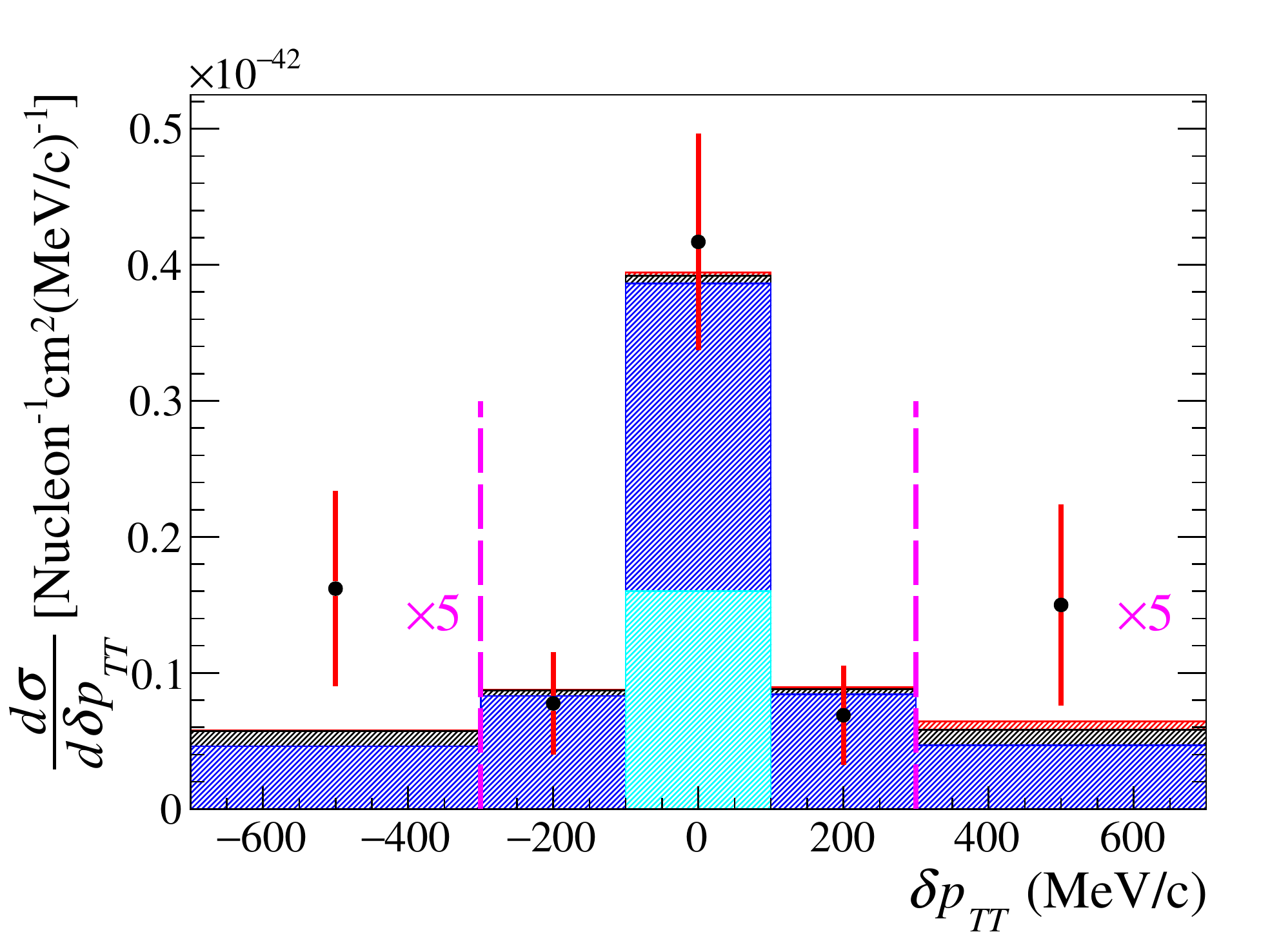}}\hfill
    \subfloat{\includegraphics[width=\linewidth]{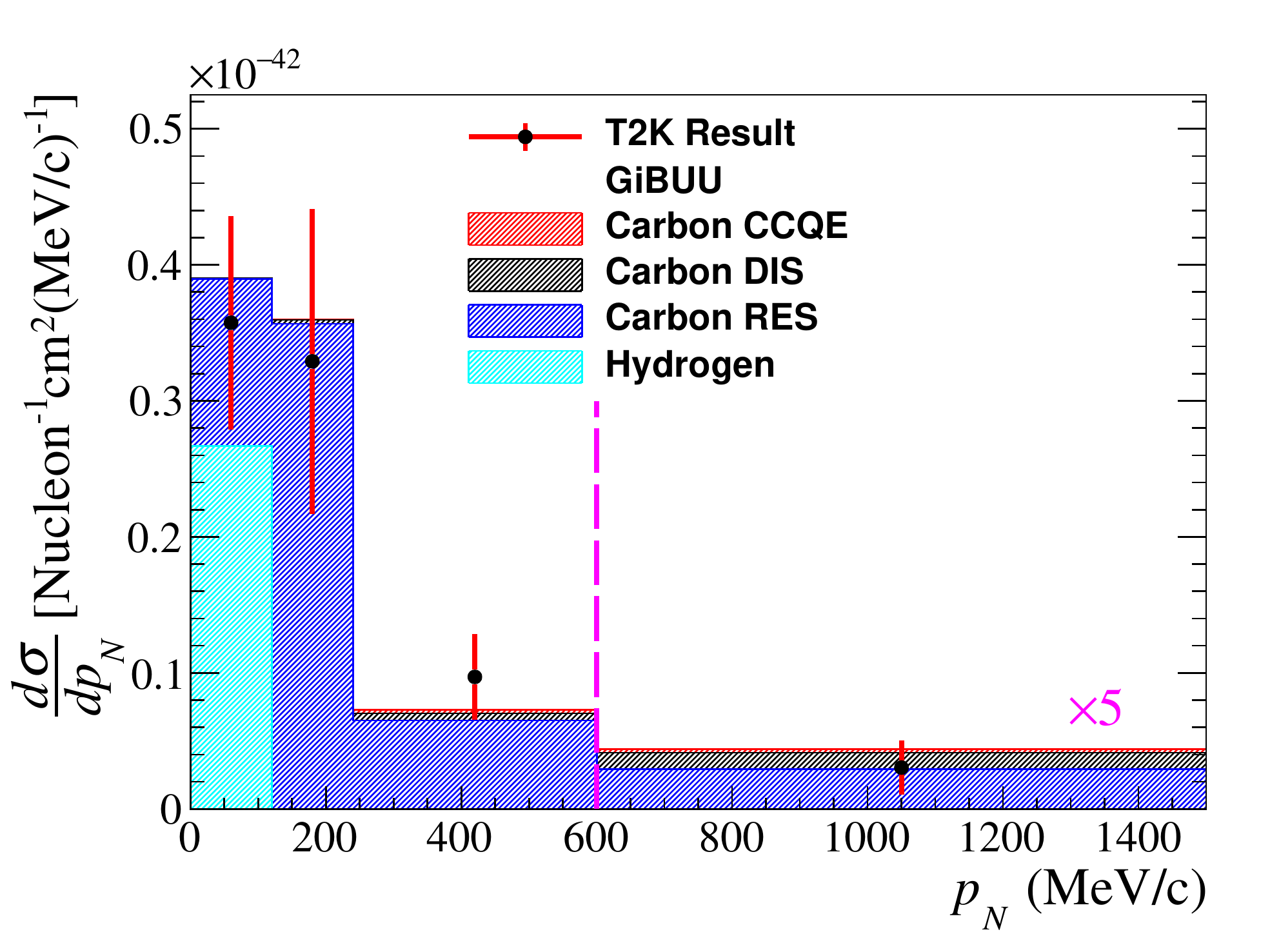}}\hfill
     \subfloat{\includegraphics[width=\linewidth]{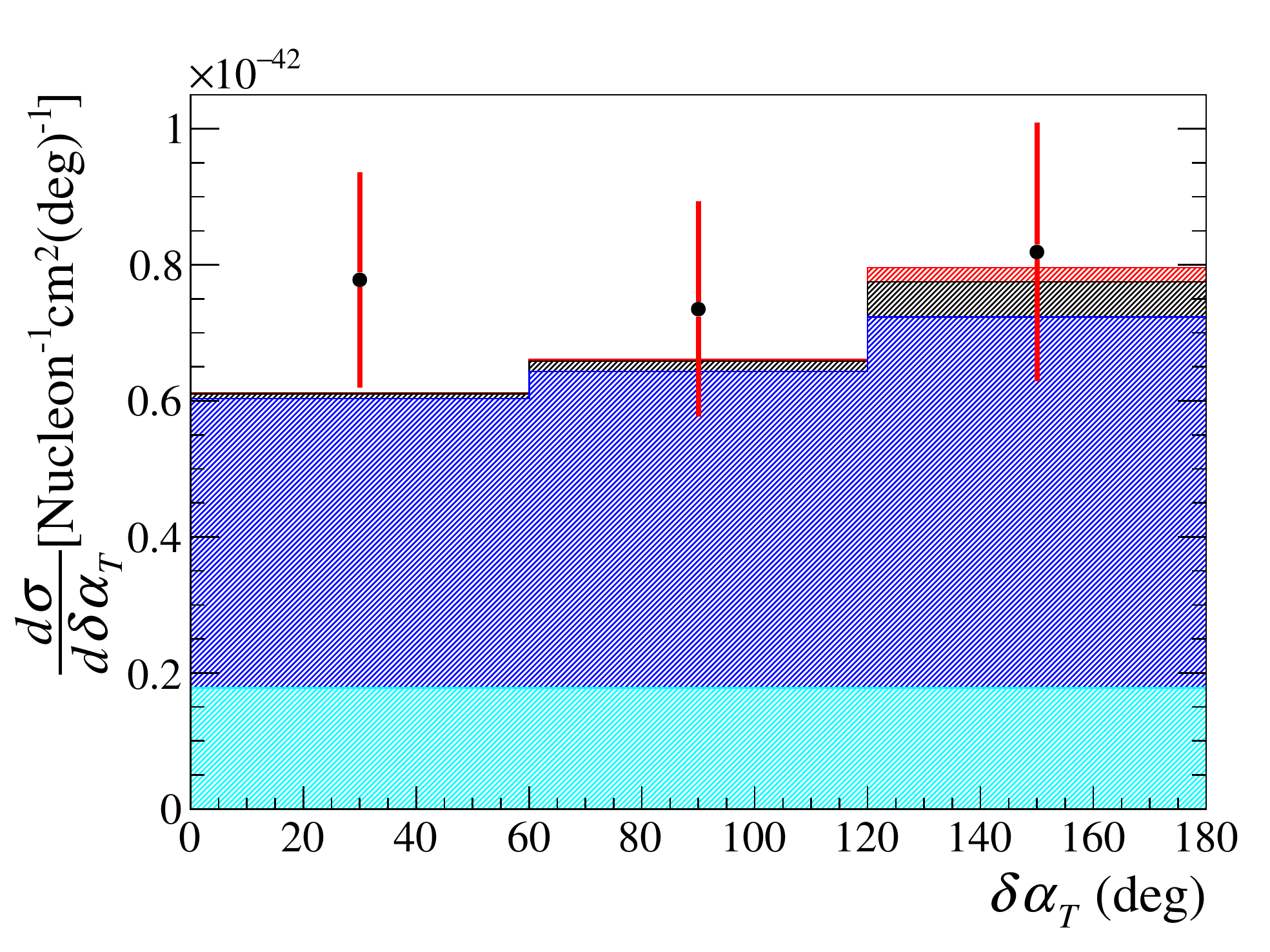}}
\caption{Measured cross sections as a function of the TKI variables compared to GiBUU predictions. The GiBUU predictions are decomposed into the contributions from carbon and hydrogen. In the tails of $\delta p_{TT}$ and $p_N$ (beyond the magenta lines), the cross sections are scaled by a factor of 5 for better visualization. } \label{fig:xsec_gibuu}
\end{figure}

It is not straight-forward to compare this study to the T2K CC0$\pi$~\cite{Abe:2018pwo} and MINER$\nu$A~\cite{Lu:2018stk,Coplowe:2020yea} TKI results, because of the different signal definition and, more importantly, a significant contribution from free nucleon targets (hydrogen) in this measurement. For example, \cref{fig:xsec_gibuu} shows the interaction target and channel breakdown of the GiBUU prediction. As explained in \cref{sec:signal_def}, the hydrogen cross section is a Dirac delta function at $\delta p_{TT}=0$ and $p_N\approx 26$ MeV/c, and is flat in $\delta\alpha_{T}$. The cross section on hydrogen is related to the carbon component via the common neutrino-nucleon cross section modelling; both components will scale similarly when the neutrino-nucleon cross section is changed. However the ratio between the hydrogen and carbon components is highly dependent on the modelling of the nuclear medium effects.

Qualitatively, almost all models are compatible with the $p_N$ tail in both T2K and MINER$\nu$A data, but have an over-prediction in the peak region. However, there are not sufficient statistics to measure the peak of $p_N$ more precisely. MINER$\nu$A also reported a mild asymmetry in $\delta p_{TT}$, and attributed it to the interference between $\Delta$ and non-resonant amplitudes~\cite{Cai:2019jzk}, but such an asymmetry is not observed within errors in this study. 
The tight phase space restrictions used in this study reduces our sensitivity to FSI modelling. 
The rather flat distribution of $\delta\alpha_{T}$ compared to MINER$\nu$A results can be attributed to the difference in phase space restrictions, where MINER$\nu$A applied no phase space restriction on the final-state $\pi^0$. The more energetic ($\sim$3~GeV) neutrino beam of MINER$\nu$A also produces more energetic final-state particles and a more curved $\delta\alpha_{T}$. 

While GiBUU has a good agreement with this CC1$\pi^+$Xp and MINER$\nu$A CC$\pi^0$ measurements, it shows an incompatibility with our CC0$\pi$ TKI results~\cite{Abe:2018pwo,Dolan:2018sbb}. This incompatibility is not in the $\delta p_{T}$ (\cref{eq:dpt}) tail or normalisation, suggesting this might be related to the nuclear ground state. 
In our previous CC0$\pi$ cross section measurements as a function of outgoing muon kinematics~\cite{PhysRevD.101.112004,PhysRevD.101.112001}, the GiBUU prediction also shows a large discrepancy, mainly in the most forward bin where the nuclear physics governing low energy and momentum transfer
interactions is the most important.

\section{Conclusion}\label{sec:conclusion}

In this paper, the CC1$\pi^+$Xp muon neutrino differential cross sections on hydrocarbon as a function of the three TKI variables, $\delta p_{TT}$, $p_N$ and $\delta\alpha_T$, have been measured independently in the ND280 tracker. $\delta p_{TT}$ and $p_N$ are most sensitive to the initial nuclear ground state, and $\delta\alpha_T$ is an independent probe of FSI. The analysis is performed with a joint fit between the signal and control samples to minimize the uncertainties on background estimation, and a maximum likelihood fit is used to unfold the detector smearing effect and extract cross sections in the truth space. The reduced flux uncertainty and better detector modelling allow to have a reduced systematic uncertainty with respect to previous T2K cross section analyses. Due to the complex and multifaceted nature of this analysis, exceptional care has been taken in mitigating sources of potential model bias in the extracted results.

An extensive comparison of the extracted results to state-of-the-art neutrino interaction models shows a slight preference for GiBUU, which uses a more realistic nuclear ground state to handle all interaction channels consistently.  
Our results are statistically limited and a large part of the model separation power comes from normalization differences. In general the simple Fermi gas models (RFG and LFG) show a large disagreement in $p_N$ with $\chi^2_\text{tot}$/ndof~$>2$, which indicates a mis-modelling of the nucleon Fermi motion. The similar data-MC comparison to the MINER$\nu$A CC$\pi^0$ results~\cite{Coplowe:2020yea} seems to confirm that the mis-modelling is general in pion production channels. While the tight phase space restrictions limit our sensitivity to FSI modelling, the relatively flat $\delta\alpha_T$ in T2K results is in strong contrast to MINER$\nu$A results, indicating a possible energy dependence of hadronic FSI. 

Future analyses will aim to unfold cross sections in multiple TKI variables simultaneously and obtain their correlations which can then be used to separate effects due to the initial nuclear state and FSI. 
The upcoming ND280 upgrade is going to expand the measurable phase space, especially in the low energy and high angle regions. Thus the ND280 upgrade is expected to increase our statistics and model sensitivity significantly. Another possible extension is to isolate hydrogen interactions from carbon ones by selecting events with small $\delta p_{TT}$ and $p_N$. With better detector resolution, this technique could better identify and separate interactions on hydrogen on an event-by-event basis, and provide the first ``free nucleon data" since the hydrogen bubble chamber experiments~\cite{Lu:2015hea, Duyang:2018lpe, Duyang:2019prb,  Hamacher-Baumann:2020ogq, Munteanu:2019llq}.

The data release for the results presented in this analysis is posted in Ref.~\cite{data}. 
It contains the analysis binning, the differential cross section best-fit values, and associated covariance matrices.

\section{Acknowledgement}

We thank the J-PARC staff for superb accelerator performance. We thank the CERN NA61/SHINE Collaboration for providing valuable particle production data. We acknowledge the support of MEXT, JSPS KAKENHI (JP16H06288, JP18K03682, JP18H03701, JP18H05537, JP19J01119, JP19J22440, JP19J22258, JP20H00162, JP20H00149, JP20J20304) and bilateral programs (JPJSBP120204806, JPJSBP120209601), Japan; NSERC, the NRC, and CFI, Canada; the CEA and CNRS/IN2P3, France; the DFG (RO 3625/2), Germany; the INFN, Italy; the Ministry of Education and Science(DIR/WK/2017/05) and the National Science Centre (UMO-2018/30/E/ST2/00441), Poland; the RSF (19-12-00325), RFBR (JSPS-RFBR 20-52-50010$\setminus$20) and the Ministry of Science and Higher Education (075-15-2020-778), Russia; MICINN (SEV-2016-0588, PID2019-107564GB-I00, PGC2018-099388-BI00) and ERDF funds and CERCA program, Spain; the SNSF and SERI (200021\_185012, 200020\_188533, 20FL21\_186178I), Switzerland; the STFC, UK; and the DOE, USA. We also thank CERN for the UA1/NOMAD magnet, DESY for the HERA-B magnet mover system, NII for SINET5, the WestGrid and SciNet consortia in Compute Canada, and GridPP in the United Kingdom. In addition, the participation of individual researchers and institutions has been further supported by funds from the ERC (FP7), "la Caixa" Foundation (ID 100010434, fellowship code LCF/BQ/IN17/11620050), the European Union's Horizon 2020 Research and Innovation Programme under the Marie Sklodowska-Curie grant agreement numbers 713673 and 754496, and H2020 grant numbers RISE-GA822070-JENNIFER2 2020 and RISE-GA872549-SK2HK; the JSPS, Japan; the Royal Society, UK; French ANR grant number ANR-19-CE31-0001; the DOE Early Career programme, USA; and RFBR, project number 20-32-70196.\\


\appendix
\section{More model comparisons}
This section shows a few more model comparisons to data with different physics configurations in the neutrino generators.

\subsection{GENIE}
\begin{figure*}
\centering
    \subfloat{\includegraphics[width=0.49\linewidth]{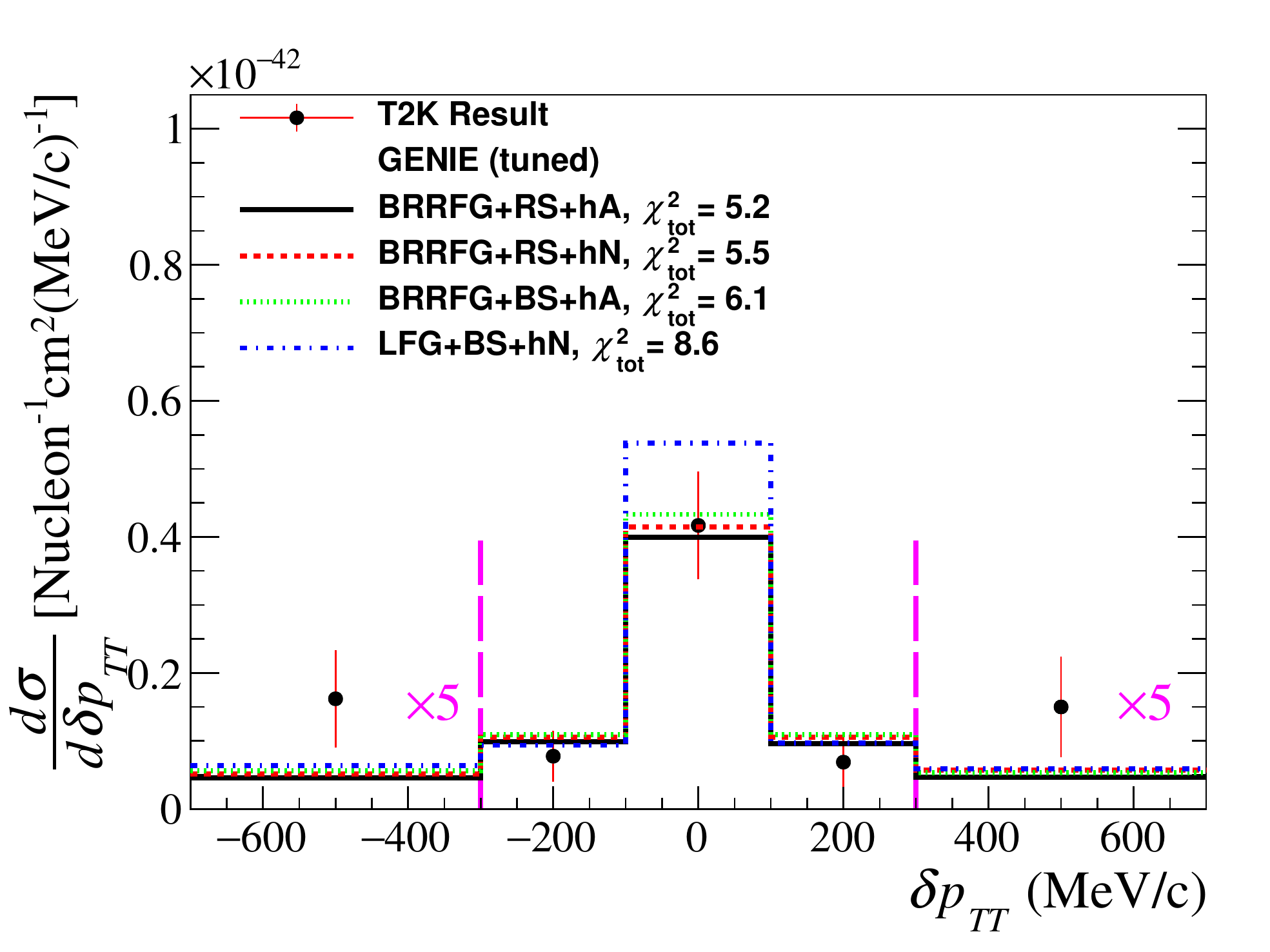}}
    \subfloat{\includegraphics[width=0.49\linewidth]{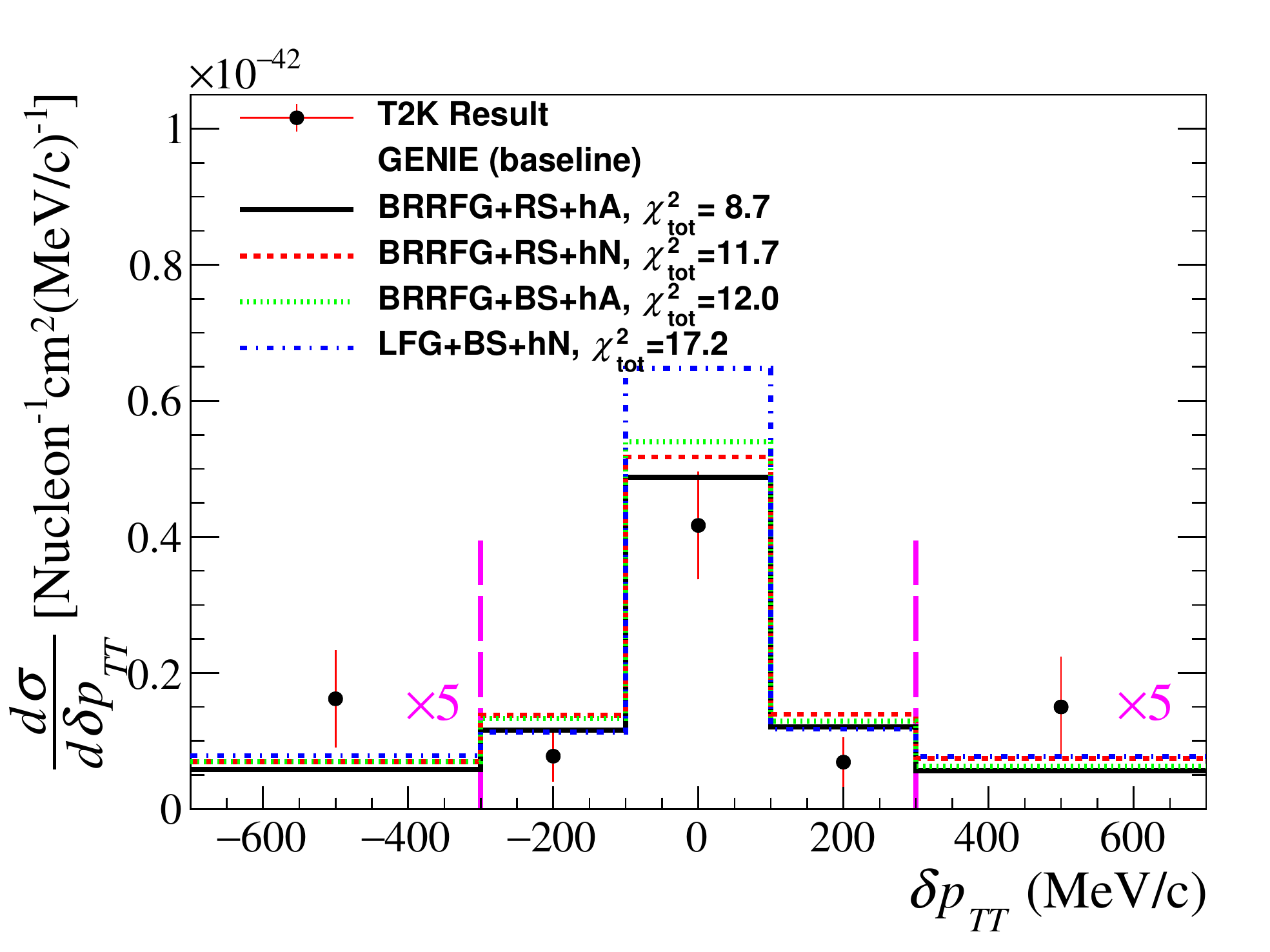}}\hfill
    \subfloat{\includegraphics[width=0.49\linewidth]{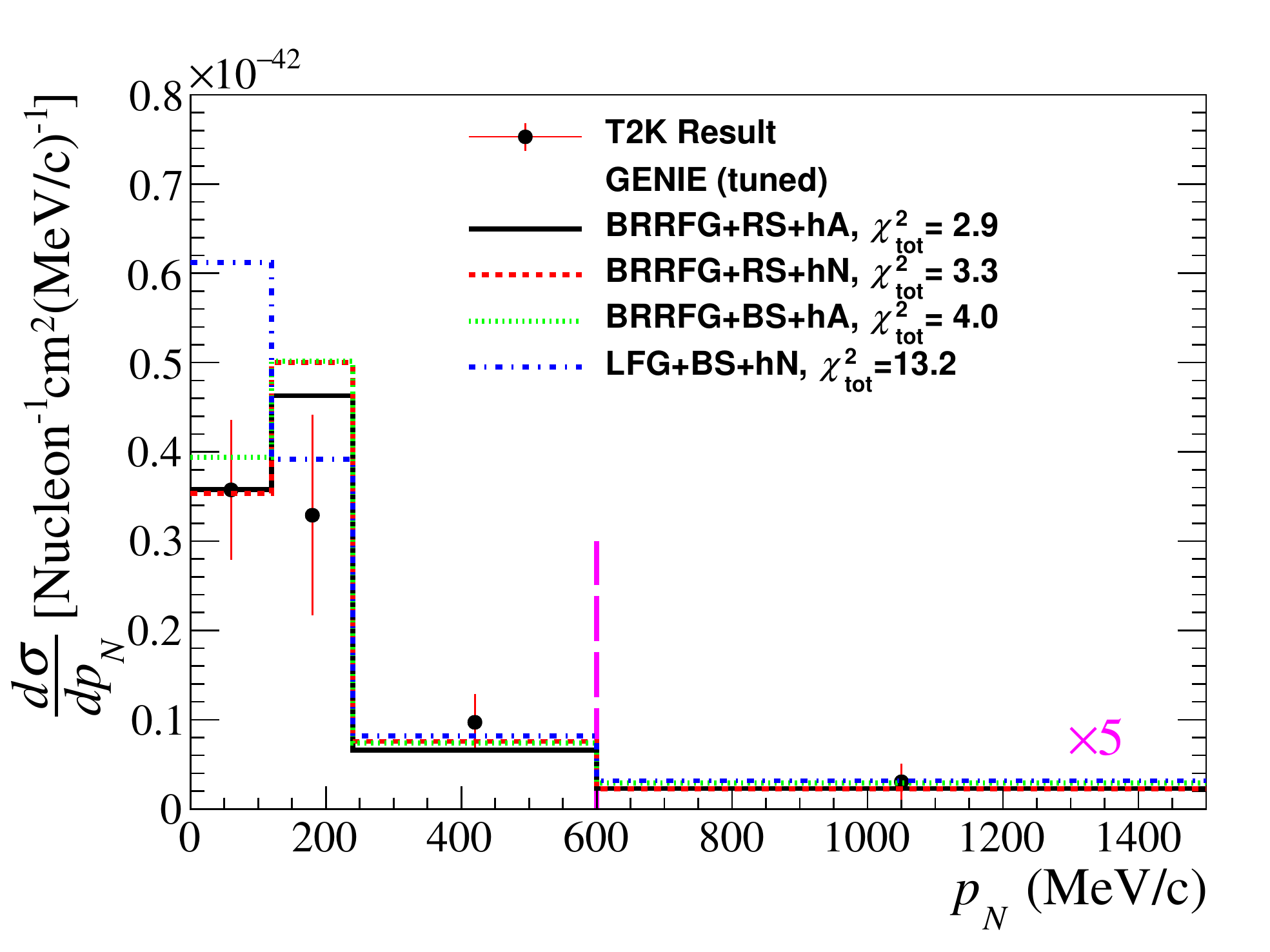}}
    \subfloat{\includegraphics[width=0.49\linewidth]{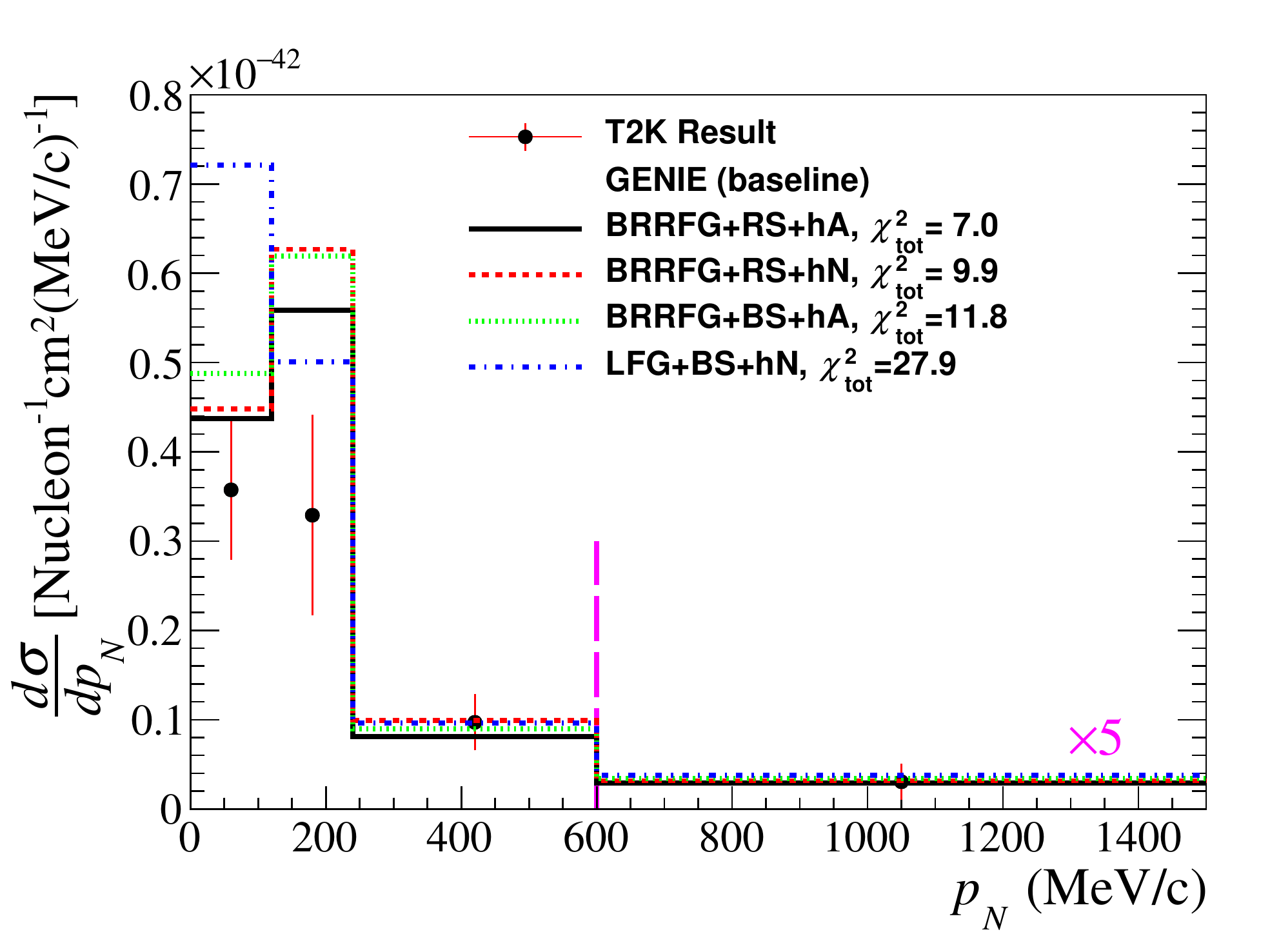}}\hfill
    \subfloat{\includegraphics[width=0.49\linewidth]{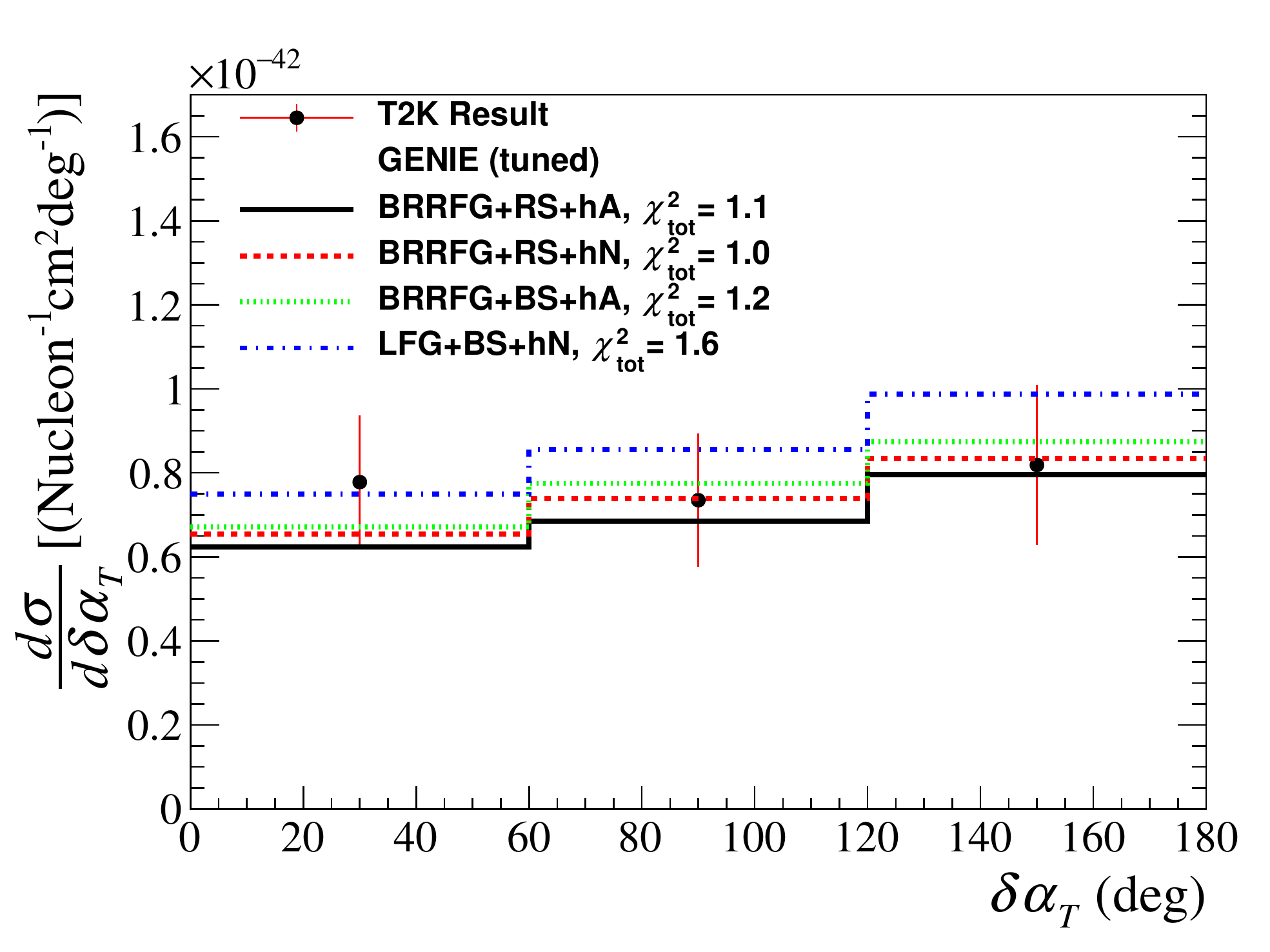}}
    \subfloat{\includegraphics[width=0.49\linewidth]{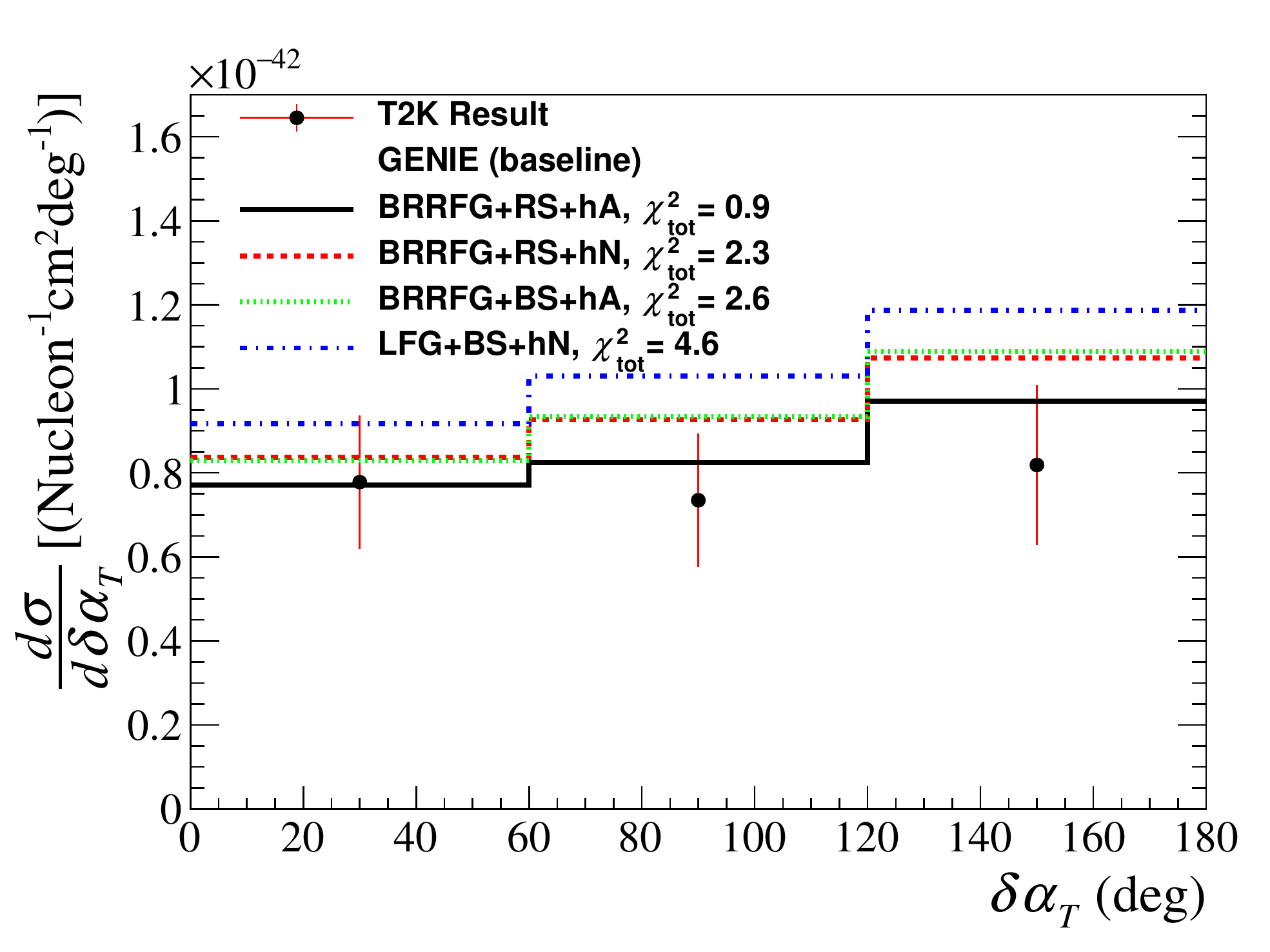}}\hfill
\caption{Measured differential cross sections per nucleon as a function of $\delta p_{TT}$ (top), $p_N$ (middle) and $\delta\alpha_{T}$ (bottom), together with predictions from the different model configurations of GENIE. The left plots use the free nucleon cross section re-tune, and the right plots use the baseline tune. In the tails of $\delta p_{TT}$ and $p_N$ (beyond the magenta lines), the cross sections are scaled by a factor of 5 for better visualization. The legend also shows the $\chi^2_\text{tot}$ from \cref{eq:chi2_tot}.} \label{fig:xsec_genie}
\end{figure*}

GENIE provides a variety of model configurations for event generation. Choices in the nuclear ground state model have a much larger effect on the TKI predictions than either the FSI models or pion production models. 
On the other hand, the GENIE 2018a free nucleon cross section model re-tune reduces the CC1$\pi$ cross sections and increases the CC2$\pi$ cross sections relative to the baseline tune. \cref{fig:xsec_genie} shows the comparison amongst these model configurations and physics tunes.

\subsection{NuWro}
\begin{figure*}
\centering
    \subfloat{\includegraphics[width=0.49\linewidth]{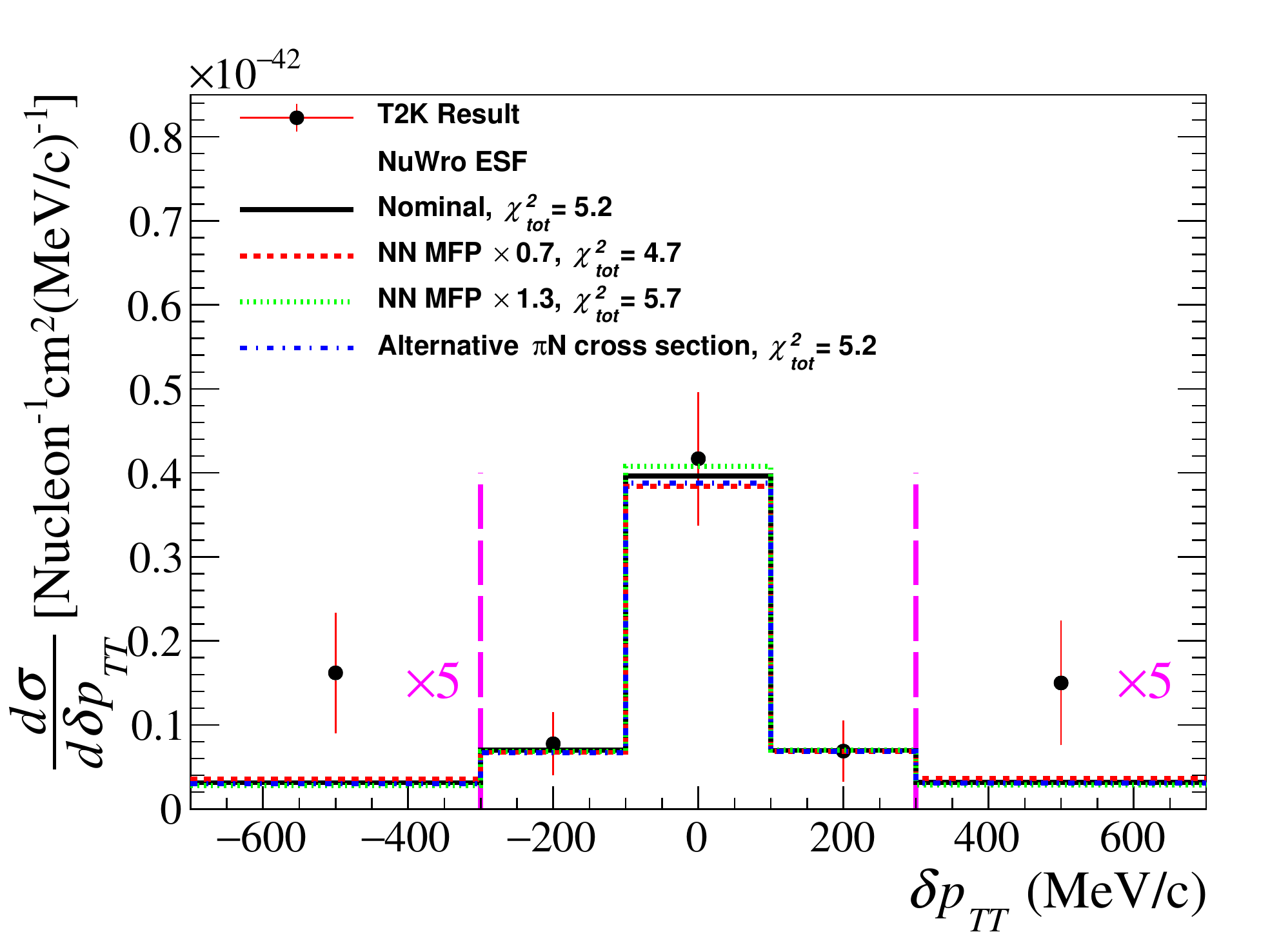}}
    \subfloat{\includegraphics[width=0.49\linewidth]{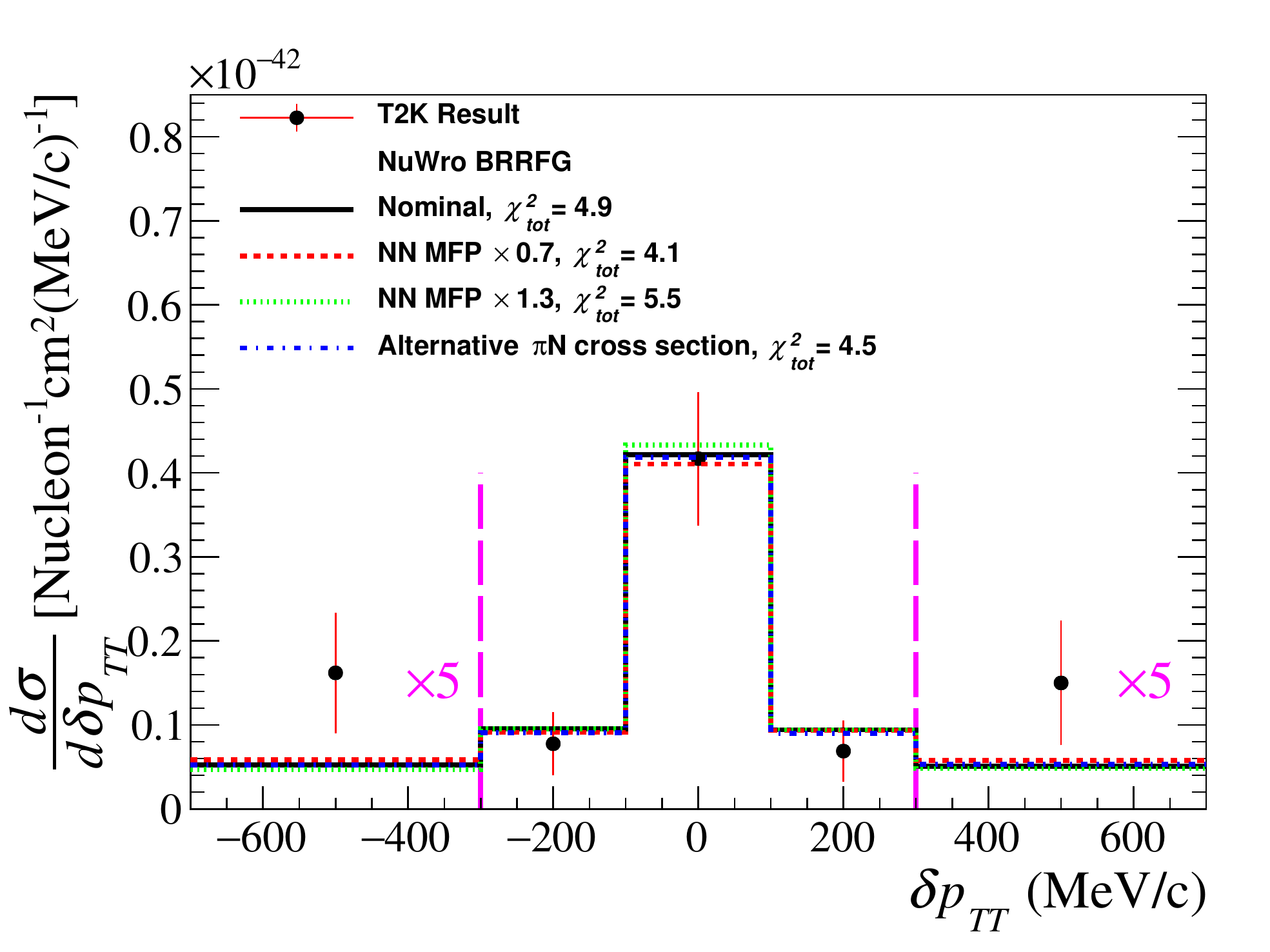}}\hfill
    \subfloat{\includegraphics[width=0.49\linewidth]{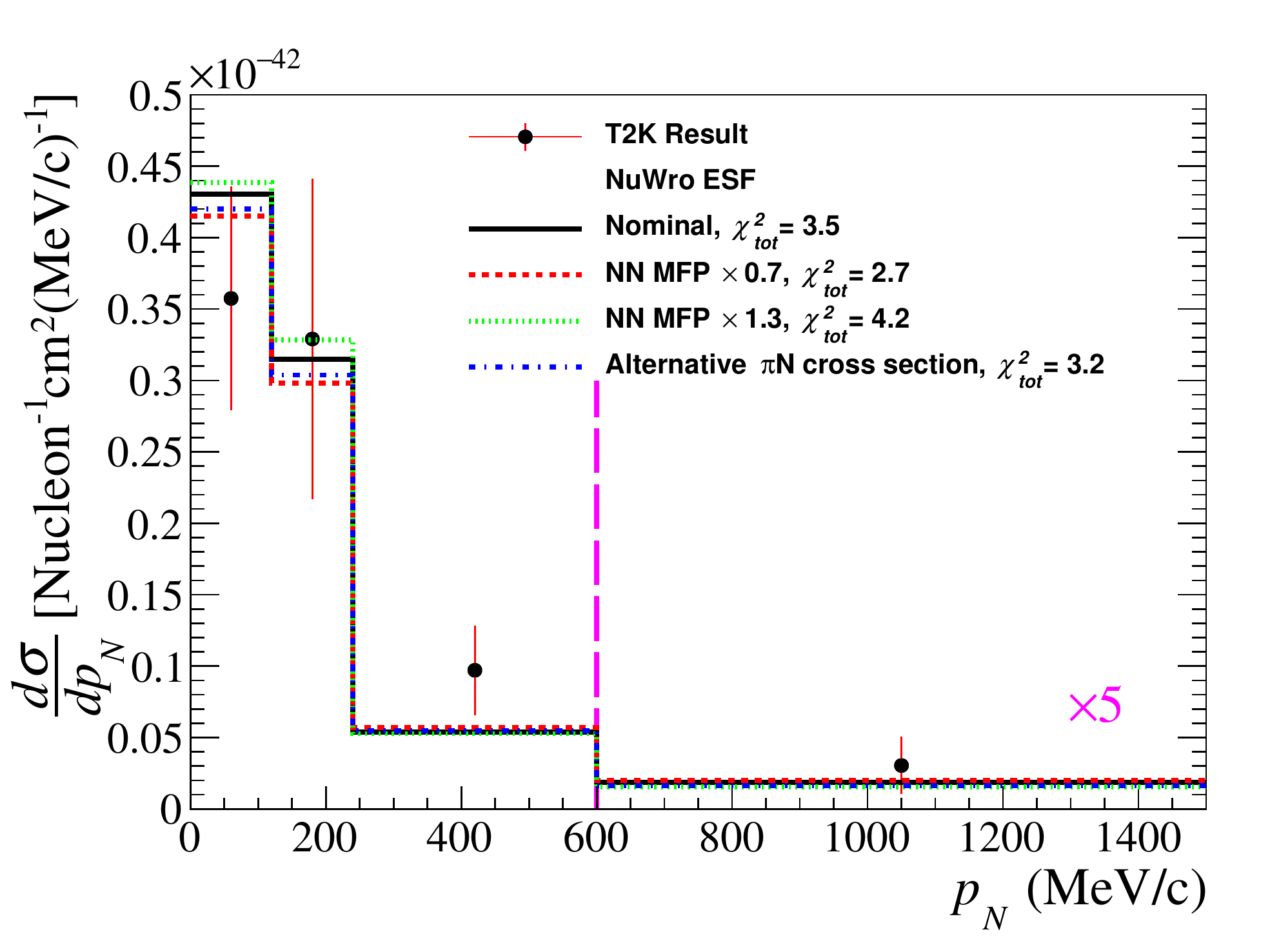}}
    \subfloat{\includegraphics[width=0.49\linewidth]{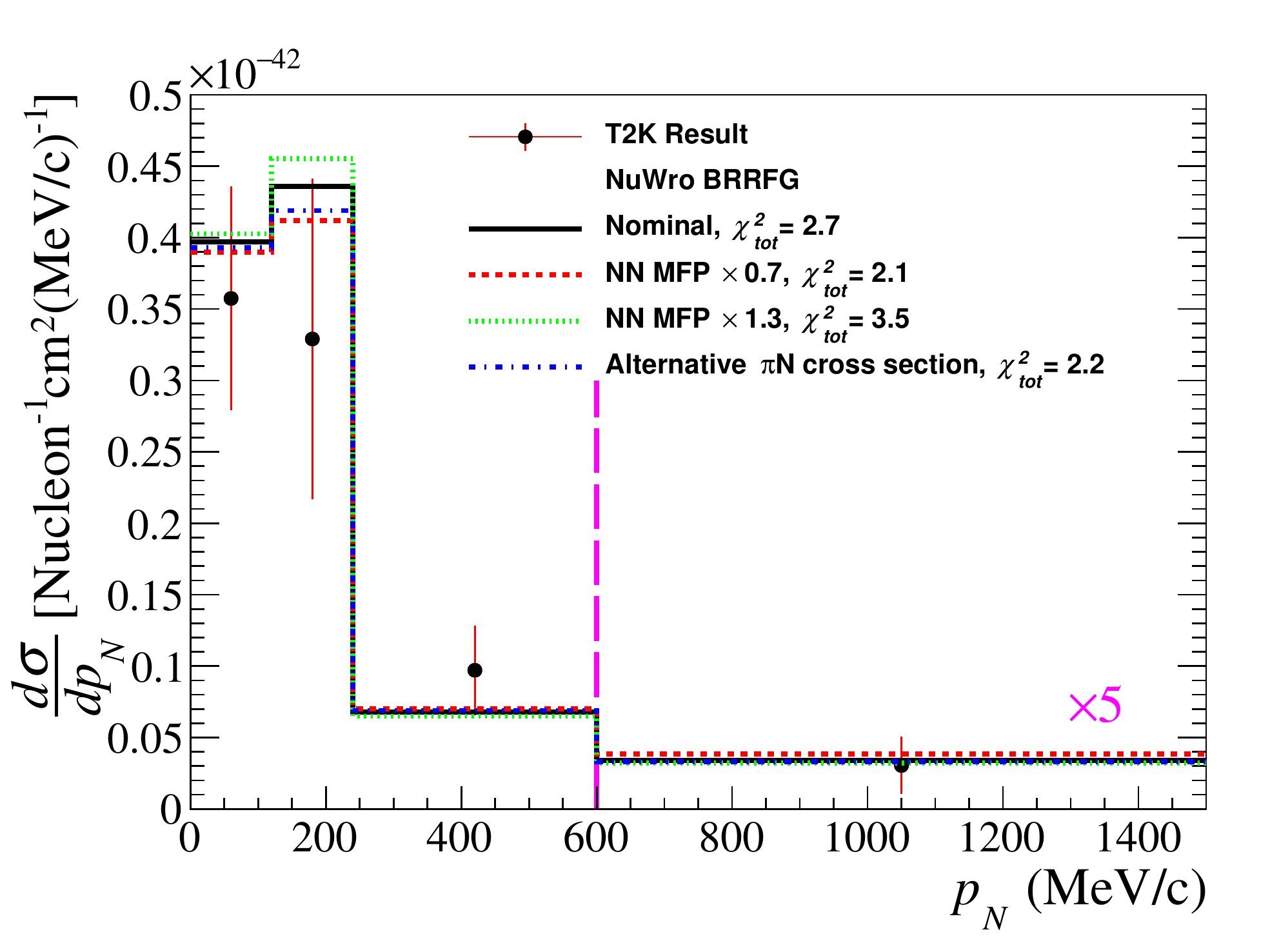}}\hfill
    \subfloat{\includegraphics[width=0.49\linewidth]{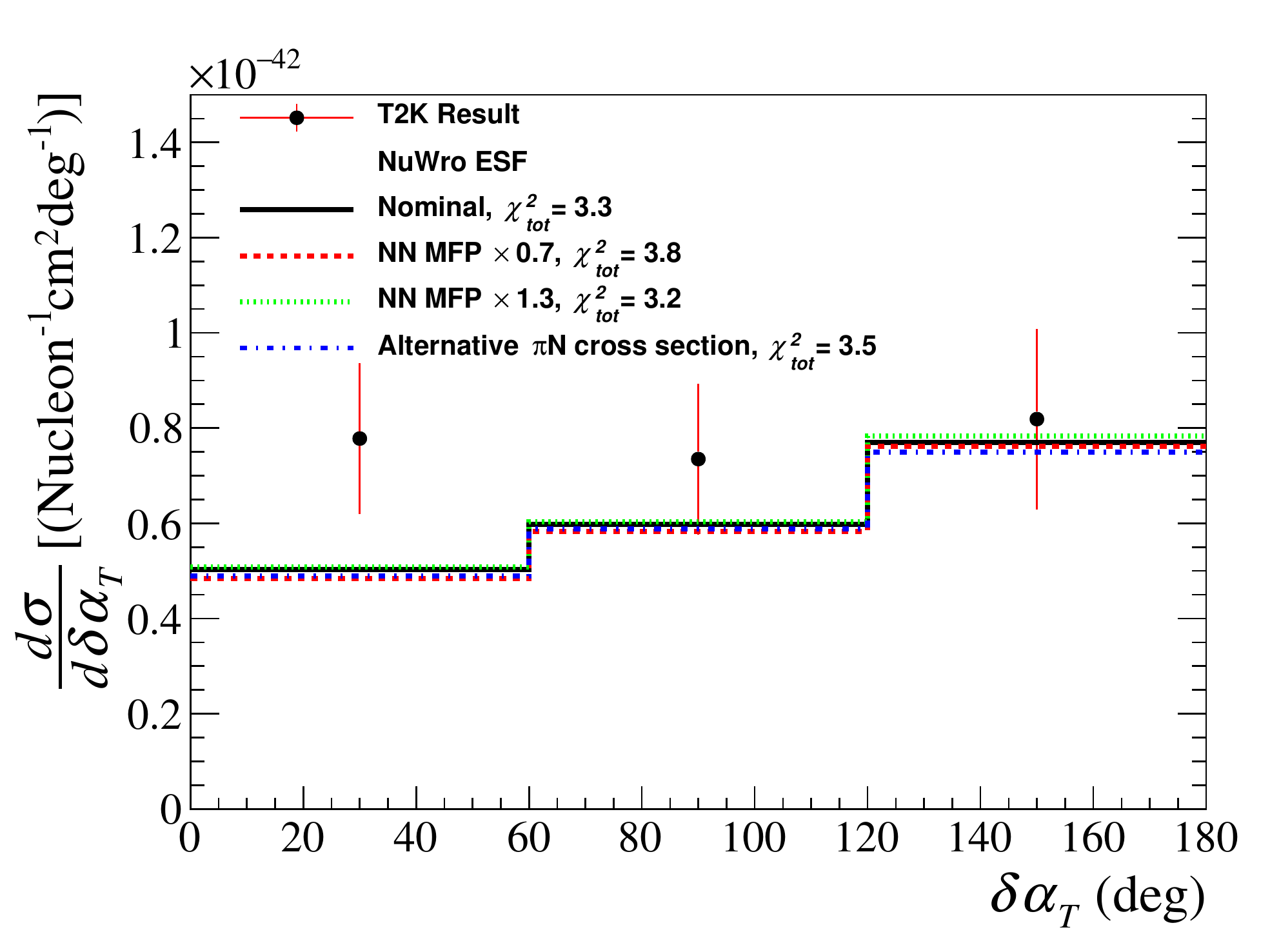}}
    \subfloat{\includegraphics[width=0.49\linewidth]{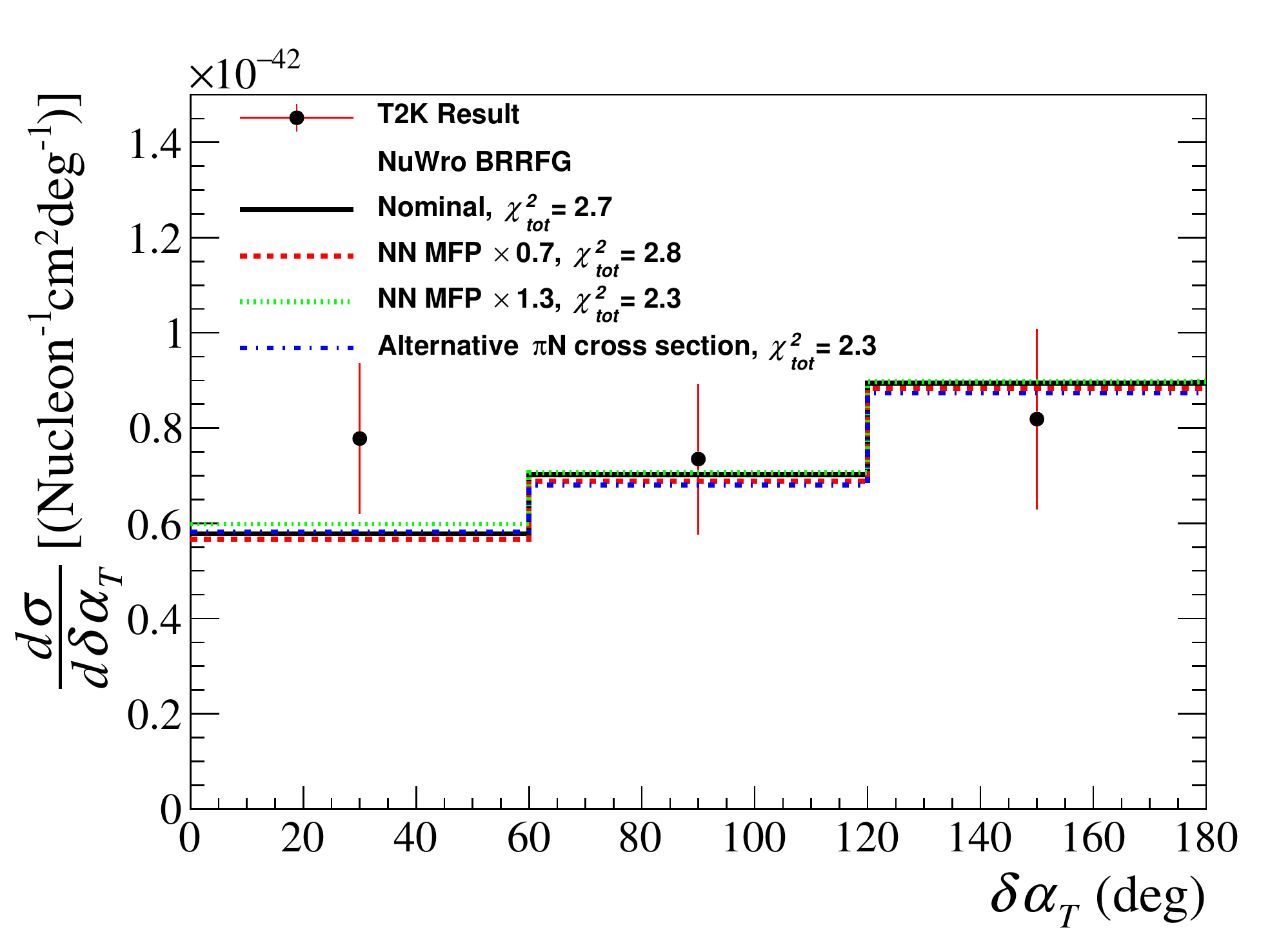}}\hfill
\caption{Measured differential cross sections per nucleon as a function of $\delta p_{TT}$ (top), $p_N$ (middle) and $\delta\alpha_{T}$ (bottom), together with predictions from the NuWro ESF (left) and BRRFG (right) models. The black solid line shows the prediction from the nominal FSI configuration, while other colors show that from a different nucleon mean free path (NN MFP) or pion-nucleon ($\pi$N) interaction model configuration. In the tails of $\delta p_{TT}$ and $p_N$ (beyond the magenta lines), the cross sections are scaled by a factor of 5 for better visualization. The legend also shows the $\chi^2_\text{tot}$ from \cref{eq:chi2_tot}.} \label{fig:xsec_nuwro}
\end{figure*}

Within NuWro, the BRRFG and ESF nuclear ground state models show the best agreement with data. The FSI configurations are varied to study their effects on the predictions. These include a global scaling of the nucleon mean free path in the cascade, or the switch of the pion-nucleon interaction model from Ref.~\cite{SALCEDO1988557} to Ref.~\cite{PhysRev.110.185}. As shown in \cref{fig:xsec_nuwro}, the change in $\chi^2_\text{tot}$ is small, indicating that there is limited sensitivity to FSI under current statistics and signal phase space restrictions.

\bibliography{ms}

\end{document}